\documentclass[acmsmall]{acmart} 

\AtBeginDocument{%
	\providecommand\BibTeX{{%
			\normalfont B\kern-0.5em{\scshape i\kern-0.25em b}\kern-0.8em\TeX}}}

\usepackage{comment}
\usepackage{graphicx}
\usepackage{amsmath}
\usepackage[american]{babel}
\usepackage{microtype}
\usepackage{subcaption}
\usepackage[ruled,,linesnumbered ]{algorithm2e}
\usepackage{tcolorbox}
\usepackage{balance} 
\usepackage{enumitem} 
\setitemize{noitemsep,topsep=0pt,parsep=0pt,partopsep=0pt,leftmargin=*}
\usepackage{multicol}
\usepackage{multirow}
\usepackage{color}
\usepackage{longtable}
\usepackage{float}
\usepackage{pifont}

\newcommand{\tool}{\textsc{NaturalCC \xspace}}%
\newcommand{\toolnospace}{\textsc{NaturalCC}}%
\newcommand{\numofall}{269\xspace}

\newcommand{\numoftasks}{4\xspace}
\newcommand{\numofmodels}{13\xspace}
\newcommand{\numofvenues}{32\xspace}

\newcommand{\forParaphrase}[2]{#1}

\usepackage{natbib}
\usepackage{multibib} 
\newcites{secondary}{APPENDIX REFERENCES}

\usepackage{tikz}
\usetikzlibrary{arrows,shapes,positioning,shadows,trees}

\usepackage{varwidth}

\newcount\DraftStatus  
\definecolor{darkgreen}{rgb}{0,0.5,0} 
\definecolor{purple}{rgb}{1,0,1} 
\definecolor{todocolor}{rgb}{0.9,0.1,0.1} 
\definecolor{fixcolor}{rgb}{0.1,0.7,0.3} 
\definecolor{wycolor}{rgb}{0.9,0.1,0.1} 
\definecolor{hycolor}{rgb}{0.7,0.7,0.3} 

\newcommand{\nbc}[3]{\ifnum\DraftStatus=1
	{\colorbox{#3}{\bfseries\sffamily\scriptsize\textcolor{white}{#1}}}
	{\textcolor{#3}{\sf\small$\blacktriangleright$\emph{#2}$\blacktriangleleft$}}
	\fi}

\newcommand{\draftnote}[2]{\ifnum\DraftStatus=1
	\marginpar{
		\tiny\raggedright
		\hbadness=10000
		\def\baselinestretch{0.8}
		\textcolor{#1}{\textsf{\hspace{0pt}#2}}}
	\fi}

\renewcommand\footnotetextcopyrightpermission[1]{} 

\begin{document}

%
%
\title{Deep Learning for Code Intelligence: Survey, Benchmark and Toolkit}

\author{Yao Wan}
\email{wanyao@hust.edu.cn}
\affiliation{%
  \department{National Engineering Research Center for Big Data Technology and System, Services Computing Technology and System Lab, Cluster and Grid Computing Lab, School of Computer Science and Technology}
	\institution{Huazhong University of Science and Technology}
	\city{Wuhan}
	\country{China}
}
\author{Yang He}
\affiliation{%
    \institution{Simon Fraser University}
    \city{Vancouver}
    \country{Canada}
 }
\email{yha244@sfu.ca}
\author{Zhangqian Bi}
\affiliation{%
  \department{School of Computer Science and Technology}
	\institution{Huazhong University of Science and Technology}
	\city{Wuhan}
	\country{China}
}
\email{zqbi@hust.edu.cn}
\author{Jianguo Zhang}
\affiliation{%
	\institution{Salesforce Research}
	\country{USA}
}
\email{jianguozhang@salesforce.com}
\author{Hongyu Zhang}
\affiliation{%
	\institution{Chongqing University}
	\country{China}}
\email{hyzhang@cqu.edu.cn}
\author{Yulei Sui}
\affiliation{%
	\institution{University of New South Wales}
	\country{Australia}}
\email{y.sui@unsw.edu.au}
\author{Guandong Xu}
\affiliation{%
	\institution{University of Technology Sydney}
	\country{Australia}}
\email{guandong.xu@uts.edu.au}
\author{Hai Jin}
\email{hjin@hust.edu.cn}
\affiliation{%
  \department{National Engineering Research Center for Big Data Technology and System, Services Computing Technology and System Lab, Cluster and Grid Computing Lab, School of Computer Science and Technology}
	\institution{Huazhong University of Science and Technology}
	\city{Wuhan}
	\country{China}
}

\author{Philip S. Yu}
\affiliation{%
	\institution{University of Illinois at Chicago}
	\city{Chicago}
	\country{USA}
}
\email{psyu@uic.edu}


\renewcommand{\shortauthors}{Wan et al.}

\begin{abstract}
Code intelligence leverages machine learning techniques to extract knowledge from extensive code corpora, with the aim of developing intelligent tools to improve the quality and productivity of computer programming.
Currently, there is already a thriving research community focusing on code intelligence, with  efforts ranging from software engineering, machine learning, data mining, natural language processing, and programming languages.
In this paper, we conduct a comprehensive literature review on deep learning for code intelligence, from the aspects of code representation learning, deep learning techniques, and application tasks.
We also benchmark several state-of-the-art neural models for code intelligence, and provide an open-source toolkit tailored for the rapid prototyping of deep-learning-based code intelligence models.
In particular, we inspect the existing code intelligence models under the basis of code representation learning, and provide a comprehensive overview to enhance comprehension of the present state of code intelligence.
Furthermore, we publicly release the source code and data resources to provide the community with a ready-to-use benchmark, which can facilitate the evaluation and comparison of existing and future code intelligence models  (\url{https://xcodemind.github.io}).
At last, we also point out several challenging and promising directions for future research.
\end{abstract}

\maketitle

\section{Introduction}\label{sec:introduction}
Software is eating the world~\cite{andreessen2011software}. With the advancement of Artificial Intelligence (AI), it is time to expand that maxim: software ate the world, and AI is eating the software.
As the software is primarily composed of code, we define the emerging concept of \textit{code intelligence} as the application of machine learning techniques to extract knowledge from large-scale
code repositories, with the aim of developing intelligent tools to improve the quality and productivity of computer programming~\cite{lu2021codexglue}. 
This concept is fueled by the ever-expanding reservoir of source code, often referred to as ``\textit{Big Code}''~\cite{allamanis2018survey}, which is harvested from platforms such as GitHub~\cite{github} and StackOverflow~\cite{stackoverflow}.
In this paper, our research scope is confined to code intelligence, with a particular focus on the application of deep learning techniques.

Achieving code intelligence necessitates a collaborative synergy in research across the domains of software engineering, machine learning, Natural Language Processing (NLP), programming language, and security. 
From our investigation, precise and reliable code representation learning or code embedding, which aims to efficiently and effectively encode the semantics of source code into distributed vector representations, is the foundation for code intelligence.
Such embedding vectors are then used in many downstream tasks, such as code completion~\cite{raychev2014code,svyatkovskiy2019pythia,kim2021code,liu2020self}, code search~\cite{gu2018deep,wan2019multi,husain2019codesearchnet}, code summarization~\cite{allamanis2016convolutional,iyer2016summarizing,wan2018improving,hu2018summarizing,zhang2020retrieval}, type inference~\cite{hellendoorn2018deep,wei2020lambdanet,pradel2020typewriter,allamanis2020typilus}, etc.
In terms of code embedding, notable advancements have been achieved by employing deep learning and NLP techniques to encode source code.

Analogous to word2vec~\cite{mikolov2013efficient} in NLP, \citet{alon2019code2vec} proposed code2vec, a distributed representation of code, based on a collection of paths extracted from the Abstract Syntax Tree (AST) of code.
In recent years, a multitude of neural networks tailored for specific tasks have been proposed and trained using supervised methods. 
As pre-trained language models (e.g., BERT~\cite{devlin2019bert} and GPT-3~\cite{brown2020language}) have been widely applied to NLP, many pre-trained language models for code have been proposed~\cite{kanade2020learning,feng2020codebert,guo2020graphcodebert} to better represent the semantics of code.
More recently, the emergence of Large Language Models (LLMs), exemplified by ChatGPT, has illuminated the pathway for further advancement of pre-trained language models, with a notable trend of increasing model sizes. 
This trend has extended to the domain of code intelligence, resulting in the development of various LLMs tailored for code, including but not limited to CodeT5~\cite{wang2023codet5+}, StarCoder~\cite{li2023starcoder}, and Code Llama~\cite{roziere2023code}.
In this paper, we examine code intelligence through the lenses of code representation learning, deep learning methods, and their applications.
\begin{table}[!t]
\centering
\small
\caption{
Comparison of the current work with previous survey efforts.
}
\vspace{-3mm}
\label{table_related_surveys}
\begin{tabular}{l|c|c|c|c|c}
\hline
\textbf{Paper}                                  & \textbf{Artifact}       & \textbf{Technique}                      & \textbf{Survey}                      & \textbf{Benchmark}                 & \textbf{Toolkit}                   \\ \hline
{\citet{allamanis2018survey}} & Software                  & Machine Learning               & \checkmark                  & $\times$                  & $\times$                  \\ \hline
\citet{watson2020systematic}           & \multirow{4}{*}{Software} & \multirow{4}{*}{Deep Learning} & \multirow{4}{*}{\checkmark} & \multirow{4}{*}{$\times$} & \multirow{4}{*}{$\times$} \\
\citet{wang2020synergy}                &                           &                                &                             &                           &                           \\
\citet{yang2021survey}                 &                           &                                &                             &                           &                           \\
{\citet{devanbu2020deep}}     &                           &                                &                             &                           &                           \\ \hline
{\citet{lu2021codexglue}}     & Software                  & Deep Learning                  & $\times$                    & \checkmark                & $\times$                  \\ \hline
Ours                                   & Code                      & Deep Learning                  & \checkmark                  & \checkmark                & \checkmark                \\ \hline
\end{tabular}
\vspace{-6mm}
\end{table}

\noindent{\textbf{Related Surveys and Differences.}}
Within our literature review, we identified several surveys related to ours. 
Notably, \citet{allamanis2018survey} conducted an exhaustive examination of machine learning approaches for modeling the naturalness of programming language.
They primarily emphasize machine learning algorithms, with a specific focus on probabilistic models, as opposed to those based on deep learning.
Recently, \citet{watson2020systematic}, \citet{wang2020synergy} and \citet{yang2021survey} 
conducted a thorough review of the literature on applications of deep learning in software engineering research.
They investigated mostly software engineering and artificial intelligence conferences and journals, focusing on various software engineering tasks (not limited to the source code) that are based on deep learning. 
\cite{devanbu2020deep} is a report that summarizes the current status of research on the subject of the intersection between deep learning and software engineering, as well as suggests several future directions.
In \cite{lu2021codexglue}, the authors established a benchmark dataset called CodeXGLUE for code representation and generation. 
In addition, several benchmark results especially based on pre-trained language models (i.e., CodeBERT) are presented. 

Table~\ref{table_related_surveys} summarizes the differences between our paper when compared with several related surveys in code intelligence.
In contrast to~\cite{allamanis2018survey} that focuses on traditional machine learning approaches, 
this paper places greater emphasis on leveraging deep learning techniques for code intelligence.
In contrast to~\cite{watson2020systematic},~\cite{wang2020synergy},~\cite{yang2021survey}, and \cite{devanbu2020deep} that cover various tasks in broad software engineering, 
our study narrows its focus to tasks associated with source code, examining them specifically from the perspective of deep learning.
In addition, we survey papers from various fields including software engineering, programming languages, machine learning, NLP, and security.
Note that, as code intelligence based on deep learning is an emerging and active research topic, we also include several high-quality unpublished papers that have been made available on arXiv.
This is because these unpublished works in arXiv can be seen as an indicator of future research.
Furthermore, existing surveys do not provide comprehensive benchmark evaluation results, nor do they develop an open-source toolkit to facilitate further research. 
This paper addresses this gap by presenting an open-source toolkit, referred to as \tool(standards for Natural Code Comprehension)~\cite{wan2022naturalcc}.
The toolkit is designed to streamline the prototyping of code intelligence models and to serve as a benchmarking platform for evaluating various state-of-the-art models.
In complement to CodeXGLUE~\cite{lu2021codexglue}, which aims to establish a benchmark dataset for code understanding and generation, particularly leveraging pre-trained code models, our focus lies in the construction of infrastructures that support diverse model implementations and provide users with the ability to conduct rapid prototyping.
Compared to CodeXGLUE, our toolkit contains a more extensive array of tools designed for the entire pipeline involved in constructing code intelligence models, offering heightened flexibility.

\noindent{\textbf{Our Contributions.}}
This paper is targeted at researchers and practitioners intrigued by the convergence of code intelligence and deep learning, with a specific emphasis on intelligent software engineering, NLP, and programming languages.
In this paper, we begin by providing a thorough review of existing research on deep learning for code intelligence. Subsequently, we advance our contribution by developing an open-source toolkit, referred to as \toolnospace, that incorporates state-of-the-art models across various downstream tasks.
Employing \toolnospace, we conduct a comprehensive performance benchmark of each model across \numoftasks downstream tasks, including code summarization, code search, code completion, and type inference.
The major contributions of this paper are summarized as follows.
\begin{itemize}
    \item 
    We conduct a comprehensive review on deep learning for code intelligence. 
    Specifically, we have collected \numofall papers from various top-tier venues and arXiv, covering multiple domains including software engineering, artificial intelligence, NLP, programming languages, and security.
    \item We benchmark the performance of \numofmodels leading models across four different tasks (i.e., code summarization, code search, code completion, and type inference). 
    All the resources, datasets and source code are publicly available.\footnote{\url{http://xcodemind.github.io}}
    \item We introduce \toolnospace, an open-source toolkit that has integrated many state-of-the-art baselines on different tasks, in order to facilitate research on code intelligence.
    Researchers in the fields of software engineering, NLP, and other fields can benefit from the toolkit for quick prototyping and replication.
\end{itemize}


\section{Survey Methodology}
\subsection{A Unified View from Code Representation Learning}\label{sec_encoder_decoder}
\begin{figure*}[t!]
	\centering
	\includegraphics[width=0.96\textwidth]{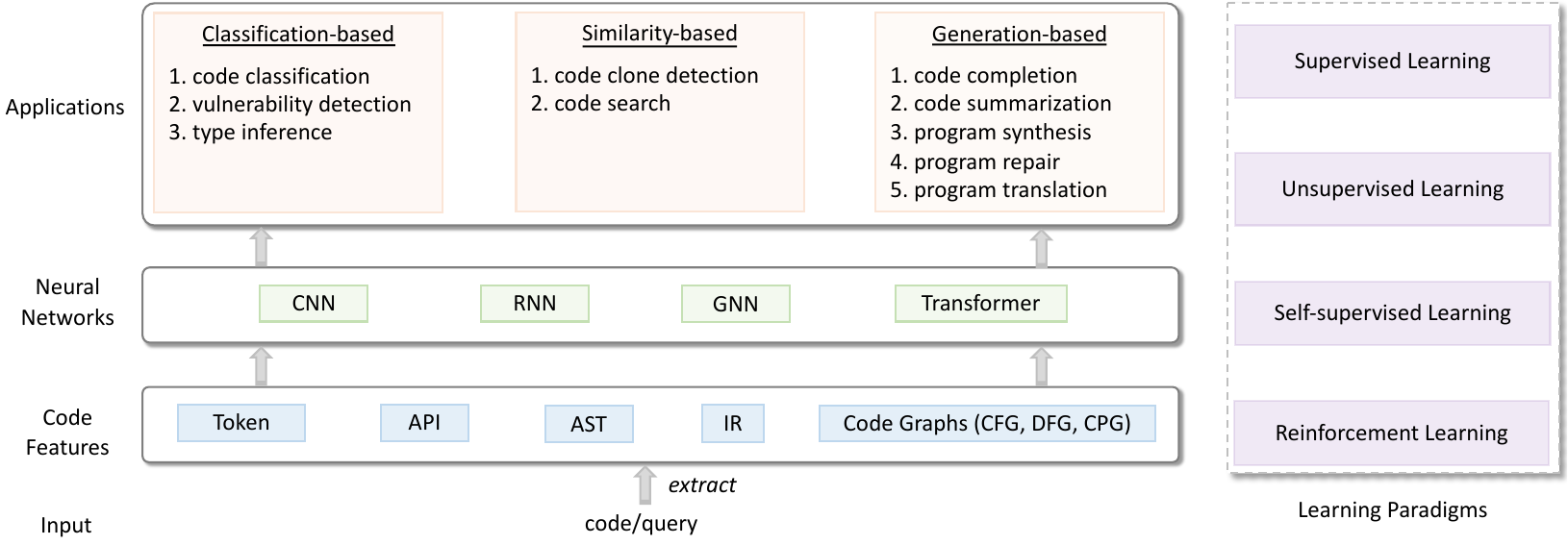}
	\vspace{-3mm}
	\caption{
 Code intelligence tasks based on code representation learning.
 }
	\label{fig_encoder_decoder}
	\vspace{-4mm}
\end{figure*}
We propose to summarize existing deep-learning-based approaches to code intelligence from the lens of code representation learning in this paper.
As shown in Figure~\ref{fig_encoder_decoder},
for code representation learning, researchers first extract features that potentially describe the semantics of code, and then design various neural networks to encode them into distributed vectors.
Code representation learning can be viewed as the foundation for different downstream applications.
Based on the characteristics of each application, the downstream applications can be divided into three groups:
(1)~\textit{Classification-based.} In these tasks (e.g., code classification, vulnerability detection, and type inference), a classifier layer (e.g., softmax) is used to map the code embeddings to labels/classes. 
(2)~\textit{Similarity-based.} In these tasks (e.g., code search and code clone detection), Siamese neural network structure~\cite{chicco2021siamese} is often adopted, where dual encoders are used to encode the source code and natural-language query into embedding vectors. 
Based on the two embeddings of code and query, a constraint (such as a triplet loss function) is always used to regularize the similarity between them.
Note that, in several approaches to code search and code clone detection, the two embeddings of code and query are also concatenated, and the task is reformulated as a classification task to determine whether the code and query are related~\cite{feng2020codebert}.
(3)~\textit{Generation-based.} In these tasks (e.g., code completion, code summarization, program translation, program synthesis, and program repair), 
{the objective is to generate source code, natural language descriptions, or programs in another programming language from a given code snippet.}
These tasks usually follow the encoder-decoder paradigm, where an encoder network is used to represent the semantics of code, and a decoder network (e.g., RNN) is designed to generate sequences, e.g., natural-language descriptions or source code. 
Additionally, we categorize the learning paradigms into four groups: supervised learning, unsupervised learning, self-supervised learning, and reinforcement learning.

\subsection{Paper Selection}
Deep learning for code intelligence has been studied in many related research communities.
In this paper, we review high-quality papers selected from top-tier conferences and journals, ranging from software engineering, programming languages, NLP, and artificial intelligence, to security.
Overall, we have identified \numofvenues publication venues, as shown in the Supplementary Materials.
We first manually check the publication list of the venues and obtain an initial collection of papers. 
Particularly, we systematically query the aforementioned venue names within the DBLP database\footnote{\url{https://dblp.uni-trier.de}} and examine the associated proceedings. Subsequently, two authors, both possessing over five years of expertise in deep learning for code intelligence, collaboratively undertake the task of manually refining the results. This involves meticulous scrutiny of titles and a brief review of abstracts to identify and filter out papers that are potentially relevant to code intelligence.
For those large conferences (e.g., AAAI and IJCAI) that accept thousands of papers per year, we first filter out those papers whose titles contain the keywords of ``code'' or ``program'', and then manually check them.

Based on this initial collection of papers, we start to augment it through keyword searching.
We systematically search DBLP and Google Scholar using the following keywords: ``code representation'', ``program comprehension'', ``code embedding'', ``code classification'', ``vulnerability detection'', ``bug finding'', ``code completion'', ``type inference'', ``code search/retrieval'', ``code clone detection'', ``code summarization'', ``program translation'', ``program synthesis'', and ``program repair'', with a combination of  ``deep'', ``learning'', ``neural'', and ``network''.

It is worth noting that, in addition to accepted papers from the aforementioned venues, we also consider some recent publications from the pre-print archive, as they reflect the most current research outputs.
We choose publications from arXiv based on 
two criteria: paper quality, author reputation.
The quality of a pre-printed paper can be assessed based on the number of citations it has garnered in recent months. 
The reputations of authors can be indicated by their Google Scholar citations. 
If a paper satisfies either of these selection criteria, we include it for consideration.
Having obtained this collection of papers, we then filter out the irrelevant papers by manual checking.
Finally, we obtained a collection of {\numofall} papers. 
To ensure transparency and accessibility, a comprehensive table of the surveyed papers and the source of papers is maintained online.\footnote{\url{https://github.com/CGCL-codes/awesome-code-intelligence}}

\subsection{Publication Trends of Code Intelligence}
\begin{figure}[!t]
\centering
	\begin{subfigure}[b][][c]{.42\textwidth}
		\centering
        \includegraphics[width=\linewidth]{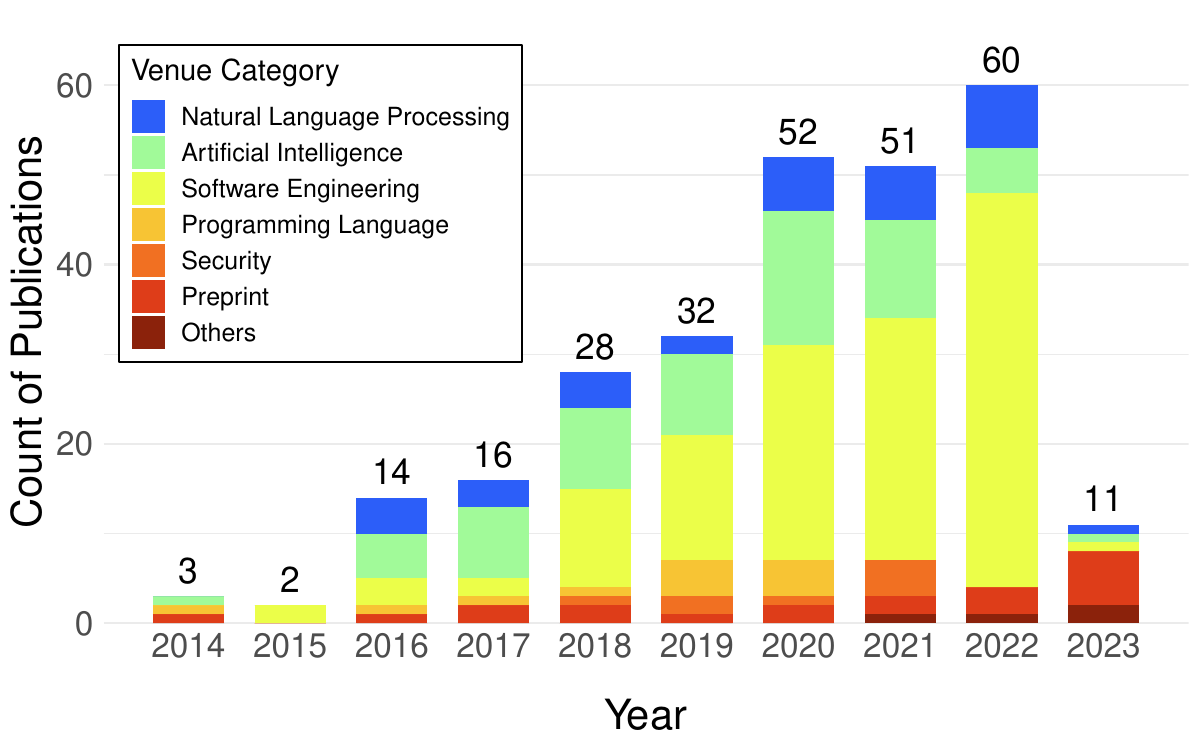}
		\caption{Number of publications in different years. 
  }
		\label{fig_trend_a}
	\end{subfigure}
	\begin{subfigure}[b][][c]{.56\textwidth}
		\centering
        \includegraphics[width=\linewidth]{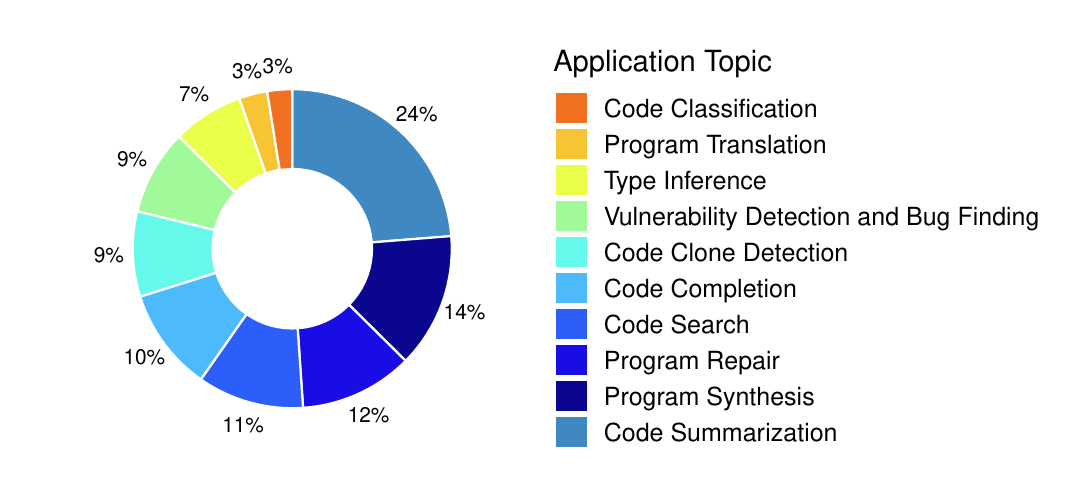}
		\caption{Publication in each application}
		\label{fig_trend_c}
	\end{subfigure}
	\vspace{-3mm}
	\caption{
            Statistics of the surveyed papers to reveal the publication trend and research topic trend.
        }
	\label{fig_trend}
	\vspace{-4mm}
\end{figure}

Figure~\ref{fig_trend} provides statistics of the surveyed papers to reveal the publication trend and research topic trend.
Figure~\ref{fig_trend_a} shows the collected papers on deep learning for code intelligence, from January 2014 to December 2022.
It is noteworthy that, 
11 papers about LLMs for code intelligence have been included, all published in 2023.
Each year, we analyze the publication trends across various communities and venues, including NLP, artificial intelligence, software engineering, programming language, security, preprint, and others. The surveyed venues for each community are detailed in Table 1 of the Supplementary Materials.
According to our statistical findings, the top five most popular publication venues are ICSE, ASE, ACL, ICLR, and FSE, from the software engineering, NLP, and AI communities. 
Notably, we can also observe that code intelligence has garnered increasing attention from the AI and NLP communities since 2018.
Although deep learning was first proposed in 2006~\cite{hinton2006fast}, it is initially used for source code modeling in 2014.
From Figure~\ref{fig_trend_a}, we can also see that the number of relevant papers for code intelligence has increased significantly since 2018, indicating that deep learning has significantly advanced code intelligence research since then.
This development can be attributed to the widespread use of deep learning in NLP since 2018, which has sparked a lot of studies on using NLP methods for tasks involving source code.

Figure~\ref{fig_trend_c} shows the distribution of papers across applications, including code classification, vulnerability detection, type inference, code search, code clone detection, code completion, code summarization, program translation, program synthesis, and program repair.
This figure shows a burgeoning interest in recent years surrounding topics of code summarization, program synthesis, program repair, vulnerability detection, and code search.


\section{Literature Review}\label{sec_review}

\subsection{Taxonomy}
\definecolor{block_dl_color}{rgb}{0.85,1,1}
\definecolor{block_features_color}{rgb}{1,1,0.9}
\definecolor{block_application_color}{rgb}{0.92,1,0.92}
\tikzstyle{block} = [rectangle, draw, fill=blue!20, 
    text width=6em, text centered, rounded corners, minimum height=2em,node distance=12em,font=\small]
\tikzstyle{block_gray} = [rectangle, draw, fill=gray!10, 
    text width=6em, text centered, rounded corners, minimum height=2em,node distance=12em,font=\small]
\tikzstyle{block_dl} = [rectangle, draw, fill=block_dl_color, 
    text width=6em, text centered, rounded corners, minimum height=2em,node distance=12em,font=\small]
\tikzstyle{block_features} = [rectangle, draw, fill=block_features_color, 
    text width=6em, text centered, rounded corners, minimum height=2em,node distance=12em,font=\small]
\tikzstyle{block_application} = [rectangle, draw, fill=block_application_color, 
    text width=6em, text centered, rounded corners, minimum height=2em,node distance=12em,font=\small]
\tikzstyle{line} = [draw, -]
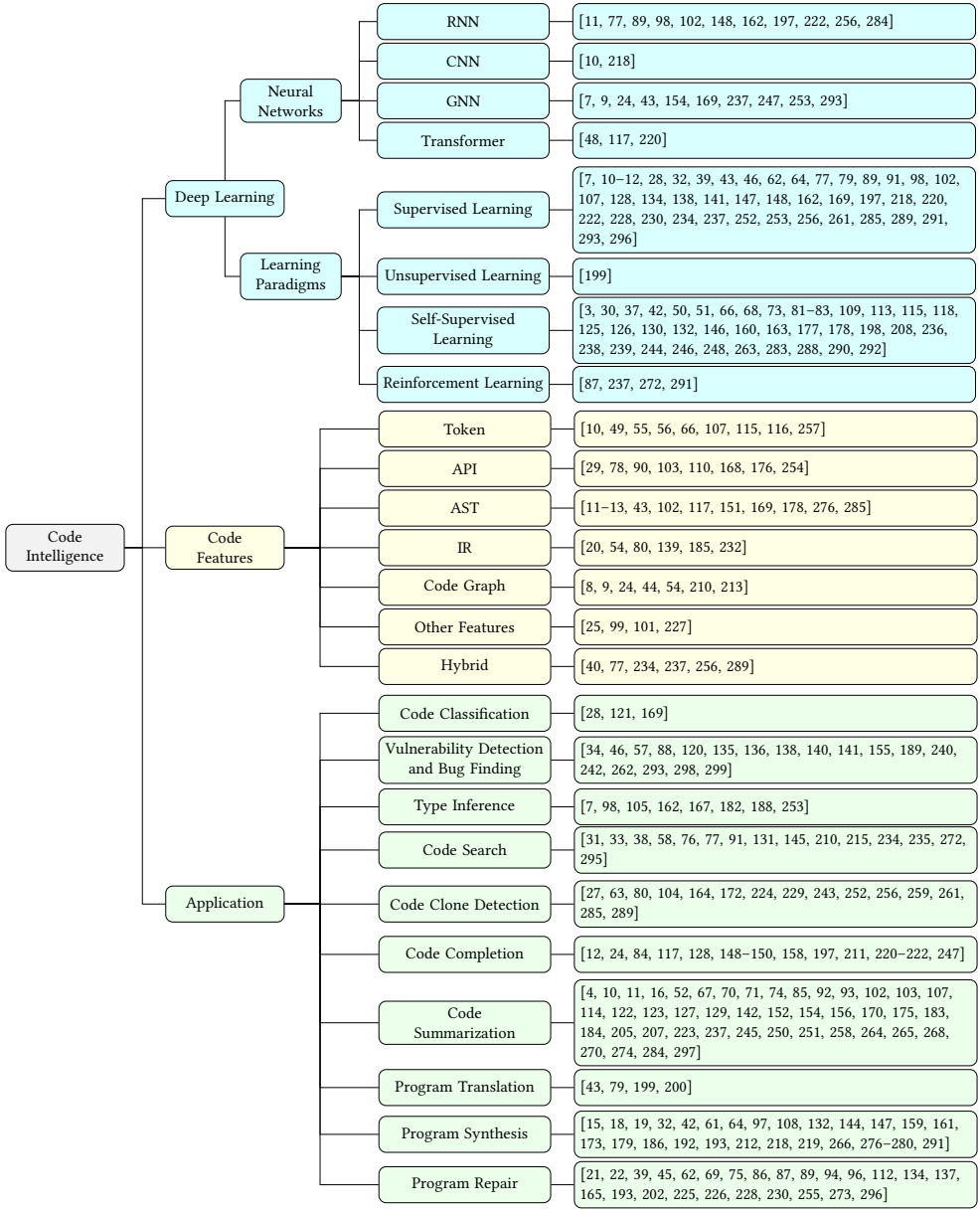
\begin{figure}
    \centering
    \resizebox{0.94\textwidth}{!}{%
    \begin{tikzpicture}
        \node [block_gray] (init) {Code\\Intelligence};
        \node [block_features, right of=init, node distance=9em] (representation) {Code\\Features};
        \node [block_dl, above of=representation, node distance=19.4em] (deeplearning) {Deep Learning};
        \node [block_application, below of=representation, node distance=19.8em] (application) {Application};
        \node [block_dl, above right=3.2em and -2.5em of deeplearning, node distance=12em, text width=5em] (neuralnetworks) {Neural Networks};
        \node [block_dl, below right=2em and -2.5em of deeplearning, node distance=12em, text width=5em] (learningparadigms) {Learning Paradigms};
        \node [block_dl, right=2em of neuralnetworks, node distance=2.2em, text width=9em] (gnn) {GNN};
        \node [block_dl, below of=gnn, node distance=2.2em, text width=9em] (transformer) {Transformer};
        \node [block_dl, above of=gnn, node distance=2.2em, text width=9em] (cnn) {CNN};
        \node [block_dl, above of=cnn, node distance=2.2em, text width=9em] (rnn) {RNN};
        \node [block_dl, right=2em of learningparadigms, node distance=3.8em, text width=9em] (unsupervised) {Unsupervised Learning};
        \node [block_dl, above of=unsupervised, node distance=3.7em, text width=9em] (supervised) {Supervised Learning};
        \node [block_dl, below of=unsupervised, node distance=3.0em, text width=9em] (selfsupervised) {Self-Supervised Learning};
        \node [block_dl, below of=selfsupervised, node distance=3.0em, text width=9em] (reinforcement) {Reinforcement Learning};
        \node [block_features, right of=representation, node distance=13.5em, text width=9em] (ir) {IR};
        \node [block_features, above of=ir, node distance=2.2em, text width=9em] (ast) {AST};
        \node [block_features, above of=ast, node distance=2.2em, text width=9em] (api) {API};
        \node [block_features, above of=api, node distance=2.2em, text width=9em] (token) {Token};
        \node [block_features, below of=ir, node distance=2.2em, text width=9em] (codegraph) {Code Graph};
        \node [block_features, below of=codegraph, node distance=2.2em, text width=9em] (otherfeature) {Other Features};
        \node [block_features, below of=otherfeature, node distance=2.2em, text width=9em] (hybrid) {Hybrid};
        \node [block_application, right of=application, node distance=13.5em, text width=9em] (clone) {Code Clone Detection};
        \node [block_application, above of=clone, node distance=3em, text width=9em] (search) {Code Search};
        \node [block_application, above of=search, node distance=2.4em, text width=9em] (typeinference) {Type Inference};
        \node [block_application, above of=typeinference, node distance=2.6em, text width=9em] (vulnerability) {Vulnerability Detection and Bug Finding};
        \node [block_application, above of=vulnerability, node distance=2.6em, text width=9em] (classification) {Code Classification};
        \node [block_application, below of=clone, node distance=2.8em, text width=9em] (completion) {Code Completion};
        \node [block_application, below of=completion, node distance=3.8em, text width=9em] (summarization) {Code\\Summarization};
        \node [block_application, below of=summarization, node distance=3.6em, text width=9em] (translation) {Program Translation};
        \node [block_application, below of=translation, node distance=2.6em, text width=9em] (synthesis) {Program Synthesis};
        \node [block_application, below of=synthesis, node distance=2.8em, text width=9em] (repair) {Program Repair};
    
        \node [block_dl, right of=rnn, node distance=17.5em, text width=22em, align=left] (rnncite) {\cite{raychev2014code,liu2016neural,gu2018deep,alon2018code2seq,hellendoorn2018deep,malik2019nl2type,svyatkovskiy2019pythia,white2016deep,hu2018deep,zhang2020retrieval,gupta2017deepfix}};
        \node [block_dl, right of=cnn, node distance=17.5em, text width=22em, align=left] (cnncite) {\cite{allamanis2016convolutional,sun2019grammar}};
        \node [block_dl, right of=gnn, node distance=17.5em, text width=22em, align=left] (gnncite) {\cite{wan2018improving,mou2016convolutional,chen2018tree,allamanis2017learning,zhou2019devign,wang2021code,brockschmidt2018generative,wei2020lambdanet,allamanis2020typilus,liu2020retrieval}};
        \node [block_dl, right of=transformer, node distance=17.5em, text width=22em, align=left] (transformercite) {\cite{kim2021code,svyatkovskiy2020intellicode,chirkova2021empirical}};
        \node [block_dl, right of=supervised, node distance=17.5em, text width=22em, align=left] (supervisedcite) {\cite{mou2016convolutional,bui2019bilateral,li2018vuldeepecker,zhou2019devign,cheng2021deepwukong,li2019improving,raychev2014code,liu2016neural,li2017code,svyatkovskiy2019pythia,alon2020structural,svyatkovskiy2020intellicode,hellendoorn2018deep,malik2019nl2type,wei2020lambdanet,allamanis2020typilus,gu2018deep,wan2019multi,haldar2020multi,white2016deep,wei2017supervised,zhao2018deepsim,wu2020scdetector,zhang2019novel,allamanis2016convolutional,iyer2016summarizing,hu2018deep,alon2018code2seq,wan2018improving,chen2018tree,gu2017deepam,dong2016language,liu2016latent,sun2019grammar,zhong2017seq2sql,cai2017encoder,gupta2017deepfix,vasic2018neural,dinella2020hoppity,chakraborty2020codit,zhu2021syntax,tufano2018empirical,li2020dlfix}};
        \node [block_dl, right of=unsupervised, node distance=17.5em, text width=22em, align=left] (unsupervisedcite) {\cite{lachaux2020unsupervised}};
        \node [block_dl, right of=selfsupervised, node distance=17.5em, text width=22em, align=left] (selfsupervisedcite) {\cite{kanade2020learning,feng2020codebert,guo2020graphcodebert,niu2022spt,jiang2021treebert,ranzato2021DOBF,ahmad2021unified,zhang2022coditt5,wang2021syncobert,guo2022unixcoder,mastropaolo2021studying,wang2021codet5,bui2021infercode,jain2021contrastive,wan2022what,zhang2022diet,shi2022compressing,zhou2021assessing,wang2022bridging,wang2022no,chen2021evaluating,li2022competition,nijkamp2022codegen,chowdhery2022palm,fried2022incoder,christopoulou2022pangu,zheng2023codegeex,chai2022ernie,li2023starcoder,roziere2023code,wang2023codet5+,gunasekar2023textbooks,liu2023improving,geng2023large,xia2023automated,li2023large,li2023enabling,luo2023wizardcoder}};
        \node [block_dl, right of=reinforcement, node distance=17.5em, text width=22em, align=left] (reinforcementcite) {\cite{wan2018improving,yao2019coacor,gupta2018deep,zhong2017seq2sql}};
        
        \node [block_features, right of=token, node distance=17.5em, text width=22em, align=left] (tokencite) {\cite{cummins2017synthesizing,
white2015toward,
iyer2016summarizing,
allamanis2016convolutional,
cvitkovic2019open,
chirkova2021simple,
kanade2020learning,
feng2020codebert,
karampatsis2020big}};
        \node [block_features, right of=api, node distance=17.5em, text width=22em, align=left] (apicite) {\cite{moreno2015can,
jiang2017unsupervised,
gu2016deep,
nguyen2017exploring,
bui2019sar,
hu2018summarizing,
wei2022clear,
hadi2022effectiveness}};
        \node [block_features, right of=ast, node distance=17.5em, text width=22em, align=left] (astcite) {\cite{mou2016convolutional,
liu2020modeling,
zhang2019novel,
kim2021code,
niu2022spt,
hu2018deep,
alon2019code2vec,
alon2018code2seq,
alon2020structural,
yin2017syntactic,
chen2018tree}};
        \node [block_features, right of=ir, node distance=17.5em, text width=22em, align=left] (ircite) {\cite{li2022unleashing,
ben2018neural,
venkatakeerthy2019ir2vec,
cummins2020programl,
peng2021how,
gui2022cross}};
        \node [block_features, right of=codegraph, node distance=17.5em, text width=22em, align=left] (codegraphcite) {\cite{cummins2020programl,
allamanis2017learning,
allamanis2017smartpaste,
brockschmidt2018generative,
sui2020flow2vec,
shi2022how,
chen2021plur}};
        \node [block_features, right of=otherfeature, node distance=17.5em, text width=22em, align=left] (otherfeaturecite) {\cite{henkel2018code,
hoang2020cc2vec,
tufano2019learning,
brody2020structural}};
        \node [block_features, right of=hybrid, node distance=17.5em, text width=22em, align=left] (hybridcite) {\cite{gu2018deep,
white2016deep,
zhao2018deepsim,
wan2018improving,
wan2019multi,
chakraborty2021multimodal}};
    
        \node [block_application, right of=classification, node distance=17.5em, text width=22em, align=left] (classificationcite) {\cite{mou2016convolutional,leclair2018adapting,bui2019bilateral}};
        \node [block_application, right of=vulnerability, node distance=17.5em, text width=22em, align=left] (vulnerabilitycite) {\cite{wang2016automatically,
dam2018automatic,
li2018vuldeepecker,
zou2019muvuldeepecker,
li2021sysevr,
le2018maximal,
zhou2019devign,
wang2020combining,
cao2022mvd,
cheng2021deepwukong,
liu2021combining,
wu2022vulcnn,
li2021vulnerability,
zou2021interpreting,
pradel2018deepbugs,
li2019improving,
gupta2019neural,
li2021fault}};
        \node [block_application, right of=typeinference, node distance=17.5em, text width=22em, align=left] (typeinferencecite) {\cite{hellendoorn2018deep,
malik2019nl2type,
pradel2020typewriter,
wei2020lambdanet,
pandi2020opttyper,
allamanis2020typilus,
mir2022type4py,
huang2022prompt}};
        \node [block_application, right of=search, node distance=17.5em, text width=22em, align=left] (searchcite) {\cite{gu2018deep,
wan2019multi,
deng2022fine,
ling2020deep,
shi2022how,
haldar2020multi,
cambronero2019deep,
bui2021selfsupervised,
li2022coderetriever,
yao2019coacor,
sun2022code,
zhu2020ocor,
chai2022cross,
wan2022you,
gu2022accelerating}};
        \node [block_application, right of=clone, node distance=17.5em, text width=22em, align=left] (clonecite) {\cite{white2016deep,
wei2017supervised,
zhao2018deepsim,
zhang2019novel,
buch2019learning,
wang2020detecting,
nair2020funcgnn,
wu2020scdetector,
hu2022treecen,
wu2022detecting,
mehrotra2021modeling,
tufano2018deep,
ding2022towards,
tao2022contrastive,
gui2022cross}};
        \node [block_application, right of=completion, node distance=17.5em, text width=22em, align=left] (completioncite) {\cite{raychev2014code,
liu2016neural,
li2017code,
svyatkovskiy2019pythia,
kim2021code,
alon2020structural,
wang2021code,
brockschmidt2018generative,
svyatkovskiy2020intellicode,
liu2020self,
liu2020multi,
lu2022reacc,
guo2022learning,
svyatkovskiy2021fast,
shrivastava2020fly}};
        \node [block_application, right of=summarization, node distance=17.5em, text width=22em, align=left] (summarizationcite) {\cite{allamanis2016convolutional,
iyer2016summarizing,
hu2018deep,
alon2018code2seq,
leclair2019neural,
fernandes2018structured,
leclair2020improved,
jin2022automatically,
guo2022modeling,
ahmad2020transformer,
wu2020sit3,
gong2022source,
shen2022ast,
wan2018improving,
shi2021cast,
yang2021multimodal,
gao2022m2ts,
wang2022gypsum,
haque2020improved,
bansal2021project,
shahbazi2021API2Com,
ciurumelea2020suggesting,
lin2021improving,
zhang2020retrieval,
wei2020retrieve,
liu2020retrieval,
li2021editsum,
zhu2022simple,
hu2018summarizing,
xie2022lowresource,
wei2019code,
yang2022dualsc,
ye2020leveraging,
mu2022automatic,
xie2021exploiting,
haque2021action,
liu2019learning,
panthaplackel2021deep,
nguyen2020suggesting,
panthaplackel2020learning,
liu2020automating,
gao2021automating,
li2022auger}};
        \node [block_application, right of=translation, node distance=17.5em, text width=22em, align=left] (translationcite) {\cite{chen2018tree,
gu2017deepam,
lachaux2020unsupervised,
roziere2022leveraging}};
        \node [block_application, right of=synthesis, node distance=17.5em, text width=22em, align=left] (synthesiscite) {\cite{dong2016language,
liu2016latent,
beltagy2016improved,
yin2017syntactic,
maddison2014structured,
ling2016latent,
rabinovich2017abstract,
sun2019grammar,
sun2020treegen,
hayati2018retrieval,
iyer2018mapping,
nan2020hisyn,
lu2021codexglue,
raffel2020exploring,
xu2020incorporating,
chen2021evaluating,
li2022competition,
poesia2022synchromesh,
shu2017neural,
balog2016deepcoder,
devlin2017robustfill,
nye2019learning,
bavishi2019autopandas,
zhong2017seq2sql,
yu2018spider,
yu2019sparc,
yu2019cosql,
cai2017encoder,
yu2018syntaxsqlnet}};
        \node [block_application, right of=repair, node distance=17.5em, text width=22em, align=left] (repaircite) {\cite{bhatia2016automated,
santos2018syntax,
gupta2017deepfix,
chen2019sequencer,
gupta2018deep,
vasic2018neural,
dinella2020hoppity,
graves2014neural,
tarlow2020learning,
mesbah2019deepdelta,
chakraborty2020codit,
zhu2021syntax,
yasunaga2020graph,
li2022DEAR,
berabi2021tfix,
fu2022vulrepair,
raffel2020exploring,
jiang2021cure,
tufano2018empirical,
hata2018learning,
harer2018learning,
gupta2020synthesize,
li2020dlfix,
white2019sorting,
tian2020evaluating}};
    
        \coordinate[right=10pt of init] (aux1);
        \coordinate[right=10pt of deeplearning] (aux2);
        \coordinate[right=10pt of neuralnetworks] (aux21);
        \coordinate[right=10pt of learningparadigms] (aux22);
        \coordinate[right=20pt of representation] (aux3);
        \coordinate[right=20pt of application] (aux4);
        \path [line] (init) -- (representation);
        \path [line] (init) -| (aux1) |- (deeplearning);
        \path [line] (init) -| (aux1) |- (application);
        \path [line] (deeplearning) |- (neuralnetworks);
        \path [line] (deeplearning) |- (learningparadigms);
        \path [line] (neuralnetworks) -| (aux21) |- (rnn);
        \path [line] (neuralnetworks) -| (aux21) |- (cnn);
        \path [line] (neuralnetworks) -| (aux21) |-(gnn);
        \path [line] (neuralnetworks) -| (aux21) |-(transformer);
        \path [line] (learningparadigms) -| (aux22) |- (supervised);
        \path [line] (learningparadigms) -| (aux22) |-(unsupervised);
        \path [line] (learningparadigms) -| (aux22) |-(selfsupervised);
        \path [line] (learningparadigms) -| (aux22) |-(reinforcement);
    
        \path [line] (representation) -| (aux3) |- (token);
        \path [line] (representation) -| (aux3) |- (api);
        \path [line] (representation) -| (aux3) |-(ast);
        \path [line] (representation) -- (ir);
        \path [line] (representation) -| (aux3) |-(codegraph);
        \path [line] (representation) -| (aux3) |-(otherfeature);
        \path [line] (representation) -| (aux3) |-(hybrid);
        
        \path [line] (application) -| (aux4) |- (classification);
        \path [line] (application) -| (aux4) |- (vulnerability);
        \path [line] (application) -| (aux4) |-(typeinference);
        \path [line] (application)-| (aux4) |-(search);
        \path [line] (application) -- (clone);
        \path [line] (application) -| (aux4) |-(completion);
        \path [line] (application) -| (aux4) |-(summarization);
        \path [line] (application) -| (aux4) |-(translation);
        \path [line] (application) -| (aux4) |-(synthesis);
        \path [line] (application) -| (aux4) |-(repair);
        
        \path [line] (rnn) -- (rnncite);
        \path [line] (cnn) -- (cnncite);
        \path [line] (gnn) -- (gnncite);
        \path [line] (transformer) -- (transformercite);
        \path [line] (supervised) -- (supervisedcite);
        \path [line] (unsupervised) -- (unsupervisedcite);
        \path [line] (selfsupervised) -- (selfsupervisedcite);
        \path [line] (reinforcement) -- (reinforcementcite);
        
        \path [line] (token) -- (tokencite);
        \path [line] (api) -- (apicite);
        \path [line] (ast) -- (astcite);
        \path [line] (ir) -- (ircite);
        \path [line] (codegraph) -- (codegraphcite);
        \path [line] (otherfeature) -- (otherfeaturecite);
        \path [line] (hybrid) -- (hybridcite);
        
        \path [line] (classification) -- (classificationcite);
        \path [line] (vulnerability) -- (vulnerabilitycite);
        \path [line] (typeinference) -- (typeinferencecite);
        \path [line] (search) -- (searchcite);
        \path [line] (clone) -- (clonecite);
        \path [line] (completion) -- (completioncite);
        \path [line] (summarization) -- (summarizationcite);
        \path [line] (translation) -- (translationcite);
        \path [line] (synthesis) -- (synthesiscite);
        \path [line] (repair) -- (repaircite);
    \end{tikzpicture}
    }
    \vspace{-3mm}
    \caption{
        The taxonomy of deep learning for code intelligence.
    }
    \label{fig_taxonomy}
    \vspace{-6mm}
\end{figure}

Figure~\ref{fig_taxonomy} illustrates the taxonomy of current studies on deep learning for code intelligence that we have surveyed in this paper. 
From our observation, the research in this field can be broken down into three distinct aspects: 
i.e., code features, deep learning techniques, and applications. 
(1) \textit{Code Features.} As the foundation of deep-learning-based code intelligence, code representation seeks to represent source code as distributed vectors.
We categorize the current code representation approaches by the features of input code that they use, such as 
code tokens, Intermediate Representations (IRs), Application Programming Interfaces (APIs), Abstract Syntax Trees (ASTs) and code graphs
(e.g., graphs that illustrate control flow and data flow).
(2) 
In the realm of deep learning techniques, we initially delve into various types of neural networks, i.e., RNNs, CNNs, Transformers, and GNNs. Subsequently, we examine the diverse learning paradigms employed for modeling source code, i.e., supervised learning, unsupervised learning, self-supervised learning, and reinforcement learning.
(3) We investigate multiple downstream applications that are based on code representation and deep learning techniques, including code classification, vulnerability detection and bug finding, type inference, code search, code clone detection, code completion, code summarization, program translation, program synthesis, and program repair.

\subsection{Code Features}
\begin{figure*}[t!]
\centering
\includegraphics[width=0.96\textwidth]{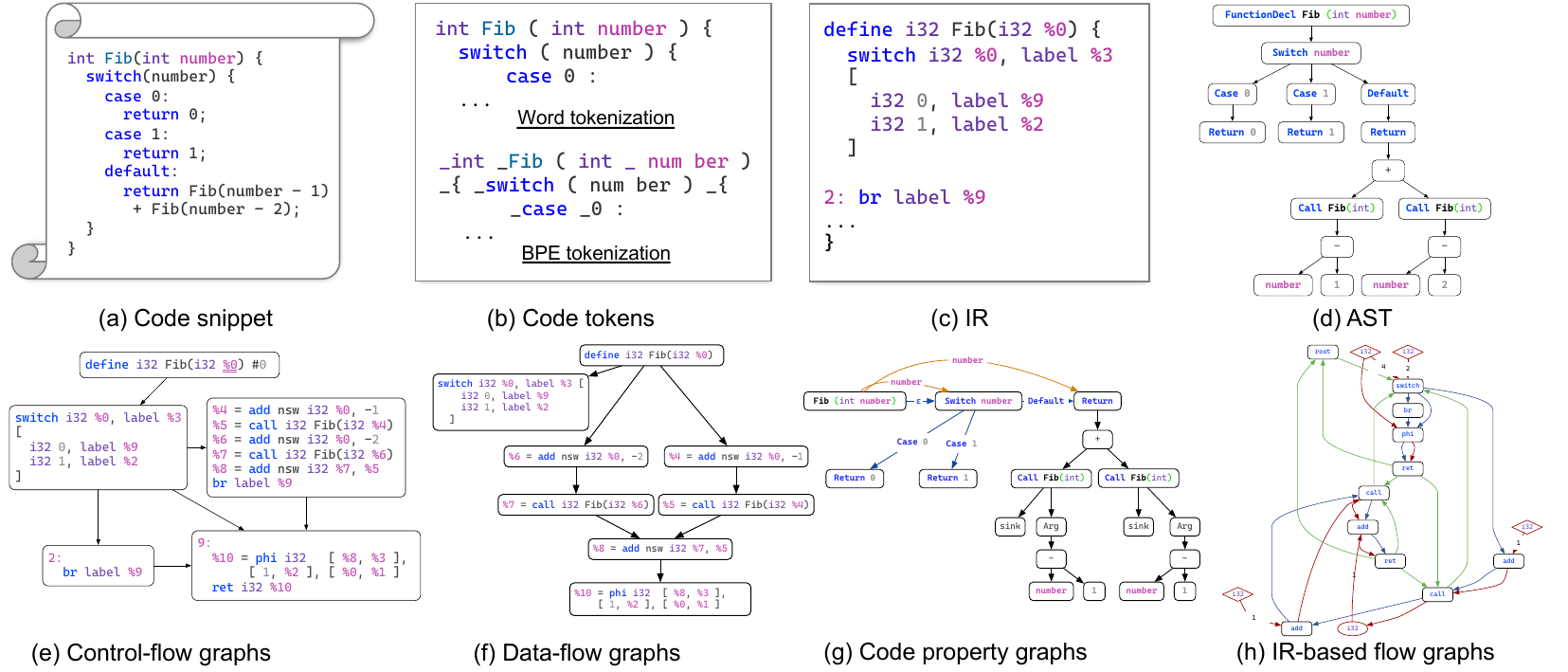}
\vspace{-3mm}
\caption{
        A detailed C code snippet with its corresponding tokens, IR, AST, and IR-based ﬂow graphs.
    }
    \label{fig_code_features}
\vspace{-4mm}
\end{figure*}
To represent source code, we need to first determine what to represent. 
Numerous works have proposed to extract the code features from diverse perspectives, including code tokens, IRs, ASTs, and various types of code graphs.
Figure~\ref{fig_code_features} shows a detailed code snippet written in C, with its corresponding code tokens, IR, AST, control-flow graph, data-flow graph, code property graph, and IR-based flow graphs.

\vspace{-2mm}
\subsubsection{Code Tokens}
Code tokens, shaping the textual appearance of source code, are composed of \textit{function name}, \textit{keywords}, and various \textit{variable identifiers}. These tokens are simple yet effective in representing the semantics of programs.
The majority of approaches for processing code involve breaking the program down into a sequence of tokens based on specific delimiters, such as spaces or the capitalization patterns in identifiers (for identifiers like \texttt{SortList} and \texttt{intArray}).
\citet{cummins2017synthesizing} introduced a character-level LSTM network to represent the sequence of code characters for program synthesis.
Since the set of characters to form a program is always a limited size, the character-level code representation does not have the problem of out-of-vocabulary. 
However, this tokenization process at the character level breaks down the meaning of the original words and also increases the length of the code sequence, which can make it challenging to understand the overall semantics of the program.

More coarsely, many word-level approaches are proposed to tokenize source code into words by separators. 
For example, \citet{white2015toward} and \citet{iyer2016summarizing} proposed to tokenize the program into words by whitespace, and designed RNNs to represent them for code summarization and code completion.
\citet{allamanis2016convolutional} designed a CNN with an attention mechanism to better represent the hierarchical structure of code over the subtokens that are simply tokenized by {camel} 
cases, to predict the function name.

\noindent{\textbf{Out-of-Vocabulary (OOV) Issue.}}
Since the variables and function names are always defined by developers without constraints, the size of vocabulary will explosively increase with the increasing training data, resulting in the \textit{out-of-vocabulary issue}, which is more severe than that in NLP.
To mitigate this issue, \citet{cvitkovic2019open} proposed a graph–structured cache, which introduces additional nodes for the encountered new words, and connects those nodes with edges based on where they occur in the code.
Recently, 
\citet{chirkova2021simple} offered a straightforward yet effective solution to mitigate the OOV issue by using identifier anonymization, and observed promising performance improvement.

Another effective approach is to tokenize the source code at a sub-word level, such as using techniques like Byte Pair Encoding (BPE),
which aims to construct a set of sub-words that can be combined to represent the entire code corpus.
Figure~\ref{fig_code_features} (b) shows the source tokens obtained by the strategy of word tokenization and BPE tokenization.
For the input variable \texttt{number}, the word tokenization will maintain the original word and consider it as a rare word, while the BPE tokenization will split it into two common sub-words, i.e., \texttt{num} and \texttt{ber}.
In the recent pre-trained language models of source code, e.g., CuBERT~\cite{kanade2020learning} and CodeBERT~\cite{feng2020codebert}, BPE has commonly been adopted for reducing the vocabulary size.
\citet{karampatsis2020big} conducted an empirical study on the granularity of word segmentation, and showed that tokenizing code by BPE can significantly reduce the vocabulary size.

\vspace{-2mm}
\subsubsection{Application Programming Interfaces (API)}
There have been multiple methods proposed to analyze the API sequences in programs.
One line of work is about mining API usage patterns from a large code corpus to demonstrate how to use an API.
For example, \citet{moreno2015can} proposed a novel approach, named Muse, to demonstrate API usage by mining and ranking the code examples in usage. 
Another line of work is API recommendation, which aims to recommend or generate a sequence of APIs for users.
\citet{jiang2017unsupervised} proposed to discover relevant tutorial fragments for APIs by calculating the correlation score based on PageRank and topic relevance.
\citet{gu2016deep} proposed a language model named DeepAPI, under the framework of sequence-to-sequence learning, to produce API sequences in response to a given natural language description.
Different from DeepAPI, \citet{nguyen2017exploring} proposed API2Vec
to represent the contextual information of API elements within an API sequence.
Likewise, they also developed a tool called API2API based on API2Vec to migrate the APIs across different programming languages, i.e., from Java to C\#, to validate the learned API embedding.
\forParaphrase{\citet{ling2021graph} introduced a method that integrated API call interactions and project structure into a single graph, and used this graph to design a graph-based collaborative filtering for making API usage recommendations.}{}
\citet{bui2019sar} proposed a cross-language API mapping approach to map APIs from Java to C\# with much less prior knowledge, through transfer learning across multiple domains. 
\citet{hu2018summarizing} 
suggested that incorporating API information as supplementary knowledge could improve code summarization.
To improve the representation of semantics in natural-language queries and API sequences, \citet{wei2022clear} proposed a contrastive learning approach for API recommendation, and \citet{hadi2022effectiveness} investigated the effectiveness of pre-trained models for generating API sequences from natural language queries.

\vspace{-2mm}
\subsubsection{Abstract Syntax Tree (AST)}
The AST is a tree-structured intermediate representation of code that describes the syntactic structure of a program.
As shown in Figure~\ref{fig_code_features} (d), in an AST, the leaf nodes (e.g., \texttt{number}, \texttt{Fib}) typically correspond to the tokens of variables and method names in the source code, while the non-leaf nodes (e.g., \texttt{FuncName}, \texttt{SwitchStmt}) represent the syntactic structure of code, like function definition, branch functions. 
As a result, this representation allows ASTs to be useful for both capturing the lexical information (e.g., variable \texttt{number}) and the syntactic structure of the source code.
In practice, we can extract ASTs using several open source tools, e.g., {\texttt{ANTLR}\footnote{\url{https://www.antlr.org}} parser}, \texttt{tree-sitter}\footnote{\url{https://tree-sitter.github.io/tree-sitter}} parser, and LLVM Clang\footnote{\url{https://clang.llvm.org}}.
To represent the ASTs, \citet{mou2016convolutional} proposed a tree structure-based CNN, and verified it in a code classification task. 
In order to handle long-distance dependencies between nodes in an AST,
\citet{liu2020modeling} proposed an improved LSTM by introducing operations such as PUSH and POP, and verified it in the tasks of code completion, code classification, and code summarization. 
To better process an AST, \citet{zhang2019novel} divided an AST into sentence-based subtrees and represented them using a two-way loop network.
Recently, \citet{kim2021code} proposed using a relative position embedding for code completion to feed the AST to Transformers.
\citet{niu2022spt} introduced a pre-trained model of source code by integrating AST information.

Another line of work~\cite{hu2018deep,alon2019code2vec,alon2018code2seq} is to represent ASTs indirectly by traversing or path sampling. 
\citet{hu2018deep} suggested traversing an AST to transform it into a linear series of nodes, and then using RNNs to represent the AST sequences for the task of code summarization.
\citet{alon2019code2vec} performed path sampling on the ASTs, and then used word2vec to represent the semantics of a program.  
Furthermore, \citet{alon2018code2seq} also applied a similar idea to the task of code summarization. 
Similarly, \citet{alon2020structural} proposed a structured code language model for code completion, by sampling paths from an incomplete AST.

In program synthesis, an AST is also incorporated to guide the synthesis of programs. 
\citet{yin2017syntactic} proposed an encoder-decoder framework for code generation, in which the encoder first encodes the natural language, then the decoder generates an AST of code, and finally, the AST is converted into source code. 
\citet{chen2018tree} proposed a Tree2Tree model for program translation, which first uses a TreeLSTM to represent the source program, and another TreeLSTM to generate the target program written in another programming language.

\vspace{-2mm}
\subsubsection{Intermediate Representation (IR)}
The IR is a well-formed structure that is independent of programming languages and machine architectures.
It is used by compilers to accurately represent the source code during the translation process from the source code to low-level machine code.
The IR can express the operations of the target machine.
It is natural to enhance the code embeddings via utilizing IRs~\cite{li2022unleashing}, with the benefit of limited vocabulary to significantly alleviate the OOV issue.
In this paper, we employ LLVM-IR, which is used in the LLVM infrastructure~\cite{lattner2004llvm}, as shown in Figure~\ref{fig_code_features} (c).
To represent IRs, \citet{ben2018neural} proposed inst2vec, which first compiles a program using LLVM Clang to obtain the LLVM intermediate representation, and then adopts skip-gram to represent the instructions.
\citet{venkatakeerthy2019ir2vec} proposed IR2Vec, which regards the intermediate code representation as triples in the knowledge graph, and then explores several knowledge graph representation methods. 
\citet{cummins2020programl} introduced ProGraML, a novel graph-based code representation based on IR. This code graph provides new opportunities to represent the semantics of source code at a low level using machine learning techniques (e.g., GNNs), for complex downstream tasks such as program optimization and analysis.
\citet{peng2021how} proposed to represent the augmented IR of source code based on pre-training and contrastive learning techniques, guided by compiler optimization.
Interestingly, \citet{gui2022cross} studied a new problem of
matching binary code and source code across languages 
by transforming both of them into LLVM-IRs.

\vspace{-2mm}
\subsubsection{Code Graphs}
Currently, many approaches have been proposed to convert programs into graphs to better represent the rich structural information within the programs, including Control-Flow Graph (CFG), Data-Flow Graph (DFG) and Code Property Graph (CPG).
As shown in Figure~\ref{fig_code_features} (e), the CFG represents the computation and control flow of a program.
In this representation, each node represents a basic block and each edge represents the transitions of control flow in the program.
As shown in Figure~\ref{fig_code_features} (f), 
the DFG is a directed graph that illustrates data relationships among various functions. Each node in the DFG has input and output data ports, and each edge links an output port to an input port on another node.
To represent multiple structural information of code using a joint data structure, \citet{yamaguchi2014modeling} proposed an innovative CPG to merge the structural information of code, including AST, CFG and Program Dependence Graph (PDG), into a single graph, as shown in Figure~\ref{fig_code_features} (g). 
In practice, we can build CFGs and DFGs using LLVM Clang, and build CPGs using Plume\footnote{\url{https://plume-oss.github.io/plume-docs}}. 
Recently, \citet{cummins2020programl} built a unified graph, termed ProGraML, which includes the CFG, DFG and call-graph, as shown in Figure~\ref{fig_code_features} (h).

To represent these code graphs,
\citet{allamanis2017learning} introduced the data flow on the top of ASTs and formed a code graph. Then, a Gated Graph Neural Network (GGNN)~\cite{li2015gated} was developed to learn the data dependencies among this code graph.
\citet{allamanis2017smartpaste} built the data flow among variables and considered the contextual information of variables for the task of automated pasting in programming. 
\citet{brockschmidt2018generative} expanded the incomplete code into a graph, and then proposed a graph neural network for code completion. 
\citet{sui2020flow2vec} made the code representation more accurate by using the value-flow graph of a program. 
\citet{shi2022how} resorted to converting the code graphs (e.g., CFG and DFG) into sequences through traversing for the task of code search.
\citet{chen2021plur} introduced a general method for transforming a code graph into a sequence of tokens and pointers.

\vspace{-2mm}
\subsubsection{Other Features of Code}
In addition to the aforementioned features of code that have already been widely explored, there also exist several kinds of features that are used in some specific scenarios.
For example, \citet{henkel2018code} introduced a novel feature for code representation learning based on abstractions of traces collected from the symbolic execution of a program.
\citet{hoang2020cc2vec} proposed using deep learning to learn distributed representations of code changes/edits that may be used to generate software patches.
In terms of code changes, several related works are also proposed to represent or predict them.
\citet{tufano2019learning} 
proposed to automate code editing through sequence-to-sequence-based neural machine translation.
\citet{brody2020structural} proposed to represent the code edits first, and then iteratively generate tree edits over the AST.

\vspace{-2mm}
\subsubsection{Hybrid Representation}
To leverage multiple code features, several approaches to representing source code in a hybrid fashion have been developed.
For instance, \citet{gu2018deep} explored using three separate RNNs for representing function names, code tokens, as well as API sequences of code, respectively.
It has also been evaluated in the code search task.
\citet{white2016deep} considered both the code tokens and AST node sequences, and used two different RNNs to represent these two sequences respectively, for the task of code cloning detection. 
\citet{zhao2018deepsim} proposed to represent the source code by incorporating the flow graphs of code into a semantic matrix.
They also developed a neural network model to assess the functional similarity between the representations of two code snippets.
Similarly, \citet{wan2018improving} and \citet{wan2019multi} developed a hybrid network consisting of an LSTM representing the code tokens, a GGNN representing the CFG of code, and a TreeLSTM representing the AST of code, for the task of code summarization and code search.
\citet{chakraborty2021multimodal} suggested leveraging three modalities of information (e.g., edit location, edit code context, and commit messages) to represent the context of programming and generate code patches automatically.

\subsection{Deep Learning Techniques}
We investigate the types of neural networks and classify the learning paradigms into four groups: \forParaphrase{supervised learning, unsupervised learning, self-supervised learning, and reinforcement learning.}{.}
\vspace{-2mm}
\subsubsection{Neural Networks}
It is natural to model source code as sequential text, and directly apply NLP techniques to represent it.
Simply, RNN~\cite{raychev2014code,liu2016neural,gu2018deep,alon2018code2seq,hellendoorn2018deep,malik2019nl2type,svyatkovskiy2019pythia,white2016deep,hu2018deep,zhang2020retrieval,gupta2017deepfix} and CNN~\cite{allamanis2016convolutional,sun2019grammar} neural networks can be easily applied to represent the sequential structure of source code.
In order to capture the syntax structure, especially the AST of source code, many tree-structured neural networks~\cite{wan2018improving,mou2016convolutional,chen2018tree} have also been designed. 
Furthermore, to represent the semantic structures (e.g., CFG and DFG) of source code, GNNs~\cite{allamanis2017learning,zhou2019devign,wang2021code,brockschmidt2018generative,wei2020lambdanet,allamanis2020typilus,liu2020retrieval} have been introduced to represent the source code.
Recently, the Transformer architecture has been utilized to represent the source code~\cite{kim2021code,svyatkovskiy2020intellicode}. 
\citet{chirkova2021empirical} conducted a comprehensive empirical study of how well Transformers can leverage syntactic information in source code for various tasks.
More preliminaries about the mentioned neural networks are referred to the Supplementary Materials.

\vspace{-2mm}
\subsubsection{Supervised Learning}
Supervised learning aims to learn a function that maps an input to an output based on a set of input-output pairs as training data.
It is a widely used learning paradigm in deep learning.
From our investigation, current deep learning approaches for code intelligence are mainly based on supervised learning. For each specific code intelligence task, such as code classification~\cite{mou2016convolutional,bui2019bilateral}, vulnerability detection and bug finding~\cite{li2018vuldeepecker,zhou2019devign,cheng2021deepwukong,li2019improving}, code completion~\cite{raychev2014code,liu2016neural,li2017code,svyatkovskiy2019pythia,alon2020structural,svyatkovskiy2020intellicode}, type inference~\cite{hellendoorn2018deep,malik2019nl2type,wei2020lambdanet,allamanis2020typilus}, code search~\cite{gu2018deep,wan2019multi,haldar2020multi}, code clone detection~\cite{white2016deep,wei2017supervised,zhao2018deepsim,wu2020scdetector,zhang2019novel}, code summarization~\cite{allamanis2016convolutional,iyer2016summarizing,hu2018deep,alon2018code2seq,wan2018improving}, program translation~\cite{chen2018tree,gu2017deepam}, program synthesis~\cite{dong2016language,liu2016latent,
sun2019grammar,zhong2017seq2sql,cai2017encoder}, and program repair~\cite{gupta2017deepfix,vasic2018neural,dinella2020hoppity,chakraborty2020codit,zhu2021syntax,tufano2018empirical,li2020dlfix}, a set of paired input-output data is collected first.
For each task, supervised learning is guided by a specific loss function. One limitation of this kind of approach is that it relies on lots of well-labeled 
input-output pairs, which are always expensive to collect in some scenarios.

\vspace{-2mm}
\subsubsection{Unsupervised Learning}
As opposed to supervised learning, unsupervised learning seeks to identify patterns from a dataset without labels.
One representative work is TransCoder~\cite{lachaux2020unsupervised}, in which a fully unsupervised neural source-to-source translator is trained based on unsupervised machine translation.
This kind of learning paradigm is challenging for code intelligence and more research work is still required. 

\vspace{-2mm}
\subsubsection{Self-Supervised Learning}
Self-supervised learning can be thought of as a blend of supervised learning and unsupervised learning. 
Different from supervised learning where data labels are available for training, self-supervised learning obtains the supervisory signals directly from the data itself, usually the underlying structure of the data. 
One common practice used by self-supervised learning is to predict any unobserved (or masked) part of input from the part that can be observed.
As a representative technique of self-supervised learning, language model pre-training has been widely studied in source code~\cite{kanade2020learning,feng2020codebert,guo2020graphcodebert}.
\citet{kanade2020learning} proposed to train a CuBERT on the Python code corpus, and verified the pre-trained model on multiple downstream tasks such as variable misuse, operator classification, and function-document matching. 
CodeBERT~\cite{feng2020codebert} is yet another pre-trained model that deals with the two different modalities of source code and natural language descriptions.
It is based on masked language modeling, and has achieved promising results in tasks such as code search and code completion.
Based on CodeBERT, GraphCodeBERT~\cite{guo2020graphcodebert}, SPT-Code~\cite{niu2022spt}, and TreeBERT~\cite{jiang2021treebert} are proposed to digest the structural information from source code.
\citet{ranzato2021DOBF} presented a pre-training objective based on deobfuscation as an alternative criterion.
Inspired by BART~\cite{lewis2020bart} which is a pre-trained deep model specially designed towards natural language understanding and generation, \citet{ahmad2021unified} trained a similar pre-trained model PLBART for tasks that are related to code generation as well as code understanding.
\forParaphrase{\citet{zhang2022coditt5} trained a model named CoditT5 on large amounts of source code and natural-language comments, for software-related editing tasks, e.g., comment updating, bug fixing, and automated code review. }{}
\citet{wang2021syncobert} and \citet{guo2022unixcoder} proposed to train a model by unifying the modality of source code and natural language with contrastive learning, 
to improve the representation of the semantics of source code.
\citet{mastropaolo2021studying} and \citet{wang2021codet5} 
explored building pre-trained models
based on the T5 (Text-To-Text Transfer Transformer) architecture, which has attained state-of-the-art results in NLP tasks.
\citet{bui2021infercode} proposed InferCode, a self-supervised 
learning method through predicting subtrees that are identified from the context of ASTs.
\citet{jain2021contrastive} proposed a contrastive learning approach for task-agnostic code representation based on program transformations in the compiler.
ERNIE-Code~\cite{chai2022ernie} is a unified pre-trained model based on 116 natural languages and 6 programming languages, with the aim of bridging the gap between multilingual natural languages and multilingual programming languages.
Given that LLMs are pre-trained on a dataset that may have a different distribution from the testing dataset, \citet{wang2022bridging} explored the fine-tuning of pre-trained code models to facilitate adaptation to downstream tasks through curriculum learning.

Instead of improving the capability of code embedding, \citet{wan2022what} investigated the explainability of pre-trained models for code intelligence, i.e., what kind of information do these models capture, through structural analysis.
\citet{zhang2022diet} and~\citet{shi2022compressing} suggested compressing pre-trained models of code, as to accelerate their efficiency in practice.
\citet{zhou2021assessing} carried out an empirical study to assess the generalizability of CodeBERT when applied to various datasets and downstream tasks.

\noindent\textbf{Large Language Models (LLMs) of Code.} 
The aforementioned pre-trained code models have demonstrated promising capabilities in comprehending the semantics of source code. More recently, with the remarkable performance achievements of LLMs in text generation and conversational dialogs, exemplified by the success of ChatGPT, a diverse array of LLMs has been specifically trained for code-related tasks, notably, code generation.
\citet{nijkamp2022codegen} 
introduced a novel code generation task that enables users to progressively express their intentions through multi-turn interactions, and further trained a family of LLMs with up to 16.1 billion parameters, called CodeGen, for this task.
PaLM~\cite{chowdhery2022palm} is a general-purpose LLM developed by Google, which is pre-trained on a substantial dataset comprising both text and code corpora, boasting a vast parameter size of up to 540 billion. Derived from PaLM, PaLM-Coder is a model specifically fine-tuned for code-related tasks, such as code generation and program translation.
InCoder~\cite{fried2022incoder} is a LLM developed by Meta, which employs a causal masking objective for the purpose of infilling code blocks based on arbitrary left and right contexts.
Pangu-Coder~\cite{christopoulou2022pangu}, introduced by Huawei, is a LLM specifically developed for code generation. Its training follows a two-stage strategy: in the first stage, it undergoes pre-training using Causal Language Modeling (CLM) on raw code corpora. 
In the second stage, it employs a combination of CLM and Masked Language Modeling (MLM) training objectives, with a focus on the downstream task of code generation from text. 
CodeGeeX~\cite{zheng2023codegeex} is a multilingual model with 13 billion parameters for code generation, pre-trained on 850 billion tokens of 23 programming languages.
StarCoder~\cite{li2023starcoder} is a LLM for code, up to 15.5 billion parameters, which is trained  on an extensive dataset consisting of 1 trillion tokens sourced from a vast collection of permissively licensed GitHub repositories with inspection tools and an opt-out process. 
Code Llama~\cite{roziere2023code} is a family of large language models for code, released by Meta, built upon the foundation of Llama 2.
These models are distinguished by their advanced infilling capabilities, extensive long-context fine-tuning, and precise instruction fine-tuning.
CodeT5+~\cite{wang2023codet5+}, released by Salesforce, represents a novel family of encoder-decoder-based LLMs explicitly tailored for a broad spectrum of tasks related to both code comprehension and code generation. This model introduces innovative pre-training objectives, including text-code contrastive learning, matching, and CLM tasks on text-code data.
phi-1~\cite{gunasekar2023textbooks} is a comparatively smaller LLM for code, consisting of 1.3 billion parameters, achieved through data set refinement, while maintaining competitive performance in code generation.

Different from traditional pre-trained code models that are designed for specific tasks, the LLMs for code are distinguished by their strong capabilities in zero-shot learning.
To unleash the zero-shot capabilities of LLMs, many techniques such as prompt tuning, in-context learning, chain-of-thought, and instruction tuning, have been developed.
Recently, numerous studies have explored the potential of LLMs in tasks such as code generation~\cite{liu2023improving}, code summarization~\cite{geng2023large}, and code repair~\cite{xia2023automated}, all achieved through the design of textual prompts.
As a specific prompting, in-context learning seeks to bolster the capabilities of LLMs by furnishing them with contextual information or illustrative examples.
\citet{li2023large} explored in-context learning for better code generation based on LLMs.
The chain-of-thought is designed to ensure the outputs of LLMs follow a logical chain.
\citet{li2023enabling} explored chain-of-thought for better code generation based on LLMs.
The instruction tuning is initially designed to enhance the generalization capabilities of LLMs across different tasks.
WizardCoder~\cite{luo2023wizardcoder} is crafted to augment the capabilities of StarCoder by creating sophisticated code instruction data via the code-specific \textit{Evol-Instruct} approach.


\subsubsection{Reinforcement Learning}
Reinforcement learning aims to learn an agent through interacting with the environment without input-output pairs.
This kind of learning paradigm has been used in code summarization~\cite{wan2018improving}, code search~\cite{yao2019coacor}, program repair~\cite{gupta2018deep}, and program synthesis~\cite{zhong2017seq2sql}.

\subsection{Classification-based Applications}\label{sec_applications}
Classification-based applications, such as code categorization, vulnerability detection, and type inference, seek to train a classifier with the objective of mapping the source code to specific labels or classes, such as identifying vulnerability status or variable types.
\subsubsection{Code Classification}
Classifying source code into different classes (e.g., different functionalities and programming languages), is important for many tasks such as code categorization, programming language identification, code prediction, and vulnerability detection.
Various studies have been conducted to classify code snippets into categories based on their functionalities. 
To represent programs in the form of ASTs, \citet{mou2016convolutional} developed a Tree-Based Convolutional Neural Network (TBCNN), which was then verified on code classification.
In the wider realm of software categorization, \citet{leclair2018adapting} devised a series of adaptations, incorporating techniques such as word embedding and neural architectures, to tailor NLP methods for text classification specifically to the domain of source code.
\citet{bui2019bilateral} presented a bilateral neural network for the cross-language algorithm classification task, where each sub-network is used to encode the semantics of code in a specific language, and an additional classification module is designed to model the connection of those bilateral programs. 


\vspace{-2mm}
\subsubsection{Vulnerability Detection and Bug Finding}
Detecting vulnerabilities or bugs in programs is essential for assuring the quality of software, as well as saves much effort and time for software development.
Although many tools have been developed for vulnerability detection, e.g., Clang Static Analyzer\footnote{\url{https://clang-analyzer.llvm.org/scan-build.html}}, Coverity\footnote{\url{https://scan.coverity.com}}, Fortify\footnote{\url{https://www.hpfod.com}}, Flawfinder\footnote{\url{https://dwheeler.com/flawfinder}}, Infer\footnote{\url{https://fbinfer.com}}, and SVF~\cite{sui2016svf}, most of them are based on static analysis.
Recently, a growing number of works employ deep learning to discover vulnerabilities.
\citet{wang2016automatically} 
made an early attempt at applying deep learning, specifically deep belief network, to predict the defects of software, which learns the semantic features of programs based on AST. 
\citet{dam2018automatic} proposed an LSTM-based method to exploit both the syntactic and semantic aspects of source code, and apply the embeddings for both within-project and cross-project vulnerability detection.
VulDeePecker~\cite{li2018vuldeepecker}, $\mu$VulDeePecker~\cite{zou2019muvuldeepecker} and SySeVR~\cite{li2021sysevr} are a series of works that preserve the semantics of program by extracting API function calls and program slices for vulnerability detection.
\citet{le2018maximal} presented a maximal divergence sequential auto-encoder network
to find vulnerabilities in binary files.
The network is designed so that the embeddings of vulnerable code and invulnerable code are encouraged to be maximally divergent.
\citet{zhou2019devign} proposed Devign for vulnerability detection, which first represents a program by fusing its AST, CFG and DFG into a unified CPG, and then designs a graph neural network to represent the CPG of code.
Similarly, \citet{wang2020combining} and \citet{cao2022mvd} proposed a flow-sensitive framework for vulnerability detection, which leverages a GNN to represent the control, data, and call dependencies of a program.
\citet{cheng2021deepwukong} introduced DeepWukong, a GNN-based model for vulnerability detection of C/C++ programs, in which the flow information of programs is preserved.
\citet{liu2021combining} introduced a GNN model with expert knowledge for detecting vulnerabilities in smart contracts, which incorporates the flow information of programs.
Inspired by image processing,
\citet{wu2022vulcnn} proposed a method to enhance the scalability of vulnerability detection
by transforming code into an image with semantics preserved, and implementing a CNN to capture them effectively.

Recently, several works have attempted to explain the results of deep learning models for vulnerability detection.
\citet{li2021vulnerability} introduced 
a GNN model for vulnerability detection that allows for interpretability, by 
providing users with 
parts of the Program Dependency Graph (PDG) that may contain the vulnerability.
Additionally, \citet{zou2021interpreting} proposed an interpretable deep-learning-based model based on heuristic searching for vulnerability detection.

In contrast to vulnerability detection which only classifies a program as vulnerable or non-vulnerable, another line of work is bug finding, which aims to pinpoint the buggy location. 
DeepBugs~\cite{pradel2018deepbugs} is an approach for name-based bug detection, which trains a classifier to distinguish buggy or non-buggy code, based on deep learning.
To enhance the accuracy of bug detection,
\citet{li2019improving} suggested a fusion method by exploiting both the PDG and DFG for better representation. Larger weights are assigned to the buggy paths using the attention mechanism to identify the possible vulnerability.
\citet{gupta2019neural} developed a tree-structured CNN to identify the vulnerabilities or faults in a flawed program with respect to a failed test. 
\citet{li2021fault} defined the fault localization problem as image recognition, and provided a deep-learning-based approach that integrates code coverage, data dependencies between statements, and source code representations.

\vspace{-2mm}

\subsubsection{Type Inference}

Programming languages with dynamic typing, like Python and JavaScript, allow for rapid prototyping for developers and can save the time of software development dramatically.
However, without the type information, unexpected run-time errors are prone to occur, which may introduce bugs and produce low-quality code.
Current works on type inference, with the aim of automatically inferring variable types, mainly fall into two categories: static-analysis-based and learning-based. 
Traditional static-analysis approaches~\cite{hassan2018maxsmt, salib2004faster} are often imprecise since the behavior of programs is always over-approximated.
In addition, static-analysis-based approaches typically analyze the dependencies of an entire program, resulting in relatively low efficiency.

Recently, many deep learning techniques have been introduced for type inference.
To the best of our knowledge, \citet{hellendoorn2018deep} was the first to employ deep learning for type inference. They proposed a neural network based on sequence-to-sequence architecture, named DeepTyper, which uses GRUs to represent the program context and predict the type annotations for TypeScript. 
Furthermore, \citet{malik2019nl2type} proposed NL2Type to predict type annotations by leveraging the natural-language information of programs.
Based on NL2Type, \citet{pradel2020typewriter} further proposed TypeWriter, which utilizes both the natural-language information and programming context (e.g., arguments usage a function).
\citet{wei2020lambdanet} proposed LambdaNet for type inference based on GNNs, which first represents the code in the form of a type dependency graph, where typed variables and logical constraints among them are preserved. 
Then a GNN is proposed to propagate and aggregate features along related type variables, and eventually, predict the type annotations.
\citet{pandi2020opttyper} presented OptTyper, which first extracts relevant logical constraints, and shapes type inference as an optimization problem. 
\citet{allamanis2020typilus} proposed Typilus for type inference in Python, which expands ASTs into a graph structure and predicts type annotations over this graph using GNNs. 
To cope with large-scale type vocabulary, \citet{mir2022type4py} presented Type4Py, a similarity-based deep learning model with type clusters, which can support the inference of rare types and user-defined classes.
Recently, \citet{huang2022prompt} formulated the type inference task as a cloze-style fill-in-blank problem and then trained a CodeBERT model based on prompt tuning.

\subsection{Similarity-based Applications}\label{sec_applications}
Similarity-based applications, such as code search and code clone detection, 
aim to assess the likeness between a query (in either natural language or programming language) and a candidate code snippet.
It is important to note that several approaches propose to reframe these tasks as a classification problem, where both the code and query are concatenated, and the goal is to determine their relatedness~\cite{feng2020codebert}.
In this paper, we differentiate between similarity-based and classification-based applications by the objects they address, namely, the query and candidate code snippet. Specifically, similarity-based applications center on tasks involving two objects.


\subsubsection{Code Search}
Code search aims to retrieve a code snippet by a natural-language query (\textit{nl-to-code}) or code query (\textit{code-to-code}).
The \textit{nl-to-code} search refers to searching code fragments that have similar semantics to the natural-language query from a codebase.
As the first solution for code search using deep learning, \citet{gu2018deep} proposed DeepCS, which simultaneously learns the source code representation (e.g., function name, parameters and API usage) and the natural-language query in a shared feature vector space, with triplet criterion as the objective function. 
On the basis of DeepCS, \citet{wan2019multi} and \citet{deng2022fine} included more structural information of source code, including the ASTs and CFGs, under a multi-modal neural network equipped with an attention mechanism for better explainability.
\citet{ling2020deep} first converted code fragments and natural-language descriptions into two different graphs, and presented a matching technique for better source code and natural-language description matching. 
Furthermore, \citet{shi2022how} suggested an improved code search method by converting code graphs (e.g., CFGs and PDGs) into sequences through traversing.
\citet{haldar2020multi} proposed a multi-perspective matching method to calculate the similarities among source code and natural-language query from multiple perspectives.
\citet{cambronero2019deep} empirically evaluated the architectures and training techniques when applying deep learning to code search.
\citet{bui2021selfsupervised} and \citet{li2022coderetriever} leveraged contrastive learning with semantics-preserving code transformations for better code representation in code search.

Similar but different to the DeepCS framework, several more works have been proposed as complements for code search.
\citet{yao2019coacor} proposed using reinforcement learning to first generate the summary of code snippet and then use the summary for better code search.
\citet{sun2022code} suggested parsing source code to machine instructions, then mapping them into natural-language descriptions based on several predefined rules, followed by an LSTM-based code search model like DeepCS.
\citet{zhu2020ocor} considered the overlapped substrings between natural-language query and source code, and developed a neural network component to represent the overlap matrix for code search.

Recently, 
\citet{chai2022cross} suggested a transfer learning method for domain-specific code search, with the aim of transferring knowledge from Python to SQL. 
\citet{wan2022you} examined the robustness of different neural code search models, and showed that some of them are vulnerable to data-poisoning-based backdoor attacks.
\citet{gu2022accelerating} proposed to optimize code search by deep hashing techniques.

In contrast to \textit{nl-to-code} search, the input of \textit{code-to-code} search is source code, rather than natural-language description.
The objective of the code-to-code search is to find code snippets that are semantically related to an input code from a codebase.
The core technique of code-to-code search is to measure the similarity index between two code snippets, which is identical to the process of identifying code clones. More related work will be investigated in the code clone detection section.

\vspace{-2mm}
\subsubsection{Code Clone Detection}
Numerous software engineering activities, including code reuse, vulnerability detection, and code search, rely on detecting similar code snippets (or code clones).
There are basically four main types of code clones: 
Type-1 code clones are ones that are identical except for spaces, blanks, and comments. 
Type-2 code clones denote identical code snippets except for the variable, type, literal, and function names. 
Type-3 code clones denote two code snippets that are almost identical except for a few statements that have been added or removed. 
Type-4 code clones denote heterogeneous code snippets with similar functionality but differing code structures or syntax.
To handle different types of code clones, various works have been proposed. 

Recently, several deep-learning-based approaches have been designed for the semantics representation of a pair of code snippets for the task of clone detection.
The core of these approaches lies in representing the source code as distributed vectors, in which the semantics are preserved.
As an example, \citet{white2016deep} proposed DLC, which comprehends the semantics of source code by considering its lexical and syntactic information, and then designs RNNs for representation. 
To improve the representation of the syntactic structure of code,
\citet{wei2017supervised} applied TreeLSTM to incorporate AST information of source code.
\citet{zhao2018deepsim} proposed encoding the CFG and DFG of code into a semantic matrix, and introduced a deep learning model to match similar code representations. 
\citet{zhang2019novel} and \citet{buch2019learning} designed approaches to better represent the ASTs of the program, and applied them for code clone detection task.
Furthermore, \citet{wang2020detecting}, \citet{nair2020funcgnn} and \citet{mehrotra2021modeling} proposed to convert source code into graphs (e.g., CFG), represent the code graphs via GNN, and then measure the similarities between them.
Instead of using GNN, \citet{wu2020scdetector}and~\citet{hu2022treecen} introduced a centrality analysis approach on the flow graph (e.g., CFG) of code for clone detection, inspired by social network analysis.
\citet{wu2022detecting} considered the nodes of an AST as distinct states and constructed a model based on Markov chain to convert the tree structure into Markov state transitions. Then, for code clone detection, a classifier model is trained on the state transitions.
\citet{tufano2018deep} empirically evaluated the effectiveness of learning representation from diverse perspectives for code clone detection, including identifiers, ASTs, CFGs, and bytecode.
Recently, \citet{ding2022towards} and~\citet{tao2022contrastive} utilized program transformation techniques to augment the training data, and then applied pre-training and contrastive learning techniques for clone detection.
\citet{gui2022cross} studied a new problem of cross-language binary-source code matching by transforming both source and binary into LLVM-IRs.

\subsection{Generation-based Applications}\label{sec_applications}
Generation-based applications, including code completion, code summarization, program translation, program synthesis, and program repair, are designed to produce source code, natural-language descriptions, or programs in an alternative programming language, in response to specific requirements presented in either natural language or (partial) code.

\subsubsection{Code Completion}
Code completion is a core feature of most modern IDEs. It offers the developers a list of possible code hints based on available information.
\citet{raychev2014code} made the first attempt to combine the program analysis with neural language models for better code completion. It first extracts the abstract histories of programs through program analysis, and then learns the probabilities of histories via an RNN-based neural language model.
Similarly, various works \cite{liu2016neural,li2017code,svyatkovskiy2019pythia} resort to inferring the next code token over the partial AST, by first traversing the AST in a depth-first order, and then introducing an RNN-based neural language model.
To better represent the structure of code, \citet{kim2021code} suggested predicting the missing partial code by feeding the ASTs to Transformers.
\citet{alon2020structural} presented a structural model for code completion, which represents code by sampling paths from an incomplete AST.
Furthermore, \citet{wang2021code} suggested a GNN-based approach for code completion, which parses the flattened sequence of an AST into a graph, and represents it using Gated Graph Neural Networks (GGNNs)~\cite{li2015gated}.
\citet{guo2022learning} modeled the problem of code completion as filling in a hole, and developed a Transformer model guided by the grammar file of a specified programming language.
\citet{brockschmidt2018generative} expanded incomplete code into a graph representation, and then proposed a GNN for code completion. 
\citet{svyatkovskiy2020intellicode} proposed IntelliCode Compose, a pre-trained language model of code based on GPT-2, providing instant code completion across different programming languages.
\citet{liu2020self,liu2020multi} proposed a multi-task learning framework that unifies the code completion and type inference tasks into one overall framework.
\citet{lu2022reacc} suggested a retrieval-augmented code completion method that retrieves similar code snippets from a code corpus and then uses them as external context.

Since instant code completion is desired, 
several studies aim to improve the efficiency and flexibility of code completion. 
\citet{svyatkovskiy2021fast} suggested improving the efficiency of neural network models for code completion by reshaping the problem from generation to ranking the candidates from static analysis.
Additionally, \citet{shrivastava2020fly} proposed a code completion approach that supports fast adaption to an unseen file based on meta-learning. 
\vspace{-2mm}

\subsubsection{Code Summarization}
Inspired by the text generation work in NLP, many approaches have been put forward to systematically generate a description or function name to summarize the semantics of source code.
To the best of our knowledge, \citet{allamanis2016convolutional} were the first to use deep learning for code summarization. They designed a CNN to represent the code and applied a hybrid breath-first search and beam search to predict the tokens of function name.
Concurrently, \citet{iyer2016summarizing} proposed an LSTM-based sequence-to-sequence network with an attention mechanism for generating descriptions for source code.
The sequence-to-sequence network~\cite{iyer2016summarizing} inspired a line of works for code summarization, distinguished in code representation learning.
To represent the AST information, \citet{hu2018deep}, \citet{alon2018code2seq}, and \citet{leclair2019neural} proposed to linearize the ASTs via traversing or path sampling, and used RNNs to represent the sequential AST traversals/paths for code summarization.
Likewise, \citet{fernandes2018structured}, \citet{leclair2020improved} and \citet{jin2022automatically} investigated representing the structure of source code via a GNN, and verified it in code summarization.
\citet{guo2022modeling} designed the triplet position to model hierarchies in the syntax structure of source code for better code summarization.
Recently, several works \cite{ahmad2020transformer,wu2020sit3,gong2022source,shen2022ast} proposed to improve code summarization by designing enhanced Transformers to better capture the structural information of code (i.e., ASTs).
\citet{wan2018improving}, \citet{shi2021cast}, \citet{yang2021multimodal}, \citet{gao2022m2ts}, and~\citet{wang2022gypsum} proposed a hybrid representation approach by combining the embeddings of sequential code tokens and structured ASTs, and feeding them into a decoder network to generate summaries. 
As a complement, \citet{haque2020improved} and \citet{bansal2021project} advanced the performance of code summarization by integrating the context of summarized code, which contains important hints for comprehending subroutines of code.
\citet{shahbazi2021API2Com} leveraged the API documentation as a knowledge resource for better code summarization.
Instead of generating a sequence of summary tokens at once, \citet{ciurumelea2020suggesting} resorted to suggesting code comment completions based on neural language modeling.
\citet{lin2021improving} proposed to improve the code summarization by splitting the AST under the guidance of CFG, which can decrease the AST size and make model training easier. 

Another line of work aims to utilize code search to enhance the quality of code summaries generated by deep learning models.
For example, 
\citet{zhang2020retrieval}, \citet{wei2020retrieve}, \citet{liu2020retrieval} and \citet{li2021editsum} suggested augmenting the provided code snippet by searching similar source code snippets together with their comments, for better code summarization.
Instead of acquiring the retrieved samples in advance, \citet{zhu2022simple} suggested a simple retrieval-based method for the task of code summarization, which estimates a probability distribution for generating each token given the current translation context.

Apart from the above approaches, several works~\cite{hu2018summarizing,xie2022lowresource,wei2019code,yang2022dualsc,ye2020leveraging} are also worthy to be mentioned.
\citet{hu2018summarizing} transferred the code API information as additional knowledge to the code summarization task. 
\citet{xie2022lowresource} studied a new task of project-specific code summarization with limited historical code summaries via meta-transfer learning.
\citet{wei2019code} and \citet{yang2022dualsc} viewed the code generation task as a dual of code summarization, and incorporated dual learning for a better summary generation. 
Similarly, \citet{ye2020leveraging} leveraged code generation for code search and code summarization through dual learning as well.
\citet{mu2022automatic} introduced a multi-pass deliberation framework for code summarization, inspired by human cognitive processes.
\citet{xie2021exploiting} proposed a multi-task learning framework by leveraging method name suggestion as an auxiliary task to improve code summarization.
\citet{haque2021action} emphasized that predicting the action word (always the first word) is an important intermediate problem in order to generate improved code summaries.
Recently, the consistency between source code and comments has also attracted much attention, which is critical to ensure the quality of software.
\citet{liu2019learning}, \citet{panthaplackel2021deep}, and \citet{nguyen2020suggesting}  trained a deep-learning-based classifier to determine whether or not the function body and function name are consistent.
\citet{panthaplackel2020learning} and \citet{liu2020automating} proposed automatically updating an existing comment when the related code is modified, as revealed in the commit histories.
\citet{gao2021automating} proposed to automate the removal of obsolete TODO comments by representing the semantic features of TODO comments, code changes, and commit messages using neural networks.
\citet{li2022auger} proposed to generate review comments automatically based on pre-trained code models.

\vspace{-2mm}
\subsubsection{Program Translation}
Translating programs from a deprecated programming language to a modern one is important for software maintenance. 
Many neural machine translation-based methods have been proposed for program translation.
In order to utilize the AST structure of code, \citet{chen2018tree} proposed Tree2Tree, a neural network with structural information preserved. It first converts ASTs into binary trees following the left-child right-sibling rule, and then feeds them into an encoder-decoder model equipped with TreeLSTM.
\citet{gu2017deepam} presented DeepAM, which can extract API mappings among programming languages without the need of bilingual projects.
Recently, \citet{lachaux2020unsupervised} proposed TransCoder, a neural program translator based on unsupervised machine translation.
Furthermore, \citet{roziere2022leveraging} leveraged the automated unit tests to filter out invalid translations for unsupervised program translation.

\vspace{-2mm}
\subsubsection{Program Synthesis}
Program synthesis 
is a task for generating source code using high-level specifications (e.g., program descriptions or input-output samples).
Given the natural-language inputs, current approaches resort to 
generating programs through machine translation.
For semantic parsing, \citet{dong2016language} proposed an attention-based encoder-decoder model, which first encodes input natural language into a vector representation using an RNN, and then incorporates another tree-based RNN to generate programs.
\citet{liu2016latent} proposed latent attention for the If-Then program synthesis, which can effectively learn the importance of words in natural-language descriptions. 
\citet{beltagy2016improved} modeled the generation of If-Then programs from natural-language descriptions as a structure prediction problem, and investigated both  neural network and logistic regression models for this problem.

Unlike synthesizing simple If-Then programs, 
\citet{yin2017syntactic} proposed a syntax-preserving model for general-purpose programming languages, which generates Python code from pseudo code, powered by a grammar model that explicitly captures the compilation rules.
\citet{maddison2014structured} proposed a probabilistic model based on probabilistic context-free grammars (PCFGs) for capturing the structure of code for code generation.
\citet{ling2016latent} collected two datasets (i.e., Hearthstone and Magic the Gathering) for code generation in trading card games, and proposed a probabilistic neural network with multiple predictors.
On the basis of~\cite{ling2016latent}, \citet{rabinovich2017abstract} proposed to incorporate the structural constraints on outputs into a decoder network for executable code generation.
Similarly, \citet{sun2019grammar} and~\citet{sun2020treegen} designed a tree-based CNN and Transformer, respectively, 
for code generation and semantic parsing tasks based on the sequence-to-sequence framework.
\citet{hayati2018retrieval} 
suggested using a neural code generation model to retrieve action subtrees at test time.

Instead of synthesizing programs from natural-language descriptions, several works resort to generating programs from the (pseudo) program in another format or language.
\citet{iyer2018mapping} proposed to synthesize the AST derivation of source code given descriptions as well as the programmatic contexts. 
The above approaches are driven by well-labeled training examples, while \citet{nan2020hisyn} proposed a novel approach to program synthesis without using any training example, inspired by how humans learn to program.

Recently, various pre-trained code models also achieved significant progress in code generation. 
CodeGPT~\cite{lu2021codexglue} is a Transformer-based model that is trained using corpus for program synthesis, following the same architecture of GPT-2.
CodeT5~\cite{wang2021codet5} is a pre-trained code model in eight programming languages based on T5~\cite{raffel2020exploring}, which incorporates an identifier-aware objective during its pre-training phase.
\citet{xu2020incorporating} 
endeavored to integrate external knowledge into the pre-training phase to enhance code generation from natural-language input.
Codex~\cite{chen2021evaluating} is a GPT model trained on a code corpus sourced from GitHub. This model has played a pivotal role as the underpinning framework for Copilot\footnote{\url{https://github.com/features/copilot}}.
\citet{li2022competition} introduced AlphaCode, a code generation system designed to produce distinctive solutions for complex problems that demand profound cognitive engagement.
\citet{poesia2022synchromesh} introduced a constrained semantic decoding mechanism into a pre-trained model, as to constrain outputs of the model in a set of valid programs.
More recently, the code generation has been dominated by the LLMs, including CodeGen~\cite{nijkamp2022codegen}, CodeT5+~\cite{wang2023codet5+}, InCoder~\cite{fried2022incoder}, GPT-3.5~\cite{ChatGPT}, StarCoder~\cite{li2023starcoder}, Code Llama~\cite{roziere2023code}, and WizardCoder~\cite{luo2023wizardcoder}.

Programming by example is another flourishing direction for program synthesis.
\citet{shu2017neural} proposed a Neural Programming By Example (NPBE) model, 
which learns to solve string manipulation problems through inducting from input-output strings.
\citet{balog2016deepcoder} proposed DeepCoder, which trains a model to predict possible functions useful in the program space, as to guide the conventional search-based synthesizer. 
\citet{devlin2017robustfill} proposed RobustFill, which is an end-to-end neural network for synthesising programs from input-output examples.
\citet{nye2019learning} developed a neuro-symbolic program synthesis system called SketchAdapt, which can build programs from input-output samples and code descriptions by intermediate sketch.
\citet{bavishi2019autopandas} proposed a program candidate generator, backed by GNNs, for program synthesis 
in large real-world API.

It is worth mentioning that there are many works on generating code from natural language for specific domain-specific programming languages, e.g., Bash and SQL.
WikiSQL~\cite{zhong2017seq2sql}, Spider~\cite{yu2018spider}, SparC~\cite{yu2019sparc}, and CoSQL~\cite{yu2019cosql} are four datasets with human annotations for the task of text-to-SQL.
Based on these datasets, many works~\cite{yu2018syntaxsqlnet,yu2019sparc,yu2019cosql} have been proposed.
For example, Seq2SQL~\cite{zhong2017seq2sql} is a neural machine translation model to generate SQL queries from natural-language descriptions with reinforcement learning.
\citet{cai2017encoder} further proposed an encoder-decoder framework to translate natural language into SQL queries, which integrates the grammar structure of SQL for better generation. 
\citet{yu2018syntaxsqlnet} proposed a neural network SyntaxSQLNet, with syntax tree preserved, for the task of text-to-SQL translation across different domains, which takes the syntax tree of SQL into account during generation.

\vspace{-2mm}
\subsubsection{Program Repair}
Automatically localizing and repairing bugs in programs can save much manual effort in software development~\cite{jiang2019manual}. 
One line of work is to learn the patterns of how programmers edit the source code, which can be used to check syntax errors while compiling.
\citet{bhatia2016automated} and \citet{santos2018syntax} 
proposed RNN-based language models for correcting syntax errors in programs.
DeepFix~\cite{gupta2017deepfix} and SequenceR~\cite{chen2019sequencer} are two sequence-to-sequence models for syntax error correction, by translating the erroneous programs into fixed ones. 
Furthermore, \citet{gupta2018deep} improved program repair by reinforcement learning.
\citet{vasic2018neural} proposed multi-headed pointer networks (one head each for localization and repair) for jointly localizing and repairing misused variables in code.
\citet{dinella2020hoppity} presented Hoppity to jointly detect and fix bugs based on neural Turing machine~\cite{graves2014neural}, where a GNN-based memory unit is designed for buggy program representation, and an LSTM-based central controller is designed to predict the operations of bug fixing, e.g., patch generation and type prediction.
\citet{tarlow2020learning} proposed Graph2Diff, which designs a GNN for representing the graph structure of programs, and a pointer network to localize the initial AST to be edited.
\citet{mesbah2019deepdelta} and \citet{chakraborty2020codit} proposed to model the modifications of ASTs, and designed a neural machine translation model to generate correct patches. 
\citet{zhu2021syntax} presented a syntax-directed decoder network with placeholder generation for program repair, which aims to generate program modifications rather than the target code.
\citet{yasunaga2020graph} proposed DrRepair, which first builds a program-feedback graph to align the corresponding symbols and diagnostic feedback, and then designs a GNN to generate repaired code. 
\citet{li2022DEAR} introduced a novel deep learning-based method for fixing general bugs, which combines spectrum-based fault localization with deep learning and flow analysis.

Benefiting from the pre-training techniques in NLP, TFix~\cite{berabi2021tfix} and VulRepair~\cite{fu2022vulrepair} directly posed program repair as a text-to-text problem and utilized a model named T5~\cite{raffel2020exploring}.
Specifically, it digests the error message and directly outputs the correct code. 
\citet{jiang2021cure} proposed CURE for program repair, which is composed of a pre-trained language model, a code-aware search method, and a sub-word tokenization technique.

Another line of work is focusing on repairing programs by generating patches.
\citet{tufano2018empirical} carried out an empirical study to evaluate the viability of applying machine translation to generate patches for program repair in real-world scenarios.
Different from~\cite{tufano2018empirical} which targets at function-level small code snippets,
\citet{hata2018learning} trained a neural machine translation model, targeting at statements, by learning from the corresponding pre- and post-correction code in previous commits. 
\citet{harer2018learning} proposed to generate the input buggy code via generative adversarial networks so that the correction model can be trained without labeled pairs.
\citet{gupta2020synthesize} embedded execution traces in order 
to predict a sequence of edits for repairing Karel programs.
\citet{li2020dlfix} treated the program repair as code transformation and introduced two neural networks, a tree-based RNN for learning the context of a bug patch, and another one designed to learn the code transformation of fixing bugs.
\citet{white2019sorting} introduced a novel approach for selecting and transforming program repair patches using deep-learning-based code similarities.
Empirically, \citet{tian2020evaluating} studied the practicality of patch generation through representation learning of code changes.

\section{Benchmark}\label{sec_benchmark}
Even though significant progress has been made in code intelligence with deep learning, two limitations remain obstacles to the development of this field.
(1) \textit{Lack of standardized implementation for reproducing the results.} It has become a common issue that deep-learning-based models are difficult to reproduce due to the sensitivity to data and hyperparameter tuning.
From our investigation, most of them are implemented independently using different toolkits (i.e., PyTorch and TensorFlow).
There is a need for a unified framework that enables developers to easily evaluate their models by utilizing some shared components.
Actually, in the artificial intelligence area (e.g., NLP and computer vision), many toolkits such as Fairseq~\cite{ott2019fairseq}, AllenNLP~\cite{gardner2018allennlp}, Detectron2~\cite{wu2019detectron2} have been developed, which significantly advance the progress of their corresponding research areas. 
(2) \textit{Lack of benchmarks for fair comparisons.}
Currently, many approaches have been proposed and each of them claims that the proposed approach has outperformed other ones.
To identify where the performance improvements come from, it is essential to create a benchmark for fair comparisons.

Based on these motivations, we propose \tool (standards for Natural Code Comprehension), a  thorough platform for evaluating source code models using deep learning techniques. Under this platform, we also benchmark four specific application tasks, including code summarization, code search, code completion, and type inference.
The implementation and usage of \tool will be introduced in Section~\ref{sec_toolkit}.

\subsection{Code Summarization}
\begin{table}[!t]
	\centering
	\scriptsize
	\caption{Performance of our model and baseline methods for code summarization over Python-Doc dataset. 
    }
 	\vspace{-3mm}
	\label{table_summarization_performance}
		\begin{tabular}{l|ccc|c}
			\hline
			& \textbf{BLEU} &\textbf{METEOR} & \textbf{ROUGE-L} & \textbf{Time Cost}\\
			\hline
			\textbf{Seq2Seq+Attn} &25.57	&14.40	&39.41 &0.09s/Batch      \\
			\textbf{Tree2Seq+Attn} &23.35	&12.59	&36.49 &0.48s/Batch      \\
			\textbf{Transformer} &30.64	&17.65	&44.59 &0.26s/Batch      \\ 
			\textbf{PLBART} &\textbf{32.71}	&\textbf{18.13}	&\textbf{46.05} &0.26s/Batch      \\
			\hline
		\end{tabular}
	\vspace{-4mm}
\end{table}

\subsubsection{Approaches}
Currently, most deep-learning-based code summarization methods use the encoder-decoder architecture.
An encoder network is used to convert the input source code into an embedding vector, and the decoder network is used to generate output summaries from the encoded vector.
In this paper, we benchmark the following representative methods for code summarization, including three different encoders (i.e., LSTM, TreeLSTM, and Transformer) as well as a pre-training-based model.
\begin{itemize}
    \item \textbf{Seq2Seq+Attn}~\cite{iyer2016summarizing,wan2018improving} is a vanilla model following sequence-to-sequence architecture with attention mechanism. 
    It is a famous method for neural machine translation.
    Unlike works that only represent the source code as token embedding~\cite{iyer2016summarizing}, we represent the source code via an LSTM network and generate the summary via another LSTM network.
    \item {\textbf{Tree2Seq+Attn}}~\cite{wan2018improving} also follows the structure of Seq2Seq. The difference is that it uses TreeLSTM as the encoder network for syntax-aware modeling of code. Moreover, an attention module is also designed to attend to different nodes of the syntax tree of code.
    \item {\textbf{Transformer}}~\cite{ahmad2020transformer} 
    is currently considered the leading approach for code summarization, which has also achieved significant improvement in neural machine translation. 
    In Transformer, a relative position embedding, rather than absolute position embedding, is introduced for modeling the positions of code tokens.
    \item \textbf{PLBART}~\cite{ahmad2021unified} is built on the top of BART~\cite{lewis2020bart}, which is originally designed for text understanding and generation. PLBART can be seen as a specific BART model pre-trained on code corpus.
\end{itemize}

\vspace{-2mm}
\subsubsection{Results}
We evaluate the performance of each model on the Python-Doc~\cite{barone2017parallel,wan2018improving} dataset using the BLEU, METEOR, and ROUGE metrics as in~\cite{wan2018improving}.
The overall performance is summarized in Table~\ref{table_summarization_performance}. 
This table shows that PLBART, which utilizes the Transformer architecture and pre-training techniques, achieves the highest performance.
It is interesting to see that the simple Seq2Seq+Attn outperforms the Tree2Seq+Attn that considers the AST of code. 
For Transformer, we find that the relative position embedding can indeed represent the relative relationships among code tokens.

\subsection{Code Search} 
\begin{table}[!t]
	\centering
	\scriptsize
	\caption{MRR of our model and baseline methods for code search over CodeSearchNet dataset. 
	}
	\vspace{-3mm}
	\label{table_retrieval_performance_all}
	\begin{tabular}{l|cccccc|c}
		\hline
		& \textbf{Go} &\textbf{Java} & \textbf{JavaScript} & \textbf{PHP} & \textbf{Python} & \textbf{Ruby} &\textbf{Time Cost} \\
		\hline
		\textbf{NBOW} &66.59	&59.92	&47.15	&54.75	 	&63.33	&42.86  &0.16s/Batch      \\
		\textbf{1D-CNN} &70.87	&60.49	&38.81	&61.92	 	&67.29	&36.53  &0.30s/Batch      \\
		\textbf{biRNN} &65.80	&48.60	&23.23	&51.36	 	&48.28	&19.35  &0.74s/Batch      \\
		\textbf{SelfAtt}  &\textbf{78.45}	&\textbf{66.55}	&\textbf{50.38}	&\textbf{65.78}	&\textbf{79.09}	&\textbf{47.96} &0.25s/Batch  \\
		\hline
	\end{tabular}
	\vspace{-4mm}
\end{table}


\subsubsection{Approaches}
\textsc{CodeSearchNet} Challenge~\cite{husain2019codesearchnet} is an open challenge designed to assess the current state of code search. In~\cite{husain2019codesearchnet}, the authors have benchmarked four code search methods.
The fundamental idea of~\cite{husain2019codesearchnet} is to learn a joint embedding of code and natural-language query in a shared vector space. 
That is, two encoders are used for representing the source code and query, respectively.
A loss function is then designed to maximize the weighted sum for paired embeddings of source code and natural-language query.
Based on different encoder networks, we have implemented the following four variant models.
\begin{itemize}
    \item \textbf{Neural Bag of Words (NBOW)}~\cite{husain2019codesearchnet} is a naive approach 
    by representing the input sequences by a bag of words. For a given code snippet or some specified query written in natural language, it represents tokens into a collection of word embeddings before feeding them into a max pooling layer for creating a sentence-level representation.
	\item \textbf{Bidirectional RNN models (biRNN)}~\cite{husain2019codesearchnet} proposes to represent the semantics of source code and query via RNN models. Specially, we adopt the two-layer bidirectional LSTM network.
	\item \textbf{1D Convolutional Neural Network (1D-CNN)}~\cite{husain2019codesearchnet} employs convolutional neural layers for code and query representation, and builds a residual connection at each layer. 
	\item \textbf{Self-Attention (SelfAtt)}~\cite{husain2019codesearchnet} adopts 
 self-attention layers to capture the semantic information of sequential source code and query.
\end{itemize}

\vspace{-2mm}
\subsubsection{Implementation Details}
We employ word-level BPE to tokenize both code snippets and natural-language descriptions in the considered methods. Subsequently, a shared vocabulary of size $50,000$ is constructed based on the sorted token frequency.
All models undergo training on a singular Nvidia RTX V100 GPU, utilizing a learning rate of $5 \times 10^{-4}$. The gradient norm is maintained at $1.0$, and a batch size of $1,000$ is specified to expedite training. The optimization process for all models is executed using the Adam optimizer.

\vspace{-2mm}
\subsubsection{Results}
We evaluate the performance of each model on the CodeSearchNet corpus using the MRR metric, as described in~\cite{husain2019codesearchnet}.
The overall performance of each model is summarized in Table~\ref{table_retrieval_performance_all}. 
As shown in the table, it is clear that the NBOW model with the simplest architecture achieves a comparable performance, at the lowest cost.
Moreover, we can also observe that the performance of biRNN is poor, in both effectiveness and efficiency. The recurrent characteristic of RNN makes it time-consuming.
The SelfAttn model obtains the best results, which may be attributed to its use of the self-attention mechanism.

\subsection{Code Completion}
\begin{table}[!t]
	\centering
	\scriptsize
	\caption{MRR of our model and baseline methods for code completion over Py150 dataset. 
	}
	\vspace{-3mm}
	\label{table_completion_performance}
	\begin{tabular}{l|ccccc|c}
		\hline
		& \textbf{Attribute} & \textbf{Number} & \textbf{Identifier} & \textbf{Parameter} & \textbf{All Tokens} & \textbf{Time Cost}\\
		\hline
		\textbf{LSTM} &51.67	&47.45	&46.52	&66.06	&73.73  &0.31s/Batch      \\
		\textbf{GPT-2} &70.37	&62.20	&63.84 &\textbf{73.54}	&82.17  &0.43s/Batch      \\
		\textbf{TravTrans} &\textbf{72.08}	&\textbf{68.55} &\textbf{76.33}	&71.08	&\textbf{83.17}  &0.43s/Batch      \\
		\hline
	\end{tabular}
	\vspace{-4mm}
\end{table}

\subsubsection{Approaches}
The code completion task aims to generate the completion text based on the given partial code.
In this paper, we investigate three representative approaches.
\begin{itemize}
	\item \textbf{LSTM}~\cite{kim2021code} denotes the model that represents the partial code by LSTM, and then predicts the missing token via a softmax layer. 
	\item \textbf{GPT-2}~\cite{kim2021code} is a pre-trained language model based on Transformer. It refers to the Transformer model that is trained by iteratively predicting the next code token. 
	\item \textbf{TravTrans}~\cite{kim2021code} is designed to preserve the syntax structure of source code while predicting the missing token. It first linearizes the code ASTs into a sequence of tokens using depth-first traversing, and afterward feeds the traversal into Transformer for representation. It also uses a softmax layer to predict the missing token.
\end{itemize}

\vspace{-2mm}
\subsubsection{Implementation Details}
For acquiring high-quality code tokens, we perform preprocessing on the code snippets by parsing them into ASTs and extracting their leaf nodes as code tokens.
We establish a unified vocabulary comprising $50,000$ tokens, organized based on token frequency.
All models undergo training utilizing four Nvidia RTX V100 GPUs, employing a learning rate of $1 \times 10^{-3}$, and a batch size of $32$.
The optimization of all models is executed using the Adam optimizer.

\vspace{-2mm}
\subsubsection{Results}
We evaluate each model on the Py150~\cite{raychev2016probabilistic} dataset using the MRR metric as used in~\cite{kim2021code}.
We divide the prediction tokens into five categories, namely attributes, numeric constants, identifier names, function parameters and all tokens.
We summarize the performance of each model in Table~\ref{table_completion_performance}.
From this table, when comparing GPT-2 with LSTM, we can observe that the Transformer architecture outperforms other models in representing the semantics of code, thus, resulting in better performance for code completion.
Furthermore, when comparing TravTrans with GPT-2, we can see that the TravTrans that incorporates the syntax structure information achieves better performance, showing that the syntax information is useful for code completion. 

\subsection{Type Inference}
\begin{table}[!t]
	\centering
	\scriptsize
	\caption{Accuracy of our model and baseline methods for type inference over Py150 dataset. 
	}
	\vspace{-3mm}
	\label{table_inference_performance}
	\begin{tabular}{l|cc|cc|c}
		\hline
		& \textbf{Accuracy@1} &\textbf{Accuracy@5} & \textbf{Accuracy@1} & \textbf{Accuracy@5} &\multirow{2}{*}{\textbf{Time Cost}} \\
		\cline{2-5}
		&\multicolumn{2}{c|}{All types}&\multicolumn{2}{c|}{Any types} \\
		\hline
		\textbf{DeepTyper}  &\textbf{0.52} &\textbf{0.67} &\textbf{0.43} &0.67 &0.42s/Batch      \\
		\textbf{Transformer} &0.34 &0.64 &0.37 &\textbf{0.75}  &0.85s/Batch      \\
		\hline
	\end{tabular}
	\vspace{-4mm}
\end{table}

\subsubsection{Approaches}
Similar to code completion, the type inference task aims to predict the types of variables based on contextual information. It first represents the contextual code into a vector, and then predicts the missing types by a softmax layer.
In our work, we employ two state-of-the-art methods for this task. 

\begin{itemize}
	\item \textbf{DeepTyper}~\cite{hellendoorn2018deep} proposes to represent the contextual code by a two-layer biGRU, and then predicts the missing variable types via a softmax layer.
	\item \textbf{Transformer}~\cite{ahmad2020transformer} proposes to represent the contextual code by a Transformer encoder network, and then predicts the missing variable types via a softmax layer.
\end{itemize}

\vspace{-2mm}
\subsubsection{Implementation Details}
We initially tokenize both the code snippets and natural-language descriptions. Subsequently, we establish a common vocabulary comprising $40,000$ tokens, determined by sorting them based on frequency.
The hardware configuration for training and the optimizer employed remains consistent with the aforementioned specifications. A batch size of $16$ and a learning rate of $1 \times 10^{-4}$ are utilized.


\vspace{-2mm}
\subsubsection{Results}
We evaluate each model on the Py150~\cite{raychev2016probabilistic}, by using the Accuracy metric as in~\cite{jain2021contrastive}.
In particular, we measure the performance under the settings of \textit{all types} and \textit{any types}.
The performance of different models is summarized in Table~\ref{table_inference_performance}.
From this table, it is interesting to see that the simple LSTM-based DeepTyper outperforms the Transformer-based approach, especially under the \textit{all types} setting, at a lower time cost.

\section{Toolkit and Demonstration}\label{sec_toolkit}
\begin{figure*}[t!]
	\centering
	\includegraphics[width=\textwidth]{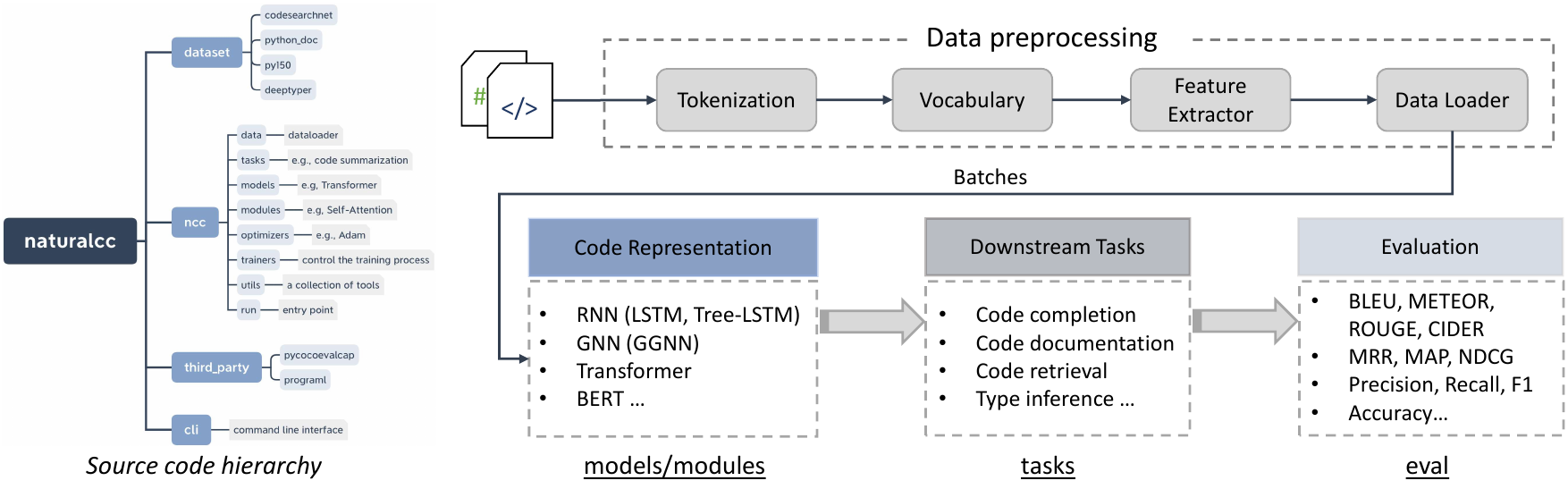}
 	\vspace{-3mm}
	\caption{The source code hierarchy and pipeline of \toolnospace.}
	\label{fig_naturalcc_pipeline}
	\vspace{-6mm}
\end{figure*}

This section introduces the design of \tool and its user interface.
Figure~\ref{fig_naturalcc_pipeline} (left) shows the code structure of \toolnospace. 
\forParaphrase{The \texttt{dataset} folder contains data preprocessing code. The \texttt{ncc} folder is the core module. The \texttt{third\_party} folder holds model evaluation packages. The \texttt{gui} folder contains graphical user interface files and assets. }{}
As shown in Figure~\ref{fig_naturalcc_pipeline} (right), \tool is composed of four components, i.e., data preprocessing, code representation, downstream tasks, and their corresponding evaluations.
At the stage of data preprocessing, we process the source code with a series of steps, including word tokenization, building vocabulary, and feature extraction.
Additionally, a data loader is used to iteratively yield batches of code samples with their features.
The resulting batches are then sent into the code representation models, which facilitate a variety of downstream tasks, including code summarization, code search, code completion, and type inference.
To evaluate the performance of each task, we also implement several corresponding metrics that have been widely adopted previously.


\subsection{Data Preprocessing Module}
In \toolnospace, we have collected and processed four datasets
including CodeSearchNet~\cite{husain2019codesearchnet}, Python-Doc~\cite{wan2018improving}, Py150~\cite{raychev2016probabilistic}, and DeepTyper~\cite{hellendoorn2018deep}.
First, we tokenize the input source code, and then build a vocabulary to map the code tokens into indexes.
Currently, we support two types of tokenizations: space tokenizer and BPE tokenizer~\cite{karampatsis2020big}.
Along with code tokens, we also explore different features of code, such as AST, IR, CFGs, and DFGs.
All the related scripts for data preprocessing have been put in the \texttt{data} and \texttt{dataset} folders.

\subsection{Code Representation Module}
As the core component of \toolnospace, we have implemented several encoders that are widely used in state-of-the-art approaches for source code representation, including RNN, GNN, and Transformer.
For example, we have implemented LSTM, TreeLSTM and Transformer networks for sequential tokens and (linearized) ASTs.
We have also implemented a GNN, i.e., GGNN, to represent the control-flow graph of source code.
It is worth mentioning that in \toolnospace, we have also incorporated the pre-training approaches for source code. We have implemented several state-of-the-art pre-trained code models, including CodeBERT~\cite{feng2020codebert}, PLBART~\cite{ahmad2021unified}, and GPT-2~\cite{lu2021codexglue}.
The \texttt{models} and \texttt{modules} folders contain all the implemented networks for code representation.


\subsection{Tool Implementation}
\tool is mainly implemented by PyTorch, and builds upon 
other successful open-source toolkits in NLP, such as Fairseq, and AllenNLP.

\noindent{\textbf{Registry Mechanism.}}
To be flexible, \tool is expected to be easily extended to different tasks and model implementations, with minimum modification.
Similar to Fairseq, 
we design a \texttt{register} decorator on instantiating a new task or model, the implementation of which is in the corresponding \texttt{\_\_init\_\_.py} in each folder. 
The registry mechanism is to create a global variable to store all the available tasks, models, and objects at the initialization stage, so that users can easily access them throughout the whole project.

\noindent{\textbf{Efficient Training.}}
\tool supports efficient training of models in a distributed way through \texttt{torch.distributed}. It can utilize multiple GPUs across different servers.
Furthermore, \tool can support calculation in mixed precision to further increase the training speed, including both FP32 and FP16 training.
Typically, the gradients are updated in FP16 while the parameters are saved in FP32.


\noindent{\textbf{Flexible Configuration.}}
Instead of employing \texttt{argparse} for managing command-line options within Fairseq, we advocate the adoption of individual \texttt{yaml} configuration files for each model's configuration. We contend that the flexibility offered by modifying these \texttt{yaml} configuration files is better suited for model exploration.

\subsection{Graphical User Interface}
We also design a Web system as a graphical user interface to help users explore the results of trained models. 
The design is based on the open-source demonstration of AllenNLP~\cite{gardner2018allennlp}.
Figure~\ref{fig_demo} shows the screenshot of our demonstration system.
Currently, we have implemented three tasks that are related to code intelligence,
i.e., code summarization, code search, and code completion.
We leave the integration of other related tasks to our future work.

\subsection{Leaderboard}
\begin{figure}[t!]
	\begin{subfigure}{.495\textwidth}
		\centering
		\includegraphics[width=\linewidth]{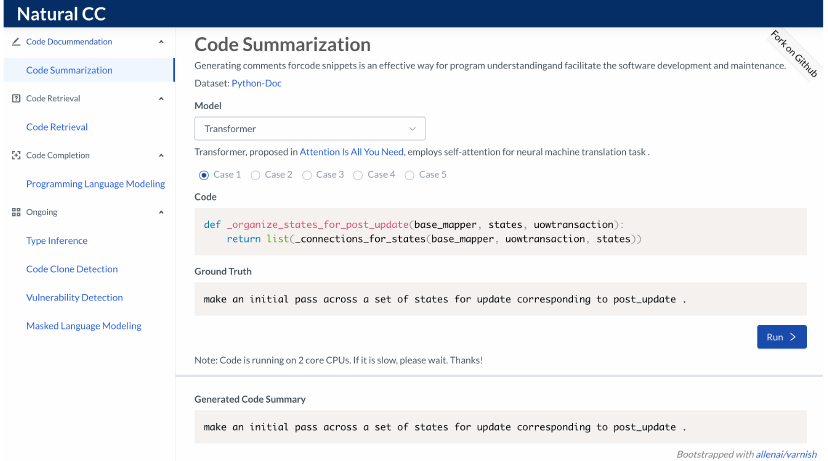}
		\caption{Demonstration}
		\label{fig_demo}
	\end{subfigure}
	\begin{subfigure}{.495\textwidth}
		\centering
		\includegraphics[width=\linewidth]{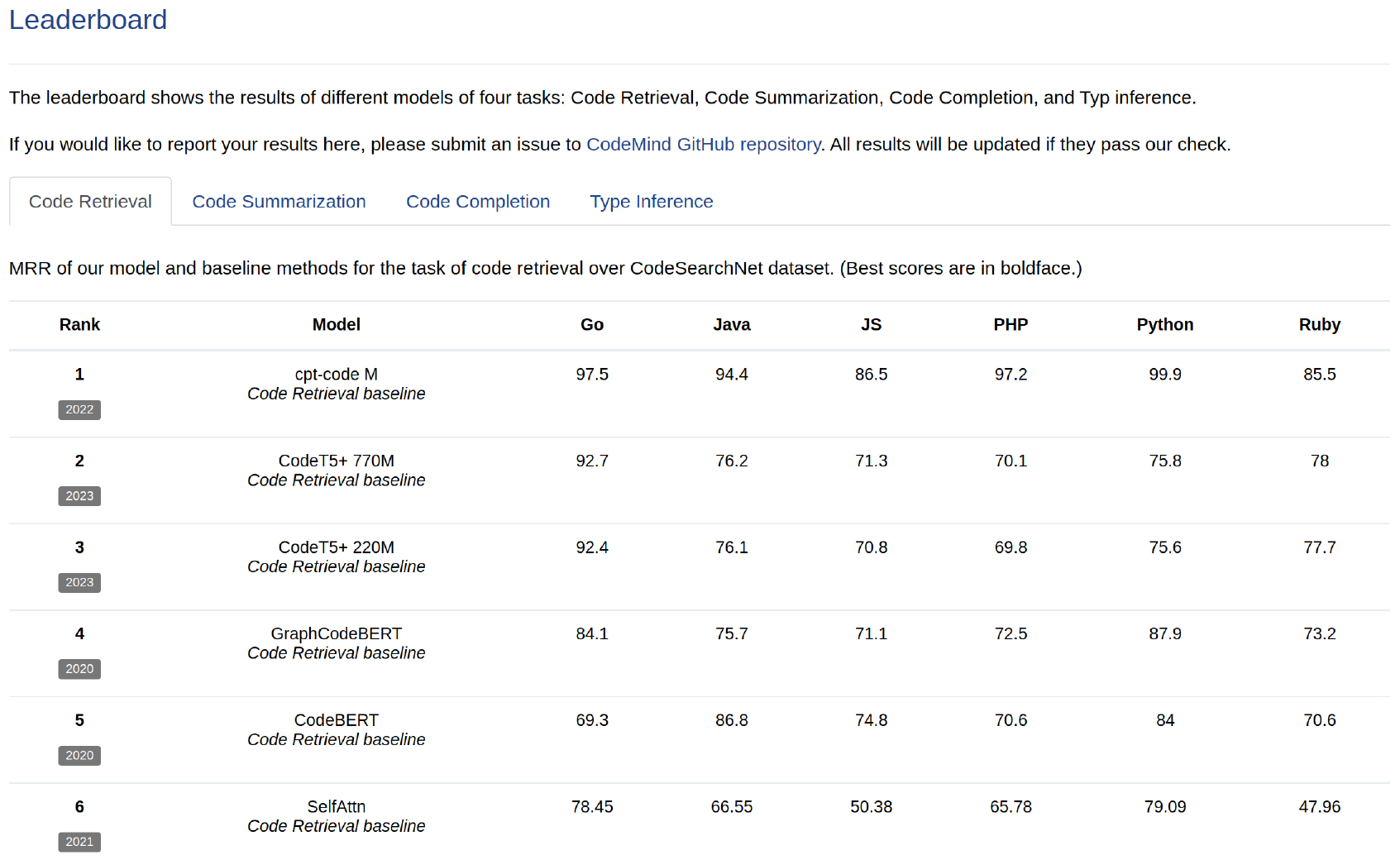}
		\caption{Leaderboard}
		\label{fig_leadboard}
	\end{subfigure}
 	\vspace{-3mm}
	\caption{Screenshots of GUI and leaderboard of \toolnospace.}
	\label{fig:fig}
	\vspace{-4mm}
\end{figure}

We also develop a leaderboard so that researchers can report the results of their own models and compete with others, as shown in Figure~\ref{fig_leadboard}.
Currently, we only support researchers and developers who use \tool to implement their approach and update the experimental results via pull requests in GitHub.
In our future work, we will build a web-based service, which allows users to upload their predicted results 
and evaluate the model performance automatically using the ground-truth labels as a reference. 

\section{Challenges and Opportunities}\label{sec_challenges}
Although much effort has been made into deep learning for code intelligence, this area of research is still in its infancy with many open challenges and opportunities.
To inspire future research, this section suggests several potential directions that are worth pursuing.

\noindent\textbf{{Comprehensive Code Representation.}}
Designing a representation approach to effectively and efficiently preserve the semantics of programs has always been a fundamental problem in code intelligence. Despite much effort 
on code representation, as mentioned in this paper, there are still three main obstacles to be overcome.
\textit{(a) Open Vocabulary.}
Building a vocabulary to index the textual tokens of code is the first step toward applying deep learning models for code intelligence.
Since the unambiguous characteristic of code, the vocabulary in code is much more open and complicated than the vocabulary in natural languages. The vocabulary of programming languages often consists of keywords, identifiers, customized method names, and variable names. The large vocabulary contains much ``noise'', making it difficult to comprehend the code.
Although many attempts~\cite{cvitkovic2019open,karampatsis2020big,chirkova2021simple} have been made towards mitigating the OOV issue, it remains a challenge to design a simple yet effective approach to map the source code into indexes while preserving the semantics. 
\textit{(b) Complex Structure of Program.}
Unlike natural language, code is written with strict grammar.
The computations described by code can be executed in an order that is different from the order in which the code was written. This is often seen in operations such as loops, recursions, and pointer manipulation.
Although many attempts to capture the structure of code from different modalities, as we surveyed in this paper, we believe that the structures of code are not sufficiently preserved, and more effort is needed here.
Inspired by the GNNs, there is potential to design specific GNNs to better represent the structure of programs.
For example, from our analysis, ASTs, CFGs, DFGs and CPGs all have high heterogeneity.
It is desirable to design some heterogeneous-information-network-based approaches~\cite{sun2022heterogeneous} to represent the heterogeneous code graph.
\textit{(c) Big Models of Code.} Despite the significant progress made by pre-trained code models in code intelligence, pre-training on a large-scale code corpus is still computationally expensive and very costly. Recently, \citet{zhang2022diet} and~\citet{shi2022compressing} proposed to improve the efficiency of the training process by model compressing. It is a promising research direction to reduce the computational resources of pre-trained code models.

\noindent\textbf{{Data Hungry and Data Quality.}}
Despite much progress achieved in deep-learning-based approaches for code intelligence,
we argue that existing approaches still suffer from the data-hungry issue. 
In other words, the effectiveness of cutting-edge techniques significantly depends on the availability of vast quantities of expensive and labor-intensive well-labeled training data.
Training the model on a small qualified dataset will result in far less imprecise results, especially for new programming languages or languages with an inadequate number of labeled samples.
Therefore, it is important to design approaches to reduce the reliance on a large quantity of labeled data.
A similar problem exists in the field of machine learning.
One promising solution for this dilemma is transfer learning, which has achieved great success in transferring knowledge to alleviate the data-hungry issue in computer vision and NLP.
Similarly, to model an emerging programming language with limited data, it is desirable to mitigate the data-hungry issue by leveraging models trained in programming languages with sufficient labeled training data~\cite{chai2022cross,cui2022zeroshot,chen2022transferability}.
Data quality is also a crucial issue for code intelligence, which may exacerbate the data-hungry problem.
From our analysis, the collected datasets from online resources, like GitHub and StackOverflow, are not quality ensured.
\citet{sun2022on} and \citet{shi2022are} investigated the importance of data quality and verified it on the tasks of code search and code summarization, respectively.

\noindent\textbf{{Multi-Lingual and Cross-Language.}}
The codebase written in multiple programming languages is can be considered a multi-lingual corpus, as in NLP.
However, the multi-lingual problem in programming languages has not been well investigated.
Different from the multi-lingual problems studied in NLP, the corpus of multiple programming languages will bring more opportunities and challenges to future research. 
Recently, several attempts have been made to learn the common knowledge shared among multiple programming languages, and transfer the knowledge across different programming languages. 
For example, 
\citet{zhang2021disentangled} proposed obtaining better interpretability and generalizability by
disentangling the semantics of source code from multiple programming languages based on variational autoencoders.
\citet{zugner2021language} introduced a language-agnostic code representation based on the features directly extracted from the AST.
\citet{ahmed2021multilingual} conducted an exploratory study and revealed the evidence that multilingual property indeed exists in the source code corpora. 
For example, it is more likely that programs that solve the same problem in different languages make use of the same or similar identifier names.
They also investigate the effect of multilingual (pre-)training for code summarization and code search.
\citet{nafi2019clcdsa} proposed CLCDSA, a cross-language clone detector with syntactical features and API documentation.
\citet{bui2019bilateral} proposed a bilateral neural network for the task of cross-language algorithm classification. 
\citet{bui2019sar} proposed SAR, which can learn cross-language API mappings with minimal knowledge.
Recently, \citet{chai2022cross} proposed a novel approach termed CDCS for domain-specific code search through transfer learning across programming languages. 
\citet{gui2022cross} proposed an approach that matches source code and binary code across different languages
based on intermediate representation.

\noindent\textbf{{Model Interpretability.}}
Lack of interpretability is a common challenge for most deep learning-based techniques for code intelligence, as deep learning is a black-box method.
New methods and studies on interpreting the working mechanisms of deep neural networks should be a potential research direction. 
Recently, several efforts have been made toward 
increasing the interpretability of deep-learning-based models.
As an example, 
\citet{li2021vulnerability} presented a novel approach to explain predicted results for GNN-based vulnerability detection by extracting sub-graphs in the program dependency graph.
In addition, \citet{zou2021interpreting} proposed interpreting a deep-learning-based model for vulnerability detection by identifying a limited number of tokens that play a significant role in the final prediction of the detectors.
\citet{zhang2021interpretable} proposed interpretable program synthesis that allows users to see the synthesis process and have control over the synthesizer.
\citet{pornprasit2021pyexplainer} proposed a local rule-based model-agnostic approach, termed PyExplainer, to explain the predictions of just-in-time defect models.
\citet{rabin2021understanding} proposed a model-agnostic explainer based on program simplification, inspired by the delta debugging algorithms.
\citet{wan2022what},~\citet{lopez2022ast}, and~\citet{sharma2022exploratory} investigated the explainability of pre-trained code models through probing the code attention and hidden representations.
We believe that it is essential to enhance the interpretability of current deep-learning-based approaches for code intelligence.


\noindent\textbf{{Robustness and Security.}}
Despite significant progress being made in the training of accurate models for code intelligence, the robustness and security of these models have rarely been explored.
As seen in the fields of NLP and CV, deep neural networks are frequently not robust~\cite{carlini2017towards}. 
Specifically, current deep learning models can be easily deceived by adversarial examples, which are created by making small changes to the inputs of the model that it would consider as benign.
There are many different ways to produce adversarial samples in the computer vision and NLP communities, particularly for image classification~\cite{eykholt2018robust,carlini2017towards,carlini2019evaluating} and sentiment classification~\cite{zhang2020adversarial}.
Similarly, for source code models, the adversarial attack also exists.
Recently, there have been several efforts to investigate the robustness and security of deep-learning-based models for code intelligence.
For example, 
\citet{ramakrishnan2020semantic} and \citet{yefet2020adversarial} investigated 
how to improve the robustness of source code models through adversarial training.
\citet{nguyen2021adversarial} empirically investigated the use of adversarial learning techniques for API recommendation.
\citet{bielik2020adversarial} introduced a novel method that incorporates adversarial training and representation refinement to create precise and robust models of source code.
\citet{zhou2022adversarial}, \citet{yang2022natural} and \citet{zhang2020generating} proposed a black-box attack for neural code models by generating adversarial examples while preserving the semantics of source code.
Based on semantics-preserving code transformations,
\citet{quiring2019misleading} and \citet{liu2021practical} developed a novel attack against authorship attribution of source code.
\citet{ramakrishnan2022backdoors} investigated the possibility of injecting a number of common backdoors into deep-learning-based models, and developed a protection approach based on spectral signatures.
\citet{schuster2021you} and \citet{wan2022you}  proposed attacking the neural code models through data poisoning, and verified it in code completion and code search, respectively.
\citet{severi2021explanation} suggested an explanation-guided backdoor approach to attack the malware classifiers.
Overall, exploring the robustness and security of code intelligence models is an interesting and important research direction.
\section{Conclusion}\label{sec_conclusion}
In this paper, we study deep learning for code intelligence by conducting a comprehensive survey, establishing a benchmark, as well as developing an open-source toolkit.
We begin by providing a thorough literature review on deep learning for code intelligence, from the perspectives of code representations, deep learning techniques, application tasks, and public datasets.
We then present an open-source toolkit for code intelligence, termed \toolnospace.
On top of \toolnospace, we have benchmarked four popular application tasks about code intelligence, i.e., code summarization, code search, code completion, and type inference. 
We hope that our study contributes to a better understanding of the current status of code intelligence. We also hope that our toolkit and benchmark will contribute to the development of better code intelligence models.
\vspace{-2mm}



\bibliographystyle{ACM-Reference-Format}
\bibliography{ref}


\begin{thebibliography}{300}


\ifx \showCODEN    \undefined \def \showCODEN     #1{\unskip}     \fi
\ifx \showDOI      \undefined \def \showDOI       #1{#1}\fi
\ifx \showISBNx    \undefined \def \showISBNx     #1{\unskip}     \fi
\ifx \showISBNxiii \undefined \def \showISBNxiii  #1{\unskip}     \fi
\ifx \showISSN     \undefined \def \showISSN      #1{\unskip}     \fi
\ifx \showLCCN     \undefined \def \showLCCN      #1{\unskip}     \fi
\ifx \shownote     \undefined \def \shownote      #1{#1}          \fi
\ifx \showarticletitle \undefined \def \showarticletitle #1{#1}   \fi
\ifx \showURL      \undefined \def \showURL       {\relax}        \fi
\providecommand\bibfield[2]{#2}
\providecommand\bibinfo[2]{#2}
\providecommand\natexlab[1]{#1}
\providecommand\showeprint[2][]{arXiv:#2}

\bibitem[git(2019)]%
        {github}
 \bibinfo{year}{2019}\natexlab{}.
\newblock \bibinfo{title}{{GitHub}}.
\newblock \bibinfo{howpublished}{\url{https://www.github.com}}.
\newblock
\newblock
\shownote{[Online; accessed 1-May-2019]}.


\bibitem[sta(2019)]%
        {stackoverflow}
 \bibinfo{year}{2019}\natexlab{}.
\newblock \bibinfo{title}{{StackOverflow}}.
\newblock \bibinfo{howpublished}{\url{https://www.stackoverflow.com}}.
\newblock
\newblock
\shownote{[Online; accessed 1-May-2019]}.


\bibitem[Ahmad et~al\mbox{.}(2021)]%
        {ahmad2021unified}
\bibfield{author}{\bibinfo{person}{Wasi Ahmad}, \bibinfo{person}{Saikat Chakraborty}, \bibinfo{person}{Baishakhi Ray}, {and} \bibinfo{person}{Kai-Wei Chang}.} \bibinfo{year}{2021}\natexlab{}.
\newblock \showarticletitle{Unified Pre-training for Program Understanding and Generation}. In \bibinfo{booktitle}{\emph{NAACL}}. \bibinfo{pages}{2655--2668}.
\newblock


\bibitem[Ahmad et~al\mbox{.}(2020)]%
        {ahmad2020transformer}
\bibfield{author}{\bibinfo{person}{Wasi~Uddin Ahmad}, \bibinfo{person}{Saikat Chakraborty}, \bibinfo{person}{Baishakhi Ray}, {and} \bibinfo{person}{Kai{-}Wei Chang}.} \bibinfo{year}{2020}\natexlab{}.
\newblock \showarticletitle{A Transformer-based Approach for Source Code Summarization}. In \bibinfo{booktitle}{\emph{ACL}}. \bibinfo{pages}{4998--5007}.
\newblock


\bibitem[Ahmed and Devanbu(2022)]%
        {ahmed2021multilingual}
\bibfield{author}{\bibinfo{person}{Toufique Ahmed} {and} \bibinfo{person}{Premkumar Devanbu}.} \bibinfo{year}{2022}\natexlab{}.
\newblock \showarticletitle{Multilingual training for Software Engineering}. In \bibinfo{booktitle}{\emph{ICSE}}.
\newblock


\bibitem[Allamanis et~al\mbox{.}(2018a)]%
        {allamanis2018survey}
\bibfield{author}{\bibinfo{person}{Miltiadis Allamanis}, \bibinfo{person}{Earl~T Barr}, \bibinfo{person}{Premkumar Devanbu}, {and} \bibinfo{person}{Charles Sutton}.} \bibinfo{year}{2018}\natexlab{a}.
\newblock \showarticletitle{A survey of machine learning for big code and naturalness}.
\newblock \bibinfo{journal}{\emph{ACM Computing Surveys (CSUR)}} \bibinfo{volume}{51}, \bibinfo{number}{4} (\bibinfo{year}{2018}), \bibinfo{pages}{1--37}.
\newblock


\bibitem[Allamanis et~al\mbox{.}(2020)]%
        {allamanis2020typilus}
\bibfield{author}{\bibinfo{person}{Miltiadis Allamanis}, \bibinfo{person}{Earl~T Barr}, \bibinfo{person}{Soline Ducousso}, {and} \bibinfo{person}{Zheng Gao}.} \bibinfo{year}{2020}\natexlab{}.
\newblock \showarticletitle{Typilus: neural type hints}. In \bibinfo{booktitle}{\emph{PLDI}}. \bibinfo{pages}{91--105}.
\newblock


\bibitem[Allamanis and Brockschmidt(2017)]%
        {allamanis2017smartpaste}
\bibfield{author}{\bibinfo{person}{Miltiadis Allamanis} {and} \bibinfo{person}{Marc Brockschmidt}.} \bibinfo{year}{2017}\natexlab{}.
\newblock \showarticletitle{Smartpaste: Learning to adapt source code}.
\newblock \bibinfo{journal}{\emph{arXiv:1705.07867}} (\bibinfo{year}{2017}).
\newblock


\bibitem[Allamanis et~al\mbox{.}(2018b)]%
        {allamanis2017learning}
\bibfield{author}{\bibinfo{person}{Miltiadis Allamanis}, \bibinfo{person}{Marc Brockschmidt}, {and} \bibinfo{person}{Mahmoud Khademi}.} \bibinfo{year}{2018}\natexlab{b}.
\newblock \showarticletitle{Learning to Represent Programs with Graphs}. In \bibinfo{booktitle}{\emph{ICLR}}.
\newblock


\bibitem[Allamanis et~al\mbox{.}(2016)]%
        {allamanis2016convolutional}
\bibfield{author}{\bibinfo{person}{Miltiadis Allamanis}, \bibinfo{person}{Hao Peng}, {and} \bibinfo{person}{Charles Sutton}.} \bibinfo{year}{2016}\natexlab{}.
\newblock \showarticletitle{A convolutional attention network for extreme summarization of source code}. In \bibinfo{booktitle}{\emph{ICML}}. \bibinfo{pages}{2091--2100}.
\newblock


\bibitem[Alon et~al\mbox{.}(2018)]%
        {alon2018code2seq}
\bibfield{author}{\bibinfo{person}{Uri Alon}, \bibinfo{person}{Shaked Brody}, \bibinfo{person}{Omer Levy}, {and} \bibinfo{person}{Eran Yahav}.} \bibinfo{year}{2018}\natexlab{}.
\newblock \showarticletitle{code2seq: Generating Sequences from Structured Representations of Code}. In \bibinfo{booktitle}{\emph{ICLR}}.
\newblock


\bibitem[Alon et~al\mbox{.}(2020)]%
        {alon2020structural}
\bibfield{author}{\bibinfo{person}{Uri Alon}, \bibinfo{person}{Roy Sadaka}, \bibinfo{person}{Omer Levy}, {and} \bibinfo{person}{Eran Yahav}.} \bibinfo{year}{2020}\natexlab{}.
\newblock \showarticletitle{Structural language models of code}. In \bibinfo{booktitle}{\emph{ICML}}. \bibinfo{pages}{245--256}.
\newblock


\bibitem[Alon et~al\mbox{.}(2019)]%
        {alon2019code2vec}
\bibfield{author}{\bibinfo{person}{Uri Alon}, \bibinfo{person}{Meital Zilberstein}, \bibinfo{person}{Omer Levy}, {and} \bibinfo{person}{Eran Yahav}.} \bibinfo{year}{2019}\natexlab{}.
\newblock \showarticletitle{code2vec: Learning distributed representations of code}.
\newblock \bibinfo{journal}{\emph{POPL}}  \bibinfo{volume}{3} (\bibinfo{year}{2019}), \bibinfo{pages}{1--29}.
\newblock


\bibitem[Andreessen(2011)]%
        {andreessen2011software}
\bibfield{author}{\bibinfo{person}{Marc Andreessen}.} \bibinfo{year}{2011}\natexlab{}.
\newblock \showarticletitle{Why software is eating the world}.
\newblock \bibinfo{journal}{\emph{Wall Street Journal}} \bibinfo{volume}{20}, \bibinfo{number}{2011} (\bibinfo{year}{2011}), \bibinfo{pages}{C2}.
\newblock


\bibitem[Balog et~al\mbox{.}(2017)]%
        {balog2016deepcoder}
\bibfield{author}{\bibinfo{person}{Matej Balog}, \bibinfo{person}{Alexander~L. Gaunt}, \bibinfo{person}{Marc Brockschmidt}, \bibinfo{person}{Sebastian Nowozin}, {and} \bibinfo{person}{Daniel Tarlow}.} \bibinfo{year}{2017}\natexlab{}.
\newblock \showarticletitle{DeepCoder: Learning to Write Programs}. In \bibinfo{booktitle}{\emph{ICLR}}.
\newblock


\bibitem[Bansal et~al\mbox{.}(2021)]%
        {bansal2021project}
\bibfield{author}{\bibinfo{person}{Aakash Bansal}, \bibinfo{person}{Sakib Haque}, {and} \bibinfo{person}{Collin McMillan}.} \bibinfo{year}{2021}\natexlab{}.
\newblock \showarticletitle{Project-Level Encoding for Neural Source Code Summarization of Subroutines}. In \bibinfo{booktitle}{\emph{ICPC}}. \bibinfo{publisher}{{IEEE}}, \bibinfo{pages}{253--264}.
\newblock


\bibitem[Barone and Sennrich(2017)]%
        {barone2017parallel}
\bibfield{author}{\bibinfo{person}{Antonio Valerio~Miceli Barone} {and} \bibinfo{person}{Rico Sennrich}.} \bibinfo{year}{2017}\natexlab{}.
\newblock \showarticletitle{A Parallel Corpus of Python Functions and Documentation Strings for Automated Code Documentation and Code Generation}. In \bibinfo{booktitle}{\emph{IJCNLP}}. \bibinfo{pages}{314--319}.
\newblock


\bibitem[Bavishi et~al\mbox{.}(2019)]%
        {bavishi2019autopandas}
\bibfield{author}{\bibinfo{person}{Rohan Bavishi}, \bibinfo{person}{Caroline Lemieux}, \bibinfo{person}{Roy Fox}, \bibinfo{person}{Koushik Sen}, {and} \bibinfo{person}{Ion Stoica}.} \bibinfo{year}{2019}\natexlab{}.
\newblock \showarticletitle{AutoPandas: neural-backed generators for program synthesis}.
\newblock \bibinfo{journal}{\emph{OOPSLA}}  \bibinfo{volume}{3} (\bibinfo{year}{2019}), \bibinfo{pages}{1--27}.
\newblock


\bibitem[Beltagy and Quirk(2016)]%
        {beltagy2016improved}
\bibfield{author}{\bibinfo{person}{Islam Beltagy} {and} \bibinfo{person}{Chris Quirk}.} \bibinfo{year}{2016}\natexlab{}.
\newblock \showarticletitle{Improved semantic parsers for if-then statements}. In \bibinfo{booktitle}{\emph{ACL}}. \bibinfo{pages}{726--736}.
\newblock


\bibitem[Ben{-}Nun et~al\mbox{.}(2018)]%
        {ben2018neural}
\bibfield{author}{\bibinfo{person}{Tal Ben{-}Nun}, \bibinfo{person}{Alice~Shoshana Jakobovits}, {and} \bibinfo{person}{Torsten Hoefler}.} \bibinfo{year}{2018}\natexlab{}.
\newblock \showarticletitle{Neural Code Comprehension: {A} Learnable Representation of Code Semantics}. In \bibinfo{booktitle}{\emph{NeurIPS}}. \bibinfo{pages}{3589--3601}.
\newblock


\bibitem[Berabi et~al\mbox{.}(2021)]%
        {berabi2021tfix}
\bibfield{author}{\bibinfo{person}{Berkay Berabi}, \bibinfo{person}{Jingxuan He}, \bibinfo{person}{Veselin Raychev}, {and} \bibinfo{person}{Martin Vechev}.} \bibinfo{year}{2021}\natexlab{}.
\newblock \showarticletitle{TFix: Learning to Fix Coding Errors with a Text-to-Text Transformer}. In \bibinfo{booktitle}{\emph{ICML}}. \bibinfo{pages}{780--791}.
\newblock


\bibitem[Bhatia and Singh(2016)]%
        {bhatia2016automated}
\bibfield{author}{\bibinfo{person}{Sahil Bhatia} {and} \bibinfo{person}{Rishabh Singh}.} \bibinfo{year}{2016}\natexlab{}.
\newblock \showarticletitle{Automated correction for syntax errors in programming assignments using recurrent neural networks}.
\newblock \bibinfo{journal}{\emph{arXiv:1603.06129}} (\bibinfo{year}{2016}).
\newblock


\bibitem[Bielik and Vechev(2020)]%
        {bielik2020adversarial}
\bibfield{author}{\bibinfo{person}{Pavol Bielik} {and} \bibinfo{person}{Martin Vechev}.} \bibinfo{year}{2020}\natexlab{}.
\newblock \showarticletitle{Adversarial robustness for code}. In \bibinfo{booktitle}{\emph{ICML}}. \bibinfo{pages}{896--907}.
\newblock


\bibitem[Brockschmidt et~al\mbox{.}(2019)]%
        {brockschmidt2018generative}
\bibfield{author}{\bibinfo{person}{Marc Brockschmidt}, \bibinfo{person}{Miltiadis Allamanis}, \bibinfo{person}{Alexander~L. Gaunt}, {and} \bibinfo{person}{Oleksandr Polozov}.} \bibinfo{year}{2019}\natexlab{}.
\newblock \showarticletitle{Generative Code Modeling with Graphs}. In \bibinfo{booktitle}{\emph{ICLR}}.
\newblock


\bibitem[Brody et~al\mbox{.}(2020)]%
        {brody2020structural}
\bibfield{author}{\bibinfo{person}{Shaked Brody}, \bibinfo{person}{Uri Alon}, {and} \bibinfo{person}{Eran Yahav}.} \bibinfo{year}{2020}\natexlab{}.
\newblock \showarticletitle{A structural model for contextual code changes}.
\newblock \bibinfo{journal}{\emph{OOPSLA}}  \bibinfo{volume}{4} (\bibinfo{year}{2020}), \bibinfo{pages}{1--28}.
\newblock


\bibitem[Brown et~al\mbox{.}(2020)]%
        {brown2020language}
\bibfield{author}{\bibinfo{person}{Tom Brown}, \bibinfo{person}{Benjamin Mann}, \bibinfo{person}{Nick Ryder}, \bibinfo{person}{Melanie Subbiah}, \bibinfo{person}{Jared~D Kaplan}, \bibinfo{person}{Prafulla Dhariwal}, {et~al\mbox{.}}} \bibinfo{year}{2020}\natexlab{}.
\newblock \showarticletitle{Language models are few-shot learners}.
\newblock \bibinfo{journal}{\emph{NeurIPS}}  \bibinfo{volume}{33} (\bibinfo{year}{2020}), \bibinfo{pages}{1877--1901}.
\newblock


\bibitem[B{\"u}ch and Andrzejak(2019)]%
        {buch2019learning}
\bibfield{author}{\bibinfo{person}{Lutz B{\"u}ch} {and} \bibinfo{person}{Artur Andrzejak}.} \bibinfo{year}{2019}\natexlab{}.
\newblock \showarticletitle{Learning-based recursive aggregation of abstract syntax trees for code clone detection}. In \bibinfo{booktitle}{\emph{SANER}}. \bibinfo{pages}{95--104}.
\newblock


\bibitem[Bui et~al\mbox{.}(2019a)]%
        {bui2019bilateral}
\bibfield{author}{\bibinfo{person}{Nghi~DQ Bui}, \bibinfo{person}{Yijun Yu}, {and} \bibinfo{person}{Lingxiao Jiang}.} \bibinfo{year}{2019}\natexlab{a}.
\newblock \showarticletitle{Bilateral dependency neural networks for cross-language algorithm classification}. In \bibinfo{booktitle}{\emph{SANER}}. \bibinfo{pages}{422--433}.
\newblock


\bibitem[Bui et~al\mbox{.}(2019b)]%
        {bui2019sar}
\bibfield{author}{\bibinfo{person}{Nghi~DQ Bui}, \bibinfo{person}{Yijun Yu}, {and} \bibinfo{person}{Lingxiao Jiang}.} \bibinfo{year}{2019}\natexlab{b}.
\newblock \showarticletitle{SAR: learning cross-language API mappings with little knowledge}. In \bibinfo{booktitle}{\emph{ESEC/FSE}}. \bibinfo{pages}{796--806}.
\newblock


\bibitem[Bui et~al\mbox{.}(2021a)]%
        {bui2021infercode}
\bibfield{author}{\bibinfo{person}{Nghi~DQ Bui}, \bibinfo{person}{Yijun Yu}, {and} \bibinfo{person}{Lingxiao Jiang}.} \bibinfo{year}{2021}\natexlab{a}.
\newblock \showarticletitle{InferCode: Self-Supervised Learning of Code Representations by Predicting Subtrees}. In \bibinfo{booktitle}{\emph{ICSE}}. \bibinfo{pages}{1186--1197}.
\newblock


\bibitem[Bui et~al\mbox{.}(2021b)]%
        {bui2021selfsupervised}
\bibfield{author}{\bibinfo{person}{Nghi D.~Q. Bui}, \bibinfo{person}{Yijun Yu}, {and} \bibinfo{person}{Lingxiao Jiang}.} \bibinfo{year}{2021}\natexlab{b}.
\newblock \showarticletitle{Self-Supervised Contrastive Learning for Code Retrieval and Summarization via Semantic-Preserving Transformations}. In \bibinfo{booktitle}{\emph{SIGIR}}. \bibinfo{publisher}{{ACM}}, \bibinfo{pages}{511--521}.
\newblock


\bibitem[Cai et~al\mbox{.}(2018)]%
        {cai2017encoder}
\bibfield{author}{\bibinfo{person}{Ruichu Cai}, \bibinfo{person}{Boyan Xu}, \bibinfo{person}{Zhenjie Zhang}, \bibinfo{person}{Xiaoyan Yang}, \bibinfo{person}{Zijian Li}, {and} \bibinfo{person}{Zhihao Liang}.} \bibinfo{year}{2018}\natexlab{}.
\newblock \showarticletitle{An Encoder-Decoder Framework Translating Natural Language to Database Queries}. In \bibinfo{booktitle}{\emph{IJCAI}}. \bibinfo{pages}{3977--3983}.
\newblock


\bibitem[Cambronero et~al\mbox{.}(2019)]%
        {cambronero2019deep}
\bibfield{author}{\bibinfo{person}{Jose Cambronero}, \bibinfo{person}{Hongyu Li}, \bibinfo{person}{Seohyun Kim}, \bibinfo{person}{Koushik Sen}, {and} \bibinfo{person}{Satish Chandra}.} \bibinfo{year}{2019}\natexlab{}.
\newblock \showarticletitle{When deep learning met code search}. In \bibinfo{booktitle}{\emph{ESEC/FSE}}. \bibinfo{pages}{964--974}.
\newblock


\bibitem[Cao et~al\mbox{.}(2022)]%
        {cao2022mvd}
\bibfield{author}{\bibinfo{person}{Sicong Cao}, \bibinfo{person}{Xiaobing Sun}, \bibinfo{person}{Lili Bo}, \bibinfo{person}{Rongxin Wu}, \bibinfo{person}{Bin Li}, {and} \bibinfo{person}{Chuanqi Tao}.} \bibinfo{year}{2022}\natexlab{}.
\newblock \showarticletitle{{MVD:} Memory-Related Vulnerability Detection Based on Flow-Sensitive Graph Neural Networks}. In \bibinfo{booktitle}{\emph{ICSE}}. \bibinfo{pages}{1456--1468}.
\newblock


\bibitem[Carlini et~al\mbox{.}(2019)]%
        {carlini2019evaluating}
\bibfield{author}{\bibinfo{person}{Nicholas Carlini}, \bibinfo{person}{Anish Athalye}, \bibinfo{person}{Nicolas Papernot}, \bibinfo{person}{Wieland Brendel}, \bibinfo{person}{Jonas Rauber}, \bibinfo{person}{Dimitris Tsipras}, \bibinfo{person}{Ian Goodfellow}, \bibinfo{person}{Aleksander Madry}, {and} \bibinfo{person}{Alexey Kurakin}.} \bibinfo{year}{2019}\natexlab{}.
\newblock \showarticletitle{On evaluating adversarial robustness}.
\newblock \bibinfo{journal}{\emph{arXiv:1902.06705}} (\bibinfo{year}{2019}).
\newblock


\bibitem[Carlini and Wagner(2017)]%
        {carlini2017towards}
\bibfield{author}{\bibinfo{person}{Nicholas Carlini} {and} \bibinfo{person}{David Wagner}.} \bibinfo{year}{2017}\natexlab{}.
\newblock \showarticletitle{Towards evaluating the robustness of neural networks}. In \bibinfo{booktitle}{\emph{S\&P}}. \bibinfo{pages}{39--57}.
\newblock


\bibitem[Chai et~al\mbox{.}(2022a)]%
        {chai2022ernie}
\bibfield{author}{\bibinfo{person}{Yekun Chai}, \bibinfo{person}{Shuohuan Wang}, \bibinfo{person}{Chao Pang}, \bibinfo{person}{Yu Sun}, \bibinfo{person}{Hao Tian}, {and} \bibinfo{person}{Hua Wu}.} \bibinfo{year}{2022}\natexlab{a}.
\newblock \showarticletitle{ERNIE-Code: Beyond English-Centric Cross-lingual Pretraining for Programming Languages}.
\newblock \bibinfo{journal}{\emph{arXiv preprint arXiv:2212.06742}} (\bibinfo{year}{2022}).
\newblock


\bibitem[Chai et~al\mbox{.}(2022b)]%
        {chai2022cross}
\bibfield{author}{\bibinfo{person}{Yitian Chai}, \bibinfo{person}{Hongyu Zhang}, \bibinfo{person}{Beijun Shen}, {and} \bibinfo{person}{Xiaodong Gu}.} \bibinfo{year}{2022}\natexlab{b}.
\newblock \showarticletitle{Cross-Domain Deep Code Search with Meta Learning}. In \bibinfo{booktitle}{\emph{ICSE}}. \bibinfo{pages}{487--498}.
\newblock


\bibitem[Chakraborty et~al\mbox{.}(2020)]%
        {chakraborty2020codit}
\bibfield{author}{\bibinfo{person}{Saikat Chakraborty}, \bibinfo{person}{Yangruibo Ding}, \bibinfo{person}{Miltiadis Allamanis}, {and} \bibinfo{person}{Baishakhi Ray}.} \bibinfo{year}{2020}\natexlab{}.
\newblock \showarticletitle{Codit: Code editing with tree-based neural models}.
\newblock \bibinfo{journal}{\emph{TSE}} (\bibinfo{year}{2020}).
\newblock


\bibitem[Chakraborty and Ray(2021)]%
        {chakraborty2021multimodal}
\bibfield{author}{\bibinfo{person}{Saikat Chakraborty} {and} \bibinfo{person}{Baishakhi Ray}.} \bibinfo{year}{2021}\natexlab{}.
\newblock \showarticletitle{On Multi-Modal Learning of Editing Source Code}. In \bibinfo{booktitle}{\emph{ASE}}. \bibinfo{publisher}{{IEEE}}, \bibinfo{pages}{443--455}.
\newblock


\bibitem[Chen et~al\mbox{.}(2022)]%
        {chen2022transferability}
\bibfield{author}{\bibinfo{person}{Fuxiang Chen}, \bibinfo{person}{Fatemeh~H. Fard}, \bibinfo{person}{David Lo}, {and} \bibinfo{person}{Timofey Bryksin}.} \bibinfo{year}{2022}\natexlab{}.
\newblock \showarticletitle{On the transferability of pre-trained language models for low-resource programming languages}. In \bibinfo{booktitle}{\emph{ICPC}}. \bibinfo{publisher}{{ACM}}, \bibinfo{pages}{401--412}.
\newblock


\bibitem[Chen et~al\mbox{.}(2021b)]%
        {chen2021evaluating}
\bibfield{author}{\bibinfo{person}{Mark Chen}, \bibinfo{person}{Jerry Tworek}, \bibinfo{person}{Heewoo Jun}, \bibinfo{person}{Qiming Yuan}, \bibinfo{person}{Henrique Ponde de~Oliveira Pinto}, \bibinfo{person}{Jared Kaplan}, \bibinfo{person}{Harri Edwards}, {et~al\mbox{.}}} \bibinfo{year}{2021}\natexlab{b}.
\newblock \showarticletitle{Evaluating large language models trained on code}.
\newblock \bibinfo{journal}{\emph{arXiv preprint arXiv:2107.03374}} (\bibinfo{year}{2021}).
\newblock


\bibitem[Chen et~al\mbox{.}(2018)]%
        {chen2018tree}
\bibfield{author}{\bibinfo{person}{Xinyun Chen}, \bibinfo{person}{Chang Liu}, {and} \bibinfo{person}{Dawn Song}.} \bibinfo{year}{2018}\natexlab{}.
\newblock \showarticletitle{Tree-to-tree Neural Networks for Program Translation}. In \bibinfo{booktitle}{\emph{NeurIPS}}. \bibinfo{pages}{2552--2562}.
\newblock


\bibitem[Chen et~al\mbox{.}(2021a)]%
        {chen2021plur}
\bibfield{author}{\bibinfo{person}{Zimin Chen}, \bibinfo{person}{Vincent Hellendoorn}, \bibinfo{person}{Pascal Lamblin}, \bibinfo{person}{Petros Maniatis}, \bibinfo{person}{Pierre-Antoine Manzagol}, {et~al\mbox{.}}} \bibinfo{year}{2021}\natexlab{a}.
\newblock \showarticletitle{PLUR: A Unifying, Graph-Based View of Program Learning, Understanding, and Repair}.
\newblock \bibinfo{journal}{\emph{NeurIPS}}  \bibinfo{volume}{34} (\bibinfo{year}{2021}).
\newblock


\bibitem[Chen et~al\mbox{.}(2019)]%
        {chen2019sequencer}
\bibfield{author}{\bibinfo{person}{Zimin Chen}, \bibinfo{person}{Steve~James Kommrusch}, \bibinfo{person}{Michele Tufano}, \bibinfo{person}{Louis-No{\"e}l Pouchet}, \bibinfo{person}{Denys Poshyvanyk}, {and} \bibinfo{person}{Martin Monperrus}.} \bibinfo{year}{2019}\natexlab{}.
\newblock \showarticletitle{Sequencer: Sequence-to-sequence learning for end-to-end program repair}.
\newblock \bibinfo{journal}{\emph{TSE}} (\bibinfo{year}{2019}).
\newblock


\bibitem[Cheng et~al\mbox{.}(2021)]%
        {cheng2021deepwukong}
\bibfield{author}{\bibinfo{person}{Xiao Cheng}, \bibinfo{person}{Haoyu Wang}, \bibinfo{person}{Jiayi Hua}, \bibinfo{person}{Guoai Xu}, {and} \bibinfo{person}{Yulei Sui}.} \bibinfo{year}{2021}\natexlab{}.
\newblock \showarticletitle{DeepWukong: Statically detecting software vulnerabilities using deep graph neural network}.
\newblock \bibinfo{journal}{\emph{TOSEM}} \bibinfo{volume}{30}, \bibinfo{number}{3} (\bibinfo{year}{2021}), \bibinfo{pages}{1--33}.
\newblock


\bibitem[Chicco(2021)]%
        {chicco2021siamese}
\bibfield{author}{\bibinfo{person}{Davide Chicco}.} \bibinfo{year}{2021}\natexlab{}.
\newblock \showarticletitle{Siamese neural networks: An overview}.
\newblock \bibinfo{journal}{\emph{Artificial Neural Networks}} (\bibinfo{year}{2021}), \bibinfo{pages}{73--94}.
\newblock


\bibitem[Chirkova and Troshin(2021a)]%
        {chirkova2021empirical}
\bibfield{author}{\bibinfo{person}{Nadezhda Chirkova} {and} \bibinfo{person}{Sergey Troshin}.} \bibinfo{year}{2021}\natexlab{a}.
\newblock \showarticletitle{Empirical study of transformers for source code}. In \bibinfo{booktitle}{\emph{ESEC/FSE}}. \bibinfo{pages}{703--715}.
\newblock


\bibitem[Chirkova and Troshin(2021b)]%
        {chirkova2021simple}
\bibfield{author}{\bibinfo{person}{Nadezhda Chirkova} {and} \bibinfo{person}{Sergey Troshin}.} \bibinfo{year}{2021}\natexlab{b}.
\newblock \showarticletitle{A Simple Approach for Handling Out-of-Vocabulary Identifiers in Deep Learning for Source Code}. In \bibinfo{booktitle}{\emph{NAACL}}. \bibinfo{pages}{278--288}.
\newblock


\bibitem[Chowdhery et~al\mbox{.}(2022)]%
        {chowdhery2022palm}
\bibfield{author}{\bibinfo{person}{Aakanksha Chowdhery}, \bibinfo{person}{Sharan Narang}, \bibinfo{person}{Jacob Devlin}, \bibinfo{person}{Maarten Bosma}, \bibinfo{person}{Gaurav Mishra}, \bibinfo{person}{Adam Roberts}, \bibinfo{person}{Paul Barham}, \bibinfo{person}{Hyung~Won Chung}, \bibinfo{person}{Charles Sutton}, \bibinfo{person}{Sebastian Gehrmann}, {et~al\mbox{.}}} \bibinfo{year}{2022}\natexlab{}.
\newblock \showarticletitle{Palm: Scaling language modeling with pathways}.
\newblock \bibinfo{journal}{\emph{arXiv preprint arXiv:2204.02311}} (\bibinfo{year}{2022}).
\newblock


\bibitem[Christopoulou et~al\mbox{.}(2022)]%
        {christopoulou2022pangu}
\bibfield{author}{\bibinfo{person}{Fenia Christopoulou}, \bibinfo{person}{Gerasimos Lampouras}, \bibinfo{person}{Milan Gritta}, \bibinfo{person}{Guchun Zhang}, \bibinfo{person}{Yinpeng Guo}, \bibinfo{person}{Zhongqi Li}, \bibinfo{person}{Qi Zhang}, \bibinfo{person}{Meng Xiao}, \bibinfo{person}{Bo Shen}, \bibinfo{person}{Lin Li}, {et~al\mbox{.}}} \bibinfo{year}{2022}\natexlab{}.
\newblock \showarticletitle{Pangu-coder: Program synthesis with function-level language modeling}.
\newblock \bibinfo{journal}{\emph{arXiv preprint arXiv:2207.11280}} (\bibinfo{year}{2022}).
\newblock


\bibitem[Ciurumelea et~al\mbox{.}(2020)]%
        {ciurumelea2020suggesting}
\bibfield{author}{\bibinfo{person}{Adelina Ciurumelea}, \bibinfo{person}{Sebastian Proksch}, {and} \bibinfo{person}{Harald~C Gall}.} \bibinfo{year}{2020}\natexlab{}.
\newblock \showarticletitle{Suggesting comment completions for python using neural language models}. In \bibinfo{booktitle}{\emph{SANER}}. \bibinfo{pages}{456--467}.
\newblock


\bibitem[Cui et~al\mbox{.}(2022)]%
        {cui2022zeroshot}
\bibfield{author}{\bibinfo{person}{Nan Cui}, \bibinfo{person}{Yuze Jiang}, \bibinfo{person}{Xiaodong Gu}, {and} \bibinfo{person}{Beijun Shen}.} \bibinfo{year}{2022}\natexlab{}.
\newblock \showarticletitle{Zero-shot program representation learning}. In \bibinfo{booktitle}{\emph{ICPC}}. \bibinfo{publisher}{{ACM}}, \bibinfo{pages}{60--70}.
\newblock


\bibitem[Cummins et~al\mbox{.}(2021)]%
        {cummins2020programl}
\bibfield{author}{\bibinfo{person}{Chris Cummins}, \bibinfo{person}{Zacharias Fisches}, \bibinfo{person}{Tal Ben-Nun}, \bibinfo{person}{Torsten Hoefler}, \bibinfo{person}{Michael O'Boyle}, {and} \bibinfo{person}{Hugh Leather}.} \bibinfo{year}{2021}\natexlab{}.
\newblock \showarticletitle{{ProGraML: A Graph-based Program Representation for Data Flow Analysis and Compiler Optimizations}}. In \bibinfo{booktitle}{\emph{ICML}}.
\newblock


\bibitem[Cummins et~al\mbox{.}(2017)]%
        {cummins2017synthesizing}
\bibfield{author}{\bibinfo{person}{Chris Cummins}, \bibinfo{person}{Pavlos Petoumenos}, \bibinfo{person}{Zheng Wang}, {and} \bibinfo{person}{Hugh Leather}.} \bibinfo{year}{2017}\natexlab{}.
\newblock \showarticletitle{Synthesizing benchmarks for predictive modeling}. In \bibinfo{booktitle}{\emph{CGO}}. \bibinfo{pages}{86--99}.
\newblock


\bibitem[Cvitkovic et~al\mbox{.}(2019)]%
        {cvitkovic2019open}
\bibfield{author}{\bibinfo{person}{Milan Cvitkovic}, \bibinfo{person}{Badal Singh}, {and} \bibinfo{person}{Animashree Anandkumar}.} \bibinfo{year}{2019}\natexlab{}.
\newblock \showarticletitle{Open vocabulary learning on source code with a graph-structured cache}. In \bibinfo{booktitle}{\emph{ICML}}. \bibinfo{pages}{1475--1485}.
\newblock


\bibitem[Dam et~al\mbox{.}(2018)]%
        {dam2018automatic}
\bibfield{author}{\bibinfo{person}{Hoa~Khanh Dam}, \bibinfo{person}{Truyen Tran}, \bibinfo{person}{Trang Thi~Minh Pham}, \bibinfo{person}{Shien~Wee Ng}, \bibinfo{person}{John Grundy}, {and} \bibinfo{person}{Aditya Ghose}.} \bibinfo{year}{2018}\natexlab{}.
\newblock \showarticletitle{Automatic feature learning for predicting vulnerable software components}.
\newblock \bibinfo{journal}{\emph{TSE}} (\bibinfo{year}{2018}).
\newblock


\bibitem[Deng et~al\mbox{.}(2022)]%
        {deng2022fine}
\bibfield{author}{\bibinfo{person}{Zhongyang Deng}, \bibinfo{person}{Ling Xu}, \bibinfo{person}{Chao Liu}, \bibinfo{person}{Meng Yan}, \bibinfo{person}{Zhou Xu}, {and} \bibinfo{person}{Yan Lei}.} \bibinfo{year}{2022}\natexlab{}.
\newblock \showarticletitle{Fine-grained Co-Attentive Representation Learning for Semantic Code Search}. In \bibinfo{booktitle}{\emph{SANER}}. \bibinfo{pages}{396--407}.
\newblock


\bibitem[Devanbu et~al\mbox{.}(2020)]%
        {devanbu2020deep}
\bibfield{author}{\bibinfo{person}{Prem Devanbu}, \bibinfo{person}{Matthew~B. Dwyer}, \bibinfo{person}{Sebastian~G. Elbaum}, \bibinfo{person}{Michael Lowry}, \bibinfo{person}{Kevin Moran}, {et~al\mbox{.}}} \bibinfo{year}{2020}\natexlab{}.
\newblock \showarticletitle{Deep Learning {\&} Software Engineering: State of Research and Future Directions}.
\newblock \bibinfo{journal}{\emph{CoRR}}  \bibinfo{volume}{abs/2009.08525} (\bibinfo{year}{2020}).
\newblock


\bibitem[Devlin et~al\mbox{.}(2019)]%
        {devlin2019bert}
\bibfield{author}{\bibinfo{person}{Jacob Devlin}, \bibinfo{person}{Ming{-}Wei Chang}, \bibinfo{person}{Kenton Lee}, {and} \bibinfo{person}{Kristina Toutanova}.} \bibinfo{year}{2019}\natexlab{}.
\newblock \showarticletitle{{BERT:} Pre-training of Deep Bidirectional Transformers for Language Understanding}. In \bibinfo{booktitle}{\emph{NAACL}}. \bibinfo{pages}{4171--4186}.
\newblock


\bibitem[Devlin et~al\mbox{.}(2017)]%
        {devlin2017robustfill}
\bibfield{author}{\bibinfo{person}{Jacob Devlin}, \bibinfo{person}{Jonathan Uesato}, \bibinfo{person}{Surya Bhupatiraju}, \bibinfo{person}{Rishabh Singh}, \bibinfo{person}{Abdel-rahman Mohamed}, {and} \bibinfo{person}{Pushmeet Kohli}.} \bibinfo{year}{2017}\natexlab{}.
\newblock \showarticletitle{Robustfill: Neural program learning under noisy i/o}. In \bibinfo{booktitle}{\emph{ICML}}. \bibinfo{pages}{990--998}.
\newblock


\bibitem[Dinella et~al\mbox{.}(2020)]%
        {dinella2020hoppity}
\bibfield{author}{\bibinfo{person}{Elizabeth Dinella}, \bibinfo{person}{Hanjun Dai}, \bibinfo{person}{Ziyang Li}, \bibinfo{person}{Mayur Naik}, \bibinfo{person}{Le Song}, {and} \bibinfo{person}{Ke Wang}.} \bibinfo{year}{2020}\natexlab{}.
\newblock \showarticletitle{Hoppity: Learning graph transformations to detect and fix bugs in programs}. In \bibinfo{booktitle}{\emph{ICLR}}.
\newblock


\bibitem[Ding et~al\mbox{.}(2022)]%
        {ding2022towards}
\bibfield{author}{\bibinfo{person}{Yangruibo Ding}, \bibinfo{person}{Luca Buratti}, \bibinfo{person}{Saurabh Pujar}, \bibinfo{person}{Alessandro Morari}, \bibinfo{person}{Baishakhi Ray}, {and} \bibinfo{person}{Saikat Chakraborty}.} \bibinfo{year}{2022}\natexlab{}.
\newblock \showarticletitle{Towards Learning (Dis)-Similarity of Source Code from Program Contrasts}. In \bibinfo{booktitle}{\emph{ACL}}. \bibinfo{pages}{6300--6312}.
\newblock


\bibitem[Dong and Lapata(2016)]%
        {dong2016language}
\bibfield{author}{\bibinfo{person}{Li Dong} {and} \bibinfo{person}{Mirella Lapata}.} \bibinfo{year}{2016}\natexlab{}.
\newblock \showarticletitle{Language to Logical Form with Neural Attention}. In \bibinfo{booktitle}{\emph{ACL}}.
\newblock


\bibitem[Eykholt et~al\mbox{.}(2018)]%
        {eykholt2018robust}
\bibfield{author}{\bibinfo{person}{Kevin Eykholt}, \bibinfo{person}{Ivan Evtimov}, \bibinfo{person}{Earlence Fernandes}, \bibinfo{person}{Bo Li}, \bibinfo{person}{Amir Rahmati}, \bibinfo{person}{Chaowei Xiao}, \bibinfo{person}{Atul Prakash}, \bibinfo{person}{Tadayoshi Kohno}, {and} \bibinfo{person}{Dawn Song}.} \bibinfo{year}{2018}\natexlab{}.
\newblock \showarticletitle{Robust physical-world attacks on deep learning visual classification}. In \bibinfo{booktitle}{\emph{CVPR}}. \bibinfo{pages}{1625--1634}.
\newblock


\bibitem[Feng et~al\mbox{.}(2020)]%
        {feng2020codebert}
\bibfield{author}{\bibinfo{person}{Zhangyin Feng}, \bibinfo{person}{Daya Guo}, \bibinfo{person}{Duyu Tang}, \bibinfo{person}{Nan Duan}, \bibinfo{person}{Xiaocheng Feng}, {et~al\mbox{.}}} \bibinfo{year}{2020}\natexlab{}.
\newblock \showarticletitle{CodeBERT: {A} Pre-Trained Model for Programming and Natural Languages}. In \bibinfo{booktitle}{\emph{Findings of EMNLP}}. \bibinfo{pages}{1536--1547}.
\newblock


\bibitem[Fernandes et~al\mbox{.}(2019)]%
        {fernandes2018structured}
\bibfield{author}{\bibinfo{person}{Patrick Fernandes}, \bibinfo{person}{Miltiadis Allamanis}, {and} \bibinfo{person}{Marc Brockschmidt}.} \bibinfo{year}{2019}\natexlab{}.
\newblock \showarticletitle{Structured Neural Summarization}. In \bibinfo{booktitle}{\emph{ICLR}}.
\newblock


\bibitem[Fried et~al\mbox{.}(2022)]%
        {fried2022incoder}
\bibfield{author}{\bibinfo{person}{Daniel Fried}, \bibinfo{person}{Armen Aghajanyan}, \bibinfo{person}{Jessy Lin}, \bibinfo{person}{Sida Wang}, \bibinfo{person}{Eric Wallace}, \bibinfo{person}{Freda Shi}, \bibinfo{person}{Ruiqi Zhong}, \bibinfo{person}{Wen-tau Yih}, \bibinfo{person}{Luke Zettlemoyer}, {and} \bibinfo{person}{Mike Lewis}.} \bibinfo{year}{2022}\natexlab{}.
\newblock \showarticletitle{Incoder: A generative model for code infilling and synthesis}.
\newblock \bibinfo{journal}{\emph{arXiv preprint arXiv:2204.05999}} (\bibinfo{year}{2022}).
\newblock


\bibitem[Fu et~al\mbox{.}(2022)]%
        {fu2022vulrepair}
\bibfield{author}{\bibinfo{person}{Michael Fu}, \bibinfo{person}{Chakkrit Tantithamthavorn}, \bibinfo{person}{Trung Le}, \bibinfo{person}{Van Nguyen}, {and} \bibinfo{person}{Dinh~Q. Phung}.} \bibinfo{year}{2022}\natexlab{}.
\newblock \showarticletitle{VulRepair: a T5-based automated software vulnerability repair}. In \bibinfo{booktitle}{\emph{ESEC/FSE}}. \bibinfo{pages}{935--947}.
\newblock


\bibitem[Gao and Lyu(2022)]%
        {gao2022m2ts}
\bibfield{author}{\bibinfo{person}{Yuexiu Gao} {and} \bibinfo{person}{Chen Lyu}.} \bibinfo{year}{2022}\natexlab{}.
\newblock \showarticletitle{{M2TS:} multi-scale multi-modal approach based on transformer for source code summarization}. In \bibinfo{booktitle}{\emph{ICPC}}. \bibinfo{publisher}{{ACM}}, \bibinfo{pages}{24--35}.
\newblock


\bibitem[Gao et~al\mbox{.}(2021)]%
        {gao2021automating}
\bibfield{author}{\bibinfo{person}{Zhipeng Gao}, \bibinfo{person}{Xin Xia}, \bibinfo{person}{David Lo}, \bibinfo{person}{John Grundy}, {and} \bibinfo{person}{Thomas Zimmermann}.} \bibinfo{year}{2021}\natexlab{}.
\newblock \showarticletitle{Automating the removal of obsolete TODO comments}. In \bibinfo{booktitle}{\emph{ESEC/FSE}}. \bibinfo{pages}{218--229}.
\newblock


\bibitem[Gardner et~al\mbox{.}(2018)]%
        {gardner2018allennlp}
\bibfield{author}{\bibinfo{person}{Matt Gardner}, \bibinfo{person}{Joel Grus}, \bibinfo{person}{Mark Neumann}, \bibinfo{person}{Oyvind Tafjord}, {et~al\mbox{.}}} \bibinfo{year}{2018}\natexlab{}.
\newblock \showarticletitle{AllenNLP: A Deep Semantic Natural Language Processing Platform}. In \bibinfo{booktitle}{\emph{Proceedings of Workshop for NLP Open Source Software (NLP-OSS)}}. \bibinfo{pages}{1--6}.
\newblock


\bibitem[Geng et~al\mbox{.}(2024)]%
        {geng2023large}
\bibfield{author}{\bibinfo{person}{Mingyang Geng}, \bibinfo{person}{Shangwen Wang}, \bibinfo{person}{Dezun Dong}, \bibinfo{person}{Haotian Wang}, \bibinfo{person}{Ge Li}, \bibinfo{person}{Zhi Jin}, \bibinfo{person}{Xiaoguang Mao}, {and} \bibinfo{person}{Xiangke Liao}.} \bibinfo{year}{2024}\natexlab{}.
\newblock \showarticletitle{Large Language Models are Few-Shot Summarizers: Multi-Intent Comment Generation via In-Context Learning}.
\newblock  (\bibinfo{year}{2024}).
\newblock


\bibitem[Gong et~al\mbox{.}(2022)]%
        {gong2022source}
\bibfield{author}{\bibinfo{person}{Zi Gong}, \bibinfo{person}{Cuiyun Gao}, \bibinfo{person}{Yasheng Wang}, \bibinfo{person}{Wenchao Gu}, \bibinfo{person}{Yun Peng}, {and} \bibinfo{person}{Zenglin Xu}.} \bibinfo{year}{2022}\natexlab{}.
\newblock \showarticletitle{Source Code Summarization with Structural Relative Position Guided Transformer}. In \bibinfo{booktitle}{\emph{SANER}}. \bibinfo{pages}{13--24}.
\newblock


\bibitem[Graves et~al\mbox{.}(2014)]%
        {graves2014neural}
\bibfield{author}{\bibinfo{person}{Alex Graves}, \bibinfo{person}{Greg Wayne}, {and} \bibinfo{person}{Ivo Danihelka}.} \bibinfo{year}{2014}\natexlab{}.
\newblock \showarticletitle{Neural turing machines}.
\newblock \bibinfo{journal}{\emph{arXiv:1410.5401}} (\bibinfo{year}{2014}).
\newblock


\bibitem[Gu et~al\mbox{.}(2022)]%
        {gu2022accelerating}
\bibfield{author}{\bibinfo{person}{Wenchao Gu}, \bibinfo{person}{Yanlin Wang}, \bibinfo{person}{Lun Du}, \bibinfo{person}{Hongyu Zhang}, \bibinfo{person}{Shi Han}, \bibinfo{person}{Dongmei Zhang}, {and} \bibinfo{person}{Michael~R. Lyu}.} \bibinfo{year}{2022}\natexlab{}.
\newblock \showarticletitle{Accelerating Code Search with Deep Hashing and Code Classification}. In \bibinfo{booktitle}{\emph{ACL}}. \bibinfo{pages}{2534--2544}.
\newblock


\bibitem[Gu et~al\mbox{.}(2018)]%
        {gu2018deep}
\bibfield{author}{\bibinfo{person}{Xiaodong Gu}, \bibinfo{person}{Hongyu Zhang}, {and} \bibinfo{person}{Sunghun Kim}.} \bibinfo{year}{2018}\natexlab{}.
\newblock \showarticletitle{Deep code search}. In \bibinfo{booktitle}{\emph{ICSE}}. \bibinfo{pages}{933--944}.
\newblock


\bibitem[Gu et~al\mbox{.}(2016)]%
        {gu2016deep}
\bibfield{author}{\bibinfo{person}{Xiaodong Gu}, \bibinfo{person}{Hongyu Zhang}, \bibinfo{person}{Dongmei Zhang}, {and} \bibinfo{person}{Sunghun Kim}.} \bibinfo{year}{2016}\natexlab{}.
\newblock \showarticletitle{Deep API learning}. In \bibinfo{booktitle}{\emph{Proceedings of the 2016 24th ACM SIGSOFT International Symposium on Foundations of Software Engineering}}. \bibinfo{pages}{631--642}.
\newblock


\bibitem[Gu et~al\mbox{.}(2017)]%
        {gu2017deepam}
\bibfield{author}{\bibinfo{person}{Xiaodong Gu}, \bibinfo{person}{Hongyu Zhang}, \bibinfo{person}{Dongmei Zhang}, {and} \bibinfo{person}{Sunghun Kim}.} \bibinfo{year}{2017}\natexlab{}.
\newblock \showarticletitle{DeepAM: Migrate APIs with Multi-Modal Sequence to Sequence Learning}. In \bibinfo{booktitle}{\emph{IJCAI}}. \bibinfo{pages}{3675–3681}.
\newblock


\bibitem[Gui et~al\mbox{.}(2022)]%
        {gui2022cross}
\bibfield{author}{\bibinfo{person}{Yi Gui}, \bibinfo{person}{Yao Wan}, \bibinfo{person}{Hongyu Zhang}, \bibinfo{person}{Huifang Huang}, \bibinfo{person}{Yulei Sui}, \bibinfo{person}{Guandong Xu}, \bibinfo{person}{Zhiyuan Shao}, {and} \bibinfo{person}{Hai Jin}.} \bibinfo{year}{2022}\natexlab{}.
\newblock \showarticletitle{Cross-Language Binary-Source Code Matching with Intermediate Representations}. In \bibinfo{booktitle}{\emph{SANER}}.
\newblock


\bibitem[Gunasekar et~al\mbox{.}(2023)]%
        {gunasekar2023textbooks}
\bibfield{author}{\bibinfo{person}{Suriya Gunasekar}, \bibinfo{person}{Yi Zhang}, \bibinfo{person}{Jyoti Aneja}, \bibinfo{person}{Caio C{\'e}sar~Teodoro Mendes}, \bibinfo{person}{Allie Del~Giorno}, \bibinfo{person}{Sivakanth Gopi}, \bibinfo{person}{Mojan Javaheripi}, \bibinfo{person}{Piero Kauffmann}, \bibinfo{person}{Gustavo de Rosa}, \bibinfo{person}{Olli Saarikivi}, {et~al\mbox{.}}} \bibinfo{year}{2023}\natexlab{}.
\newblock \showarticletitle{Textbooks Are All You Need}.
\newblock \bibinfo{journal}{\emph{arXiv preprint arXiv:2306.11644}} (\bibinfo{year}{2023}).
\newblock


\bibitem[Guo et~al\mbox{.}(2022b)]%
        {guo2022unixcoder}
\bibfield{author}{\bibinfo{person}{Daya Guo}, \bibinfo{person}{Shuai Lu}, \bibinfo{person}{Nan Duan}, \bibinfo{person}{Yanlin Wang}, \bibinfo{person}{Ming Zhou}, {and} \bibinfo{person}{Jian Yin}.} \bibinfo{year}{2022}\natexlab{b}.
\newblock \showarticletitle{UniXcoder: Unified Cross-Modal Pre-training for Code Representation}. In \bibinfo{booktitle}{\emph{ACL}}. \bibinfo{pages}{7212--7225}.
\newblock


\bibitem[Guo et~al\mbox{.}(2021)]%
        {guo2020graphcodebert}
\bibfield{author}{\bibinfo{person}{Daya Guo}, \bibinfo{person}{Shuo Ren}, \bibinfo{person}{Shuai Lu}, \bibinfo{person}{Zhangyin Feng}, \bibinfo{person}{Duyu Tang}, \bibinfo{person}{Shujie Liu}, \bibinfo{person}{Long Zhou}, {et~al\mbox{.}}} \bibinfo{year}{2021}\natexlab{}.
\newblock \showarticletitle{GraphCodeBERT: Pre-training Code Representations with Data Flow}. In \bibinfo{booktitle}{\emph{ICLR}}.
\newblock


\bibitem[Guo et~al\mbox{.}(2022c)]%
        {guo2022learning}
\bibfield{author}{\bibinfo{person}{Daya Guo}, \bibinfo{person}{Alexey Svyatkovskiy}, \bibinfo{person}{Jian Yin}, \bibinfo{person}{Nan Duan}, \bibinfo{person}{Marc Brockschmidt}, {and} \bibinfo{person}{Miltiadis Allamanis}.} \bibinfo{year}{2022}\natexlab{c}.
\newblock \showarticletitle{Learning to Complete Code with Sketches}. In \bibinfo{booktitle}{\emph{ICLR}}.
\newblock


\bibitem[Guo et~al\mbox{.}(2022a)]%
        {guo2022modeling}
\bibfield{author}{\bibinfo{person}{Juncai Guo}, \bibinfo{person}{Jin Liu}, \bibinfo{person}{Yao Wan}, \bibinfo{person}{Li Li}, {and} \bibinfo{person}{Pingyi Zhou}.} \bibinfo{year}{2022}\natexlab{a}.
\newblock \showarticletitle{Modeling Hierarchical Syntax Structure with Triplet Position for Source Code Summarization}. In \bibinfo{booktitle}{\emph{ACL}}. \bibinfo{pages}{486--500}.
\newblock


\bibitem[Gupta et~al\mbox{.}(2020)]%
        {gupta2020synthesize}
\bibfield{author}{\bibinfo{person}{Kavi Gupta}, \bibinfo{person}{Peter~Ebert Christensen}, \bibinfo{person}{Xinyun Chen}, {and} \bibinfo{person}{Dawn Song}.} \bibinfo{year}{2020}\natexlab{}.
\newblock \showarticletitle{Synthesize, Execute and Debug: Learning to Repair for Neural Program Synthesis}. In \bibinfo{booktitle}{\emph{NeurIPS}}.
\newblock


\bibitem[Gupta et~al\mbox{.}(2018)]%
        {gupta2018deep}
\bibfield{author}{\bibinfo{person}{Rahul Gupta}, \bibinfo{person}{Aditya Kanade}, {and} \bibinfo{person}{Shirish Shevade}.} \bibinfo{year}{2018}\natexlab{}.
\newblock \showarticletitle{Deep reinforcement learning for programming language correction}.
\newblock \bibinfo{journal}{\emph{arXiv:1801.10467}} (\bibinfo{year}{2018}).
\newblock


\bibitem[Gupta et~al\mbox{.}(2019)]%
        {gupta2019neural}
\bibfield{author}{\bibinfo{person}{R Gupta}, \bibinfo{person}{A Kanade}, {and} \bibinfo{person}{S Shevade}.} \bibinfo{year}{2019}\natexlab{}.
\newblock \showarticletitle{Neural attribution for semantic bug-localization in student programs}.
\newblock \bibinfo{journal}{\emph{NeurIPS}}  \bibinfo{volume}{32} (\bibinfo{year}{2019}).
\newblock


\bibitem[Gupta et~al\mbox{.}(2017)]%
        {gupta2017deepfix}
\bibfield{author}{\bibinfo{person}{Rahul Gupta}, \bibinfo{person}{Soham Pal}, \bibinfo{person}{Aditya Kanade}, {and} \bibinfo{person}{Shirish Shevade}.} \bibinfo{year}{2017}\natexlab{}.
\newblock \showarticletitle{Deepfix: Fixing common c language errors by deep learning}. In \bibinfo{booktitle}{\emph{AAAI}}.
\newblock


\bibitem[Hadi et~al\mbox{.}(2022)]%
        {hadi2022effectiveness}
\bibfield{author}{\bibinfo{person}{Mohammad~Abdul Hadi}, \bibinfo{person}{Imam Nur~Bani Yusuf}, \bibinfo{person}{Ferdian Thung}, \bibinfo{person}{Kien~Gia Luong}, \bibinfo{person}{Lingxiao Jiang}, \bibinfo{person}{Fatemeh~H. Fard}, {and} \bibinfo{person}{David Lo}.} \bibinfo{year}{2022}\natexlab{}.
\newblock \showarticletitle{On the effectiveness of pretrained models for {API} learning}. In \bibinfo{booktitle}{\emph{ICPC}}. \bibinfo{publisher}{{ACM}}, \bibinfo{pages}{309--320}.
\newblock


\bibitem[Haldar et~al\mbox{.}(2020)]%
        {haldar2020multi}
\bibfield{author}{\bibinfo{person}{Rajarshi Haldar}, \bibinfo{person}{Lingfei Wu}, \bibinfo{person}{JinJun Xiong}, {and} \bibinfo{person}{Julia Hockenmaier}.} \bibinfo{year}{2020}\natexlab{}.
\newblock \showarticletitle{A Multi-Perspective Architecture for Semantic Code Search}. In \bibinfo{booktitle}{\emph{ACL}}. \bibinfo{pages}{8563--8568}.
\newblock


\bibitem[Haque et~al\mbox{.}(2021)]%
        {haque2021action}
\bibfield{author}{\bibinfo{person}{Sakib Haque}, \bibinfo{person}{Aakash Bansal}, \bibinfo{person}{Lingfei Wu}, {and} \bibinfo{person}{Collin McMillan}.} \bibinfo{year}{2021}\natexlab{}.
\newblock \showarticletitle{Action Word Prediction for Neural Source Code Summarization}. In \bibinfo{booktitle}{\emph{SANER}}. \bibinfo{publisher}{{IEEE}}, \bibinfo{pages}{330--341}.
\newblock


\bibitem[Haque et~al\mbox{.}(2020)]%
        {haque2020improved}
\bibfield{author}{\bibinfo{person}{Sakib Haque}, \bibinfo{person}{Alexander LeClair}, \bibinfo{person}{Lingfei Wu}, {and} \bibinfo{person}{Collin McMillan}.} \bibinfo{year}{2020}\natexlab{}.
\newblock \showarticletitle{Improved automatic summarization of subroutines via attention to file context}. In \bibinfo{booktitle}{\emph{MSR}}. \bibinfo{pages}{300--310}.
\newblock


\bibitem[Harer et~al\mbox{.}(2018)]%
        {harer2018learning}
\bibfield{author}{\bibinfo{person}{Jacob Harer}, \bibinfo{person}{Onur Ozdemir}, \bibinfo{person}{Tomo Lazovich}, \bibinfo{person}{Christopher~P. Reale}, \bibinfo{person}{Rebecca~L. Russell}, \bibinfo{person}{Louis~Y. Kim}, {and} \bibinfo{person}{Sang~Peter Chin}.} \bibinfo{year}{2018}\natexlab{}.
\newblock \showarticletitle{Learning to Repair Software Vulnerabilities with Generative Adversarial Networks}. In \bibinfo{booktitle}{\emph{NeurIPS}}. \bibinfo{pages}{7944--7954}.
\newblock


\bibitem[Hassan et~al\mbox{.}(2018)]%
        {hassan2018maxsmt}
\bibfield{author}{\bibinfo{person}{Mostafa Hassan}, \bibinfo{person}{Caterina Urban}, \bibinfo{person}{Marco Eilers}, {and} \bibinfo{person}{Peter M{\"u}ller}.} \bibinfo{year}{2018}\natexlab{}.
\newblock \showarticletitle{MaxSMT-based type inference for Python 3}. In \bibinfo{booktitle}{\emph{International Conference on Computer Aided Verification}}. \bibinfo{pages}{12--19}.
\newblock


\bibitem[Hata et~al\mbox{.}(2018)]%
        {hata2018learning}
\bibfield{author}{\bibinfo{person}{Hideaki Hata}, \bibinfo{person}{Emad Shihab}, {and} \bibinfo{person}{Graham Neubig}.} \bibinfo{year}{2018}\natexlab{}.
\newblock \showarticletitle{Learning to generate corrective patches using neural machine translation}.
\newblock \bibinfo{journal}{\emph{arXiv:1812.07170}} (\bibinfo{year}{2018}).
\newblock


\bibitem[Hayati et~al\mbox{.}(2018)]%
        {hayati2018retrieval}
\bibfield{author}{\bibinfo{person}{Shirley~Anugrah Hayati}, \bibinfo{person}{Raphael Olivier}, \bibinfo{person}{Pravalika Avvaru}, \bibinfo{person}{Pengcheng Yin}, \bibinfo{person}{Anthony Tomasic}, {and} \bibinfo{person}{Graham Neubig}.} \bibinfo{year}{2018}\natexlab{}.
\newblock \showarticletitle{Retrieval-Based Neural Code Generation}. In \bibinfo{booktitle}{\emph{EMNLP}}. \bibinfo{pages}{925--930}.
\newblock


\bibitem[Hellendoorn et~al\mbox{.}(2018)]%
        {hellendoorn2018deep}
\bibfield{author}{\bibinfo{person}{Vincent~J Hellendoorn}, \bibinfo{person}{Christian Bird}, \bibinfo{person}{Earl~T Barr}, {and} \bibinfo{person}{Miltiadis Allamanis}.} \bibinfo{year}{2018}\natexlab{}.
\newblock \showarticletitle{Deep learning type inference}. In \bibinfo{booktitle}{\emph{ESEC/FSE}}. \bibinfo{pages}{152--162}.
\newblock


\bibitem[Henkel et~al\mbox{.}(2018)]%
        {henkel2018code}
\bibfield{author}{\bibinfo{person}{Jordan Henkel}, \bibinfo{person}{Shuvendu~K Lahiri}, \bibinfo{person}{Ben Liblit}, {and} \bibinfo{person}{Thomas Reps}.} \bibinfo{year}{2018}\natexlab{}.
\newblock \showarticletitle{Code vectors: Understanding programs through embedded abstracted symbolic traces}. In \bibinfo{booktitle}{\emph{ESEC/FSE}}. \bibinfo{pages}{163--174}.
\newblock


\bibitem[Hinton et~al\mbox{.}(2006)]%
        {hinton2006fast}
\bibfield{author}{\bibinfo{person}{Geoffrey~E Hinton}, \bibinfo{person}{Simon Osindero}, {and} \bibinfo{person}{Yee-Whye Teh}.} \bibinfo{year}{2006}\natexlab{}.
\newblock \showarticletitle{A fast learning algorithm for deep belief nets}.
\newblock \bibinfo{journal}{\emph{Neural computation}} \bibinfo{volume}{18}, \bibinfo{number}{7} (\bibinfo{year}{2006}), \bibinfo{pages}{1527--1554}.
\newblock


\bibitem[Hoang et~al\mbox{.}(2020)]%
        {hoang2020cc2vec}
\bibfield{author}{\bibinfo{person}{Thong Hoang}, \bibinfo{person}{Hong~Jin Kang}, \bibinfo{person}{David Lo}, {and} \bibinfo{person}{Julia Lawall}.} \bibinfo{year}{2020}\natexlab{}.
\newblock \showarticletitle{Cc2vec: Distributed representations of code changes}. In \bibinfo{booktitle}{\emph{ICSE}}. \bibinfo{pages}{518--529}.
\newblock


\bibitem[Hu et~al\mbox{.}(2018a)]%
        {hu2018deep}
\bibfield{author}{\bibinfo{person}{Xing Hu}, \bibinfo{person}{Ge Li}, \bibinfo{person}{Xin Xia}, \bibinfo{person}{David Lo}, {and} \bibinfo{person}{Zhi Jin}.} \bibinfo{year}{2018}\natexlab{a}.
\newblock \showarticletitle{Deep code comment generation}. In \bibinfo{booktitle}{\emph{ICPC}}. \bibinfo{pages}{200--20010}.
\newblock


\bibitem[Hu et~al\mbox{.}(2018b)]%
        {hu2018summarizing}
\bibfield{author}{\bibinfo{person}{Xing Hu}, \bibinfo{person}{Ge Li}, \bibinfo{person}{Xin Xia}, \bibinfo{person}{David Lo}, \bibinfo{person}{Shuai Lu}, {and} \bibinfo{person}{Zhi Jin}.} \bibinfo{year}{2018}\natexlab{b}.
\newblock \showarticletitle{Summarizing source code with transferred api knowledge.(2018)}. In \bibinfo{booktitle}{\emph{IJCAI}}, Vol.~\bibinfo{volume}{19}. \bibinfo{pages}{2269--2275}.
\newblock


\bibitem[Hu et~al\mbox{.}(2022)]%
        {hu2022treecen}
\bibfield{author}{\bibinfo{person}{Yutao Hu}, \bibinfo{person}{Deqing Zou}, \bibinfo{person}{Junru Peng}, \bibinfo{person}{Yueming Wu}, \bibinfo{person}{Junjie Shan}, {and} \bibinfo{person}{Hai Jin}.} \bibinfo{year}{2022}\natexlab{}.
\newblock \showarticletitle{TreeCen: Building Tree Graph for Scalable Semantic Code Clone Detection}. In \bibinfo{booktitle}{\emph{ASE}}. \bibinfo{publisher}{{ACM}}, \bibinfo{pages}{109:1--109:12}.
\newblock


\bibitem[Huang et~al\mbox{.}(2022)]%
        {huang2022prompt}
\bibfield{author}{\bibinfo{person}{Qing Huang}, \bibinfo{person}{Zhiqiang Yuan}, \bibinfo{person}{Zhenchang Xing}, \bibinfo{person}{Xiwei Xu}, \bibinfo{person}{Liming Zhu}, {and} \bibinfo{person}{Qinghua Lu}.} \bibinfo{year}{2022}\natexlab{}.
\newblock \showarticletitle{Prompt-tuned Code Language Model as a Neural Knowledge Base for Type Inference in Statically-Typed Partial Code}. In \bibinfo{booktitle}{\emph{ASE}}. \bibinfo{publisher}{{ACM}}, \bibinfo{pages}{79:1--79:13}.
\newblock


\bibitem[Husain et~al\mbox{.}(2019)]%
        {husain2019codesearchnet}
\bibfield{author}{\bibinfo{person}{Hamel Husain}, \bibinfo{person}{Ho-Hsiang Wu}, \bibinfo{person}{Tiferet Gazit}, \bibinfo{person}{Miltiadis Allamanis}, {and} \bibinfo{person}{Marc Brockschmidt}.} \bibinfo{year}{2019}\natexlab{}.
\newblock \showarticletitle{Codesearchnet challenge: Evaluating the state of semantic code search}.
\newblock \bibinfo{journal}{\emph{arXiv:1909.09436}} (\bibinfo{year}{2019}).
\newblock


\bibitem[Iyer et~al\mbox{.}(2016)]%
        {iyer2016summarizing}
\bibfield{author}{\bibinfo{person}{Srinivasan Iyer}, \bibinfo{person}{Ioannis Konstas}, \bibinfo{person}{Alvin Cheung}, {and} \bibinfo{person}{Luke Zettlemoyer}.} \bibinfo{year}{2016}\natexlab{}.
\newblock \showarticletitle{Summarizing source code using a neural attention model}. In \bibinfo{booktitle}{\emph{ACL}}. \bibinfo{pages}{2073--2083}.
\newblock


\bibitem[Iyer et~al\mbox{.}(2018)]%
        {iyer2018mapping}
\bibfield{author}{\bibinfo{person}{Srinivasan Iyer}, \bibinfo{person}{Ioannis Konstas}, \bibinfo{person}{Alvin Cheung}, {and} \bibinfo{person}{Luke Zettlemoyer}.} \bibinfo{year}{2018}\natexlab{}.
\newblock \showarticletitle{Mapping Language to Code in Programmatic Context}. In \bibinfo{booktitle}{\emph{EMNLP}}. \bibinfo{pages}{1643--1652}.
\newblock


\bibitem[Jain et~al\mbox{.}(2021)]%
        {jain2021contrastive}
\bibfield{author}{\bibinfo{person}{Paras Jain}, \bibinfo{person}{Ajay Jain}, \bibinfo{person}{Tianjun Zhang}, \bibinfo{person}{Pieter Abbeel}, \bibinfo{person}{Joseph Gonzalez}, {and} \bibinfo{person}{Ion Stoica}.} \bibinfo{year}{2021}\natexlab{}.
\newblock \showarticletitle{Contrastive Code Representation Learning}. In \bibinfo{booktitle}{\emph{EMNLP}}. \bibinfo{pages}{5954--5971}.
\newblock


\bibitem[Jiang et~al\mbox{.}(2017)]%
        {jiang2017unsupervised}
\bibfield{author}{\bibinfo{person}{He Jiang}, \bibinfo{person}{Jingxuan Zhang}, \bibinfo{person}{Zhilei Ren}, {and} \bibinfo{person}{Tao Zhang}.} \bibinfo{year}{2017}\natexlab{}.
\newblock \showarticletitle{An unsupervised approach for discovering relevant tutorial fragments for APIs}. In \bibinfo{booktitle}{\emph{ICSE}}. \bibinfo{pages}{38--48}.
\newblock


\bibitem[Jiang et~al\mbox{.}(2019)]%
        {jiang2019manual}
\bibfield{author}{\bibinfo{person}{Jiajun Jiang}, \bibinfo{person}{Yingfei Xiong}, {and} \bibinfo{person}{Xin Xia}.} \bibinfo{year}{2019}\natexlab{}.
\newblock \showarticletitle{A manual inspection of Defects4J bugs and its implications for automatic program repair}.
\newblock \bibinfo{journal}{\emph{Sci. China Inf. Sci.}} \bibinfo{volume}{62}, \bibinfo{number}{10} (\bibinfo{year}{2019}), \bibinfo{pages}{200102:1--200102:16}.
\newblock


\bibitem[Jiang et~al\mbox{.}(2021a)]%
        {jiang2021cure}
\bibfield{author}{\bibinfo{person}{Nan Jiang}, \bibinfo{person}{Thibaud Lutellier}, {and} \bibinfo{person}{Lin Tan}.} \bibinfo{year}{2021}\natexlab{a}.
\newblock \showarticletitle{CURE: Code-Aware Neural Machine Translation for Automatic Program Repair}. In \bibinfo{booktitle}{\emph{ICSE}}. \bibinfo{pages}{1161--1173}.
\newblock


\bibitem[Jiang et~al\mbox{.}(2021b)]%
        {jiang2021treebert}
\bibfield{author}{\bibinfo{person}{Xue Jiang}, \bibinfo{person}{Zhuoran Zheng}, \bibinfo{person}{Chen Lyu}, \bibinfo{person}{Liang Li}, {and} \bibinfo{person}{Lei Lyu}.} \bibinfo{year}{2021}\natexlab{b}.
\newblock \showarticletitle{TreeBERT: A tree-based pre-trained model for programming language}. In \bibinfo{booktitle}{\emph{Uncertainty in Artificial Intelligence}}. \bibinfo{pages}{54--63}.
\newblock


\bibitem[Jin et~al\mbox{.}(2022)]%
        {jin2022automatically}
\bibfield{author}{\bibinfo{person}{Dun Jin}, \bibinfo{person}{Peiyu Liu}, {and} \bibinfo{person}{Zhenfang Zhu}.} \bibinfo{year}{2022}\natexlab{}.
\newblock \showarticletitle{Automatically Generating Code Comment Using Heterogeneous Graph Neural Networks}. In \bibinfo{booktitle}{\emph{SANER}}. \bibinfo{pages}{1078--1088}.
\newblock


\bibitem[Kanade et~al\mbox{.}(2020)]%
        {kanade2020learning}
\bibfield{author}{\bibinfo{person}{Aditya Kanade}, \bibinfo{person}{Petros Maniatis}, \bibinfo{person}{Gogul Balakrishnan}, {and} \bibinfo{person}{Kensen Shi}.} \bibinfo{year}{2020}\natexlab{}.
\newblock \showarticletitle{Learning and Evaluating Contextual Embedding of Source Code}. In \bibinfo{booktitle}{\emph{ICML}}. \bibinfo{pages}{5110--5121}.
\newblock


\bibitem[Karampatsis et~al\mbox{.}(2020)]%
        {karampatsis2020big}
\bibfield{author}{\bibinfo{person}{Rafael-Michael Karampatsis}, \bibinfo{person}{Hlib Babii}, \bibinfo{person}{Romain Robbes}, \bibinfo{person}{Charles Sutton}, {and} \bibinfo{person}{Andrea Janes}.} \bibinfo{year}{2020}\natexlab{}.
\newblock \showarticletitle{Big code!= big vocabulary: Open-vocabulary models for source code}. In \bibinfo{booktitle}{\emph{ICSE}}. \bibinfo{pages}{1073--1085}.
\newblock


\bibitem[Kim et~al\mbox{.}(2021)]%
        {kim2021code}
\bibfield{author}{\bibinfo{person}{Seohyun Kim}, \bibinfo{person}{Jinman Zhao}, \bibinfo{person}{Yuchi Tian}, {and} \bibinfo{person}{Satish Chandra}.} \bibinfo{year}{2021}\natexlab{}.
\newblock \showarticletitle{Code prediction by feeding trees to transformers}. In \bibinfo{booktitle}{\emph{ICSE}}. \bibinfo{pages}{150--162}.
\newblock


\bibitem[Lachaux et~al\mbox{.}(2021)]%
        {ranzato2021DOBF}
\bibfield{author}{\bibinfo{person}{Marie{-}Anne Lachaux}, \bibinfo{person}{Baptiste Rozi{\`{e}}re}, \bibinfo{person}{Marc Szafraniec}, {and} \bibinfo{person}{Guillaume Lample}.} \bibinfo{year}{2021}\natexlab{}.
\newblock \showarticletitle{{DOBF:} {A} Deobfuscation Pre-Training Objective for Programming Languages}. In \bibinfo{booktitle}{\emph{NeurIPS}}. \bibinfo{pages}{14967--14979}.
\newblock


\bibitem[Lattner and Adve(2004)]%
        {lattner2004llvm}
\bibfield{author}{\bibinfo{person}{Chris Lattner} {and} \bibinfo{person}{Vikram Adve}.} \bibinfo{year}{2004}\natexlab{}.
\newblock \showarticletitle{LLVM: A compilation framework for lifelong program analysis \& transformation}. In \bibinfo{booktitle}{\emph{CGO}}. \bibinfo{pages}{75--86}.
\newblock


\bibitem[Le et~al\mbox{.}(2018)]%
        {le2018maximal}
\bibfield{author}{\bibinfo{person}{Tue Le}, \bibinfo{person}{Tuan Nguyen}, \bibinfo{person}{Trung Le}, \bibinfo{person}{Dinh Phung}, \bibinfo{person}{Paul Montague}, \bibinfo{person}{Olivier De~Vel}, {and} \bibinfo{person}{Lizhen Qu}.} \bibinfo{year}{2018}\natexlab{}.
\newblock \showarticletitle{Maximal divergence sequential autoencoder for binary software vulnerability detection}. In \bibinfo{booktitle}{\emph{ICLR}}.
\newblock


\bibitem[LeClair et~al\mbox{.}(2018)]%
        {leclair2018adapting}
\bibfield{author}{\bibinfo{person}{Alexander LeClair}, \bibinfo{person}{Zachary Eberhart}, {and} \bibinfo{person}{Collin McMillan}.} \bibinfo{year}{2018}\natexlab{}.
\newblock \showarticletitle{Adapting neural text classification for improved software categorization}. In \bibinfo{booktitle}{\emph{ICSME}}. \bibinfo{pages}{461--472}.
\newblock


\bibitem[LeClair et~al\mbox{.}(2020)]%
        {leclair2020improved}
\bibfield{author}{\bibinfo{person}{Alexander LeClair}, \bibinfo{person}{Sakib Haque}, \bibinfo{person}{Lingfei Wu}, {and} \bibinfo{person}{Collin McMillan}.} \bibinfo{year}{2020}\natexlab{}.
\newblock \showarticletitle{Improved code summarization via a graph neural network}. In \bibinfo{booktitle}{\emph{ICPC}}. \bibinfo{pages}{184--195}.
\newblock


\bibitem[LeClair et~al\mbox{.}(2019)]%
        {leclair2019neural}
\bibfield{author}{\bibinfo{person}{Alexander LeClair}, \bibinfo{person}{Siyuan Jiang}, {and} \bibinfo{person}{Collin McMillan}.} \bibinfo{year}{2019}\natexlab{}.
\newblock \showarticletitle{A neural model for generating natural language summaries of program subroutines}. In \bibinfo{booktitle}{\emph{ICSE}}. \bibinfo{pages}{795--806}.
\newblock


\bibitem[Lewis et~al\mbox{.}(2020)]%
        {lewis2020bart}
\bibfield{author}{\bibinfo{person}{Mike Lewis}, \bibinfo{person}{Yinhan Liu}, \bibinfo{person}{Naman Goyal}, \bibinfo{person}{Marjan Ghazvininejad}, {et~al\mbox{.}}} \bibinfo{year}{2020}\natexlab{}.
\newblock \showarticletitle{BART: Denoising Sequence-to-Sequence Pre-training for Natural Language Generation, Translation, and Comprehension}. In \bibinfo{booktitle}{\emph{ACL}}. \bibinfo{pages}{7871--7880}.
\newblock


\bibitem[Li et~al\mbox{.}(2023b)]%
        {li2023enabling}
\bibfield{author}{\bibinfo{person}{Jia Li}, \bibinfo{person}{Ge Li}, \bibinfo{person}{Yongmin Li}, {and} \bibinfo{person}{Zhi Jin}.} \bibinfo{year}{2023}\natexlab{b}.
\newblock \showarticletitle{Enabling Programming Thinking in Large Language Models Toward Code Generation}.
\newblock \bibinfo{journal}{\emph{arXiv preprint arXiv:2305.06599}} (\bibinfo{year}{2023}).
\newblock


\bibitem[Li et~al\mbox{.}(2023c)]%
        {li2023large}
\bibfield{author}{\bibinfo{person}{Jia Li}, \bibinfo{person}{Ge Li}, \bibinfo{person}{Chongyang Tao}, \bibinfo{person}{Huangzhao Zhang}, \bibinfo{person}{Fang Liu}, {and} \bibinfo{person}{Zhi Jin}.} \bibinfo{year}{2023}\natexlab{c}.
\newblock \showarticletitle{Large Language Model-Aware In-Context Learning for Code Generation}.
\newblock \bibinfo{journal}{\emph{arXiv preprint arXiv:2310.09748}} (\bibinfo{year}{2023}).
\newblock


\bibitem[Li et~al\mbox{.}(2021a)]%
        {li2021editsum}
\bibfield{author}{\bibinfo{person}{Jia Li}, \bibinfo{person}{Yongmin Li}, \bibinfo{person}{Ge Li}, \bibinfo{person}{Xing Hu}, \bibinfo{person}{Xin Xia}, {and} \bibinfo{person}{Zhi Jin}.} \bibinfo{year}{2021}\natexlab{a}.
\newblock \showarticletitle{EditSum: A Retrieve-and-Edit Framework for Source Code Summarization}. In \bibinfo{booktitle}{\emph{ASE}}. \bibinfo{pages}{155--166}.
\newblock


\bibitem[Li et~al\mbox{.}(2018a)]%
        {li2017code}
\bibfield{author}{\bibinfo{person}{Jian Li}, \bibinfo{person}{Yue Wang}, \bibinfo{person}{Michael~R. Lyu}, {and} \bibinfo{person}{Irwin King}.} \bibinfo{year}{2018}\natexlab{a}.
\newblock \showarticletitle{Code Completion with Neural Attention and Pointer Networks}. In \bibinfo{booktitle}{\emph{IJCAI}}. \bibinfo{pages}{4159--4165}.
\newblock


\bibitem[Li et~al\mbox{.}(2022e)]%
        {li2022auger}
\bibfield{author}{\bibinfo{person}{Lingwei Li}, \bibinfo{person}{Li Yang}, \bibinfo{person}{Huaxi Jiang}, \bibinfo{person}{Jun Yan}, \bibinfo{person}{Tiejian Luo}, \bibinfo{person}{Zihan Hua}, \bibinfo{person}{Geng Liang}, {and} \bibinfo{person}{Chun Zuo}.} \bibinfo{year}{2022}\natexlab{e}.
\newblock \showarticletitle{{AUGER:} automatically generating review comments with pre-training models}. In \bibinfo{booktitle}{\emph{ESEC/FSE}}. \bibinfo{pages}{1009--1021}.
\newblock


\bibitem[Li et~al\mbox{.}(2023a)]%
        {li2023starcoder}
\bibfield{author}{\bibinfo{person}{Raymond Li}, \bibinfo{person}{Loubna~Ben Allal}, \bibinfo{person}{Yangtian Zi}, \bibinfo{person}{Niklas Muennighoff}, \bibinfo{person}{Denis Kocetkov}, \bibinfo{person}{Chenghao Mou}, \bibinfo{person}{Marc Marone}, \bibinfo{person}{Christopher Akiki}, {et~al\mbox{.}}} \bibinfo{year}{2023}\natexlab{a}.
\newblock \showarticletitle{StarCoder: may the source be with you!}
\newblock \bibinfo{journal}{\emph{arXiv preprint arXiv:2305.06161}} (\bibinfo{year}{2023}).
\newblock


\bibitem[Li et~al\mbox{.}(2022b)]%
        {li2022coderetriever}
\bibfield{author}{\bibinfo{person}{Xiaonan Li}, \bibinfo{person}{Yeyun Gong}, \bibinfo{person}{Yelong Shen}, {et~al\mbox{.}}} \bibinfo{year}{2022}\natexlab{b}.
\newblock \showarticletitle{CodeRetriever: Unimodal and Bimodal Contrastive Learning}. In \bibinfo{booktitle}{\emph{EMNLP}}.
\newblock


\bibitem[Li et~al\mbox{.}(2022a)]%
        {li2022competition}
\bibfield{author}{\bibinfo{person}{Yujia Li}, \bibinfo{person}{David Choi}, \bibinfo{person}{Junyoung Chung}, \bibinfo{person}{Nate Kushman}, \bibinfo{person}{Julian Schrittwieser}, \bibinfo{person}{Rémi Leblond}, \bibinfo{person}{Tom Eccles}, {et~al\mbox{.}}} \bibinfo{year}{2022}\natexlab{a}.
\newblock \showarticletitle{Competition-Level Code Generation with AlphaCode}.
\newblock \bibinfo{journal}{\emph{Science}} \bibinfo{volume}{378}, \bibinfo{number}{6624} (\bibinfo{year}{2022}), \bibinfo{pages}{1092--1097}.
\newblock


\bibitem[Li et~al\mbox{.}(2016)]%
        {li2015gated}
\bibfield{author}{\bibinfo{person}{Yujia Li}, \bibinfo{person}{Daniel Tarlow}, \bibinfo{person}{Marc Brockschmidt}, {and} \bibinfo{person}{Richard~S. Zemel}.} \bibinfo{year}{2016}\natexlab{}.
\newblock \showarticletitle{Gated Graph Sequence Neural Networks}. In \bibinfo{booktitle}{\emph{ICLR}}.
\newblock


\bibitem[Li et~al\mbox{.}(2020)]%
        {li2020dlfix}
\bibfield{author}{\bibinfo{person}{Yi Li}, \bibinfo{person}{Shaohua Wang}, {and} \bibinfo{person}{Tien~N Nguyen}.} \bibinfo{year}{2020}\natexlab{}.
\newblock \showarticletitle{Dlfix: Context-based code transformation learning for automated program repair}. In \bibinfo{booktitle}{\emph{ICSE}}. \bibinfo{pages}{602--614}.
\newblock


\bibitem[Li et~al\mbox{.}(2021b)]%
        {li2021fault}
\bibfield{author}{\bibinfo{person}{Yi Li}, \bibinfo{person}{Shaohua Wang}, {and} \bibinfo{person}{Tien~N Nguyen}.} \bibinfo{year}{2021}\natexlab{b}.
\newblock \showarticletitle{Fault Localization with Code Coverage Representation Learning}. In \bibinfo{booktitle}{\emph{ICSE}}. \bibinfo{pages}{661--673}.
\newblock


\bibitem[Li et~al\mbox{.}(2021c)]%
        {li2021vulnerability}
\bibfield{author}{\bibinfo{person}{Yi Li}, \bibinfo{person}{Shaohua Wang}, {and} \bibinfo{person}{Tien~N. Nguyen}.} \bibinfo{year}{2021}\natexlab{c}.
\newblock \showarticletitle{Vulnerability detection with fine-grained interpretations}. In \bibinfo{booktitle}{\emph{ESEC/FSE}}. \bibinfo{pages}{292--303}.
\newblock


\bibitem[Li et~al\mbox{.}(2022d)]%
        {li2022DEAR}
\bibfield{author}{\bibinfo{person}{Yi Li}, \bibinfo{person}{Shaohua Wang}, {and} \bibinfo{person}{Tien~N. Nguyen}.} \bibinfo{year}{2022}\natexlab{d}.
\newblock \showarticletitle{{DEAR:} {A} Novel Deep Learning-based Approach for Automated Program Repair}. In \bibinfo{booktitle}{\emph{ICSE}}. \bibinfo{pages}{511--523}.
\newblock


\bibitem[Li et~al\mbox{.}(2019)]%
        {li2019improving}
\bibfield{author}{\bibinfo{person}{Yi Li}, \bibinfo{person}{Shaohua Wang}, \bibinfo{person}{Tien~N Nguyen}, {and} \bibinfo{person}{Son Van~Nguyen}.} \bibinfo{year}{2019}\natexlab{}.
\newblock \showarticletitle{Improving bug detection via context-based code representation learning and attention-based neural networks}.
\newblock \bibinfo{journal}{\emph{OOPSLA}}  \bibinfo{volume}{3} (\bibinfo{year}{2019}), \bibinfo{pages}{1--30}.
\newblock


\bibitem[Li et~al\mbox{.}(2022c)]%
        {li2022unleashing}
\bibfield{author}{\bibinfo{person}{Zongjie Li}, \bibinfo{person}{Pingchuan Ma}, \bibinfo{person}{Huaijin Wang}, \bibinfo{person}{Shuai Wang}, \bibinfo{person}{Qiyi Tang}, \bibinfo{person}{Sen Nie}, {and} \bibinfo{person}{Shi Wu}.} \bibinfo{year}{2022}\natexlab{c}.
\newblock \showarticletitle{Unleashing the Power of Compiler Intermediate Representation to Enhance Neural Program Embeddings}. In \bibinfo{booktitle}{\emph{ICSE}}. \bibinfo{pages}{2253--2265}.
\newblock


\bibitem[Li et~al\mbox{.}(2021d)]%
        {li2021sysevr}
\bibfield{author}{\bibinfo{person}{Zhen Li}, \bibinfo{person}{Deqing Zou}, \bibinfo{person}{Shouhuai Xu}, \bibinfo{person}{Hai Jin}, \bibinfo{person}{Yawei Zhu}, {and} \bibinfo{person}{Zhaoxuan Chen}.} \bibinfo{year}{2021}\natexlab{d}.
\newblock \showarticletitle{SySeVR: A framework for using deep learning to detect software vulnerabilities}.
\newblock \bibinfo{journal}{\emph{TDSC}} (\bibinfo{year}{2021}).
\newblock


\bibitem[Li et~al\mbox{.}(2018b)]%
        {li2018vuldeepecker}
\bibfield{author}{\bibinfo{person}{Zhen Li}, \bibinfo{person}{Deqing Zou}, \bibinfo{person}{Shouhuai Xu}, \bibinfo{person}{Xinyu Ou}, \bibinfo{person}{Hai Jin}, \bibinfo{person}{Sujuan Wang}, \bibinfo{person}{Zhijun Deng}, {and} \bibinfo{person}{Yuyi Zhong}.} \bibinfo{year}{2018}\natexlab{b}.
\newblock \showarticletitle{VulDeePecker: {A} Deep Learning-Based System for Vulnerability Detection}. In \bibinfo{booktitle}{\emph{NDSS}}.
\newblock


\bibitem[Lin et~al\mbox{.}(2021)]%
        {lin2021improving}
\bibfield{author}{\bibinfo{person}{Chen Lin}, \bibinfo{person}{Zhichao Ouyang}, \bibinfo{person}{Junqing Zhuang}, \bibinfo{person}{Jianqiang Chen}, \bibinfo{person}{Hui Li}, {and} \bibinfo{person}{Rongxin Wu}.} \bibinfo{year}{2021}\natexlab{}.
\newblock \showarticletitle{Improving Code Summarization with Block-wise Abstract Syntax Tree Splitting}. In \bibinfo{booktitle}{\emph{ICPC}}. \bibinfo{publisher}{{IEEE}}, \bibinfo{pages}{184--195}.
\newblock


\bibitem[Ling et~al\mbox{.}(2021b)]%
        {ling2021graph}
\bibfield{author}{\bibinfo{person}{Chunyang Ling}, \bibinfo{person}{Yanzhen Zou}, {and} \bibinfo{person}{Bing Xie}.} \bibinfo{year}{2021}\natexlab{b}.
\newblock \showarticletitle{Graph Neural Network Based Collaborative Filtering for {API} Usage Recommendation}. In \bibinfo{booktitle}{\emph{SANER}}. \bibinfo{publisher}{{IEEE}}, \bibinfo{pages}{36--47}.
\newblock


\bibitem[Ling et~al\mbox{.}(2016)]%
        {ling2016latent}
\bibfield{author}{\bibinfo{person}{Wang Ling}, \bibinfo{person}{Phil Blunsom}, \bibinfo{person}{Edward Grefenstette}, \bibinfo{person}{Karl~Moritz Hermann}, \bibinfo{person}{Tom{\'a}{\v{s}} Ko{\v{c}}isk{\`y}}, \bibinfo{person}{Fumin Wang}, {and} \bibinfo{person}{Andrew Senior}.} \bibinfo{year}{2016}\natexlab{}.
\newblock \showarticletitle{Latent Predictor Networks for Code Generation}. In \bibinfo{booktitle}{\emph{ACL}}. \bibinfo{pages}{599--609}.
\newblock


\bibitem[Ling et~al\mbox{.}(2021a)]%
        {ling2020deep}
\bibfield{author}{\bibinfo{person}{Xiang Ling}, \bibinfo{person}{Lingfei Wu}, \bibinfo{person}{Saizhuo Wang}, \bibinfo{person}{Gaoning Pan}, \bibinfo{person}{Tengfei Ma}, \bibinfo{person}{Fangli Xu}, \bibinfo{person}{Alex~X. Liu}, \bibinfo{person}{Chunming Wu}, {and} \bibinfo{person}{Shouling Ji}.} \bibinfo{year}{2021}\natexlab{a}.
\newblock \showarticletitle{Deep Graph Matching and Searching for Semantic Code Retrieval}.
\newblock \bibinfo{journal}{\emph{TKDD}} \bibinfo{volume}{15}, \bibinfo{number}{5} (\bibinfo{year}{2021}), \bibinfo{pages}{88:1--88:21}.
\newblock


\bibitem[Liu et~al\mbox{.}(2023)]%
        {liu2023improving}
\bibfield{author}{\bibinfo{person}{Chao Liu}, \bibinfo{person}{Xuanlin Bao}, \bibinfo{person}{Hongyu Zhang}, \bibinfo{person}{Neng Zhang}, \bibinfo{person}{Haibo Hu}, \bibinfo{person}{Xiaohong Zhang}, {and} \bibinfo{person}{Meng Yan}.} \bibinfo{year}{2023}\natexlab{}.
\newblock \showarticletitle{Improving ChatGPT Prompt for Code Generation}.
\newblock \bibinfo{journal}{\emph{arXiv preprint arXiv:2305.08360}} (\bibinfo{year}{2023}).
\newblock


\bibitem[Liu et~al\mbox{.}(2016a)]%
        {liu2016latent}
\bibfield{author}{\bibinfo{person}{Chang Liu}, \bibinfo{person}{Xinyun Chen}, \bibinfo{person}{Eui~Chul Shin}, \bibinfo{person}{Mingcheng Chen}, {and} \bibinfo{person}{Dawn Song}.} \bibinfo{year}{2016}\natexlab{a}.
\newblock \showarticletitle{Latent attention for if-then program synthesis}.
\newblock \bibinfo{journal}{\emph{NeurIPS}}  \bibinfo{volume}{29} (\bibinfo{year}{2016}), \bibinfo{pages}{4574--4582}.
\newblock


\bibitem[Liu et~al\mbox{.}(2016b)]%
        {liu2016neural}
\bibfield{author}{\bibinfo{person}{Chang Liu}, \bibinfo{person}{Xin Wang}, \bibinfo{person}{Richard Shin}, \bibinfo{person}{Joseph~E Gonzalez}, {and} \bibinfo{person}{Dawn Song}.} \bibinfo{year}{2016}\natexlab{b}.
\newblock \showarticletitle{Neural code completion}.
\newblock  (\bibinfo{year}{2016}).
\newblock


\bibitem[Liu et~al\mbox{.}(2020b)]%
        {liu2020self}
\bibfield{author}{\bibinfo{person}{Fang Liu}, \bibinfo{person}{Ge Li}, \bibinfo{person}{Bolin Wei}, \bibinfo{person}{Xin Xia}, \bibinfo{person}{Zhiyi Fu}, {and} \bibinfo{person}{Zhi Jin}.} \bibinfo{year}{2020}\natexlab{b}.
\newblock \showarticletitle{A Self-Attentional Neural Architecture for Code Completion with Multi-Task Learning}. In \bibinfo{booktitle}{\emph{ICPC}}. \bibinfo{pages}{37--47}.
\newblock


\bibitem[Liu et~al\mbox{.}(2020c)]%
        {liu2020multi}
\bibfield{author}{\bibinfo{person}{Fang Liu}, \bibinfo{person}{Ge Li}, \bibinfo{person}{Yunfei Zhao}, {and} \bibinfo{person}{Zhi Jin}.} \bibinfo{year}{2020}\natexlab{c}.
\newblock \showarticletitle{Multi-task learning based pre-trained language model for code completion}. In \bibinfo{booktitle}{\emph{ASE}}. \bibinfo{pages}{473--485}.
\newblock


\bibitem[Liu et~al\mbox{.}(2020e)]%
        {liu2020modeling}
\bibfield{author}{\bibinfo{person}{Fang Liu}, \bibinfo{person}{Lu Zhang}, {and} \bibinfo{person}{Zhi Jin}.} \bibinfo{year}{2020}\natexlab{e}.
\newblock \showarticletitle{Modeling programs hierarchically with stack-augmented LSTM}.
\newblock \bibinfo{journal}{\emph{Journal of Systems and Software}}  \bibinfo{volume}{164} (\bibinfo{year}{2020}), \bibinfo{pages}{110547}.
\newblock


\bibitem[Liu et~al\mbox{.}(2019)]%
        {liu2019learning}
\bibfield{author}{\bibinfo{person}{Kui Liu}, \bibinfo{person}{Dongsun Kim}, \bibinfo{person}{Tegawend{\'e}~F Bissyand{\'e}}, \bibinfo{person}{Taeyoung Kim}, \bibinfo{person}{Kisub Kim}, \bibinfo{person}{Anil Koyuncu}, \bibinfo{person}{Suntae Kim}, {and} \bibinfo{person}{Yves Le~Traon}.} \bibinfo{year}{2019}\natexlab{}.
\newblock \showarticletitle{Learning to spot and refactor inconsistent method names}. In \bibinfo{booktitle}{\emph{ICSE}}. \bibinfo{pages}{1--12}.
\newblock


\bibitem[Liu et~al\mbox{.}(2021a)]%
        {liu2021practical}
\bibfield{author}{\bibinfo{person}{Qianjun Liu}, \bibinfo{person}{Shouling Ji}, \bibinfo{person}{Changchang Liu}, {and} \bibinfo{person}{Chunming Wu}.} \bibinfo{year}{2021}\natexlab{a}.
\newblock \showarticletitle{A Practical Black-box Attack on Source Code Authorship Identification Classifiers}.
\newblock \bibinfo{journal}{\emph{TIFS}} (\bibinfo{year}{2021}).
\newblock


\bibitem[Liu et~al\mbox{.}(2020a)]%
        {liu2020retrieval}
\bibfield{author}{\bibinfo{person}{Shangqing Liu}, \bibinfo{person}{Yu Chen}, \bibinfo{person}{Xiaofei Xie}, \bibinfo{person}{Jing~Kai Siow}, {and} \bibinfo{person}{Yang Liu}.} \bibinfo{year}{2020}\natexlab{a}.
\newblock \showarticletitle{Retrieval-Augmented Generation for Code Summarization via Hybrid GNN}. In \bibinfo{booktitle}{\emph{ICLR}}.
\newblock


\bibitem[Liu et~al\mbox{.}(2021b)]%
        {liu2021combining}
\bibfield{author}{\bibinfo{person}{Zhenguang Liu}, \bibinfo{person}{Peng Qian}, \bibinfo{person}{Xiaoyang Wang}, \bibinfo{person}{Yuan Zhuang}, \bibinfo{person}{Lin Qiu}, {and} \bibinfo{person}{Xun Wang}.} \bibinfo{year}{2021}\natexlab{b}.
\newblock \showarticletitle{Combining Graph Neural Networks with Expert Knowledge for Smart Contract Vulnerability Detection}.
\newblock \bibinfo{journal}{\emph{TKDE}} (\bibinfo{year}{2021}).
\newblock


\bibitem[Liu et~al\mbox{.}(2020d)]%
        {liu2020automating}
\bibfield{author}{\bibinfo{person}{Zhongxin Liu}, \bibinfo{person}{Xin Xia}, \bibinfo{person}{Meng Yan}, {and} \bibinfo{person}{Shanping Li}.} \bibinfo{year}{2020}\natexlab{d}.
\newblock \showarticletitle{Automating just-in-time comment updating}. In \bibinfo{booktitle}{\emph{ASE}}. \bibinfo{pages}{585--597}.
\newblock


\bibitem[L{\'{o}}pez et~al\mbox{.}(2022)]%
        {lopez2022ast}
\bibfield{author}{\bibinfo{person}{Jos{\'{e}} Antonio~Hern{\'{a}}ndez L{\'{o}}pez}, \bibinfo{person}{Martin Weyssow}, \bibinfo{person}{Jes{\'{u}}s~S{\'{a}}nchez Cuadrado}, {and} \bibinfo{person}{Houari~A. Sahraoui}.} \bibinfo{year}{2022}\natexlab{}.
\newblock \showarticletitle{AST-Probe: Recovering abstract syntax trees from hidden representations of pre-trained language models}. In \bibinfo{booktitle}{\emph{ASE}}.
\newblock


\bibitem[Lu et~al\mbox{.}(2022)]%
        {lu2022reacc}
\bibfield{author}{\bibinfo{person}{Shuai Lu}, \bibinfo{person}{Nan Duan}, \bibinfo{person}{Hojae Han}, \bibinfo{person}{Daya Guo}, \bibinfo{person}{Seung{-}won Hwang}, {and} \bibinfo{person}{Alexey Svyatkovskiy}.} \bibinfo{year}{2022}\natexlab{}.
\newblock \showarticletitle{ReACC: {A} Retrieval-Augmented Code Completion Framework}. In \bibinfo{booktitle}{\emph{ACL}}. \bibinfo{pages}{6227--6240}.
\newblock


\bibitem[Lu et~al\mbox{.}(2021)]%
        {lu2021codexglue}
\bibfield{author}{\bibinfo{person}{Shuai Lu}, \bibinfo{person}{Daya Guo}, \bibinfo{person}{Shuo Ren}, \bibinfo{person}{Junjie Huang}, \bibinfo{person}{Alexey Svyatkovskiy}, {et~al\mbox{.}}} \bibinfo{year}{2021}\natexlab{}.
\newblock \showarticletitle{CodeXGLUE: {A} Machine Learning Benchmark Dataset for Code Understanding and Generation}. In \bibinfo{booktitle}{\emph{NeurIPS Datasets and Benchmarks}}.
\newblock


\bibitem[Luo et~al\mbox{.}(2023)]%
        {luo2023wizardcoder}
\bibfield{author}{\bibinfo{person}{Ziyang Luo}, \bibinfo{person}{Can Xu}, \bibinfo{person}{Pu Zhao}, \bibinfo{person}{Qingfeng Sun}, \bibinfo{person}{Xiubo Geng}, \bibinfo{person}{Wenxiang Hu}, \bibinfo{person}{Chongyang Tao}, \bibinfo{person}{Jing Ma}, \bibinfo{person}{Qingwei Lin}, {and} \bibinfo{person}{Daxin Jiang}.} \bibinfo{year}{2023}\natexlab{}.
\newblock \showarticletitle{WizardCoder: Empowering Code Large Language Models with Evol-Instruct}.
\newblock \bibinfo{journal}{\emph{arXiv preprint arXiv:2306.08568}} (\bibinfo{year}{2023}).
\newblock


\bibitem[Maddison and Tarlow(2014)]%
        {maddison2014structured}
\bibfield{author}{\bibinfo{person}{Chris Maddison} {and} \bibinfo{person}{Daniel Tarlow}.} \bibinfo{year}{2014}\natexlab{}.
\newblock \showarticletitle{Structured generative models of natural source code}. In \bibinfo{booktitle}{\emph{ICML}}. \bibinfo{pages}{649--657}.
\newblock


\bibitem[Malik et~al\mbox{.}(2019)]%
        {malik2019nl2type}
\bibfield{author}{\bibinfo{person}{Rabee~Sohail Malik}, \bibinfo{person}{Jibesh Patra}, {and} \bibinfo{person}{Michael Pradel}.} \bibinfo{year}{2019}\natexlab{}.
\newblock \showarticletitle{NL2Type: inferring JavaScript function types from natural language information}. In \bibinfo{booktitle}{\emph{ICSE}}. \bibinfo{pages}{304--315}.
\newblock


\bibitem[Mastropaolo et~al\mbox{.}(2021)]%
        {mastropaolo2021studying}
\bibfield{author}{\bibinfo{person}{Antonio Mastropaolo}, \bibinfo{person}{Simone Scalabrino}, \bibinfo{person}{Nathan Cooper}, \bibinfo{person}{David~Nader Palacio}, \bibinfo{person}{Denys Poshyvanyk}, {et~al\mbox{.}}} \bibinfo{year}{2021}\natexlab{}.
\newblock \showarticletitle{Studying the usage of text-to-text transfer transformer to support code-related tasks}. In \bibinfo{booktitle}{\emph{ICSE}}. \bibinfo{pages}{336--347}.
\newblock


\bibitem[Mehrotra et~al\mbox{.}(2021)]%
        {mehrotra2021modeling}
\bibfield{author}{\bibinfo{person}{Nikita Mehrotra}, \bibinfo{person}{Navdha Agarwal}, \bibinfo{person}{Piyush Gupta}, \bibinfo{person}{Saket Anand}, \bibinfo{person}{David Lo}, {and} \bibinfo{person}{Rahul Purandare}.} \bibinfo{year}{2021}\natexlab{}.
\newblock \showarticletitle{Modeling Functional Similarity in Source Code with Graph-Based Siamese Networks}.
\newblock \bibinfo{journal}{\emph{TSE}} (\bibinfo{year}{2021}).
\newblock


\bibitem[Mesbah et~al\mbox{.}(2019)]%
        {mesbah2019deepdelta}
\bibfield{author}{\bibinfo{person}{Ali Mesbah}, \bibinfo{person}{Andrew Rice}, \bibinfo{person}{Emily Johnston}, \bibinfo{person}{Nick Glorioso}, {and} \bibinfo{person}{Edward Aftandilian}.} \bibinfo{year}{2019}\natexlab{}.
\newblock \showarticletitle{DeepDelta: learning to repair compilation errors}. In \bibinfo{booktitle}{\emph{ESEC/FSE}}. \bibinfo{pages}{925--936}.
\newblock


\bibitem[Mikolov et~al\mbox{.}(2013)]%
        {mikolov2013efficient}
\bibfield{author}{\bibinfo{person}{Tom{\'{a}}s Mikolov}, \bibinfo{person}{Kai Chen}, \bibinfo{person}{Greg Corrado}, {and} \bibinfo{person}{Jeffrey Dean}.} \bibinfo{year}{2013}\natexlab{}.
\newblock \showarticletitle{Efficient Estimation of Word Representations in Vector Space}. In \bibinfo{booktitle}{\emph{ICLR}}.
\newblock


\bibitem[Mir et~al\mbox{.}(2022)]%
        {mir2022type4py}
\bibfield{author}{\bibinfo{person}{Amir~M. Mir}, \bibinfo{person}{Evaldas Latoskinas}, \bibinfo{person}{Sebastian Proksch}, {and} \bibinfo{person}{Georgios Gousios}.} \bibinfo{year}{2022}\natexlab{}.
\newblock \showarticletitle{Type4Py: Practical Deep Similarity Learning-Based Type Inference for Python}. In \bibinfo{booktitle}{\emph{ICSE}}. \bibinfo{pages}{2241--2252}.
\newblock


\bibitem[Moreno et~al\mbox{.}(2015)]%
        {moreno2015can}
\bibfield{author}{\bibinfo{person}{Laura Moreno}, \bibinfo{person}{Gabriele Bavota}, \bibinfo{person}{Massimiliano Di~Penta}, \bibinfo{person}{Rocco Oliveto}, {and} \bibinfo{person}{Andrian Marcus}.} \bibinfo{year}{2015}\natexlab{}.
\newblock \showarticletitle{How can I use this method?}. In \bibinfo{booktitle}{\emph{ICSE}}, Vol.~\bibinfo{volume}{1}. \bibinfo{pages}{880--890}.
\newblock


\bibitem[Mou et~al\mbox{.}(2016)]%
        {mou2016convolutional}
\bibfield{author}{\bibinfo{person}{Lili Mou}, \bibinfo{person}{Ge Li}, \bibinfo{person}{Lu Zhang}, \bibinfo{person}{Tao Wang}, {and} \bibinfo{person}{Zhi Jin}.} \bibinfo{year}{2016}\natexlab{}.
\newblock \showarticletitle{Convolutional neural networks over tree structures for programming language processing}. In \bibinfo{booktitle}{\emph{AAAI}}, Vol.~\bibinfo{volume}{30}.
\newblock


\bibitem[Mu et~al\mbox{.}(2022)]%
        {mu2022automatic}
\bibfield{author}{\bibinfo{person}{Fangwen Mu}, \bibinfo{person}{Xiao Chen}, \bibinfo{person}{Lin Shi}, \bibinfo{person}{Song Wang}, {and} \bibinfo{person}{Qing Wang}.} \bibinfo{year}{2022}\natexlab{}.
\newblock \showarticletitle{Automatic Comment Generation via Multi-Pass Deliberation}. In \bibinfo{booktitle}{\emph{ASE}}. \bibinfo{publisher}{{ACM}}, \bibinfo{pages}{14:1--14:12}.
\newblock


\bibitem[Nafi et~al\mbox{.}(2019)]%
        {nafi2019clcdsa}
\bibfield{author}{\bibinfo{person}{Kawser~Wazed Nafi}, \bibinfo{person}{Tonny~Shekha Kar}, \bibinfo{person}{Banani Roy}, \bibinfo{person}{Chanchal~K Roy}, {and} \bibinfo{person}{Kevin~A Schneider}.} \bibinfo{year}{2019}\natexlab{}.
\newblock \showarticletitle{Clcdsa: cross language code clone detection using syntactical features and api documentation}. In \bibinfo{booktitle}{\emph{ASE}}. \bibinfo{pages}{1026--1037}.
\newblock


\bibitem[Nair et~al\mbox{.}(2020)]%
        {nair2020funcgnn}
\bibfield{author}{\bibinfo{person}{Aravind Nair}, \bibinfo{person}{Avijit Roy}, {and} \bibinfo{person}{Karl Meinke}.} \bibinfo{year}{2020}\natexlab{}.
\newblock \showarticletitle{funcGNN: A Graph Neural Network Approach to Program Similarity}. In \bibinfo{booktitle}{\emph{ESEM}}. \bibinfo{pages}{1--11}.
\newblock


\bibitem[Nan et~al\mbox{.}(2020)]%
        {nan2020hisyn}
\bibfield{author}{\bibinfo{person}{Zifan Nan}, \bibinfo{person}{Hui Guan}, {and} \bibinfo{person}{Xipeng Shen}.} \bibinfo{year}{2020}\natexlab{}.
\newblock \showarticletitle{HISyn: human learning-inspired natural language programming}. In \bibinfo{booktitle}{\emph{ESEC/FSE}}. \bibinfo{pages}{75--86}.
\newblock


\bibitem[Nguyen et~al\mbox{.}(2021)]%
        {nguyen2021adversarial}
\bibfield{author}{\bibinfo{person}{Phuong~T Nguyen}, \bibinfo{person}{Claudio Di~Sipio}, \bibinfo{person}{Juri Di~Rocco}, \bibinfo{person}{Massimiliano Di~Penta}, {and} \bibinfo{person}{Davide Di~Ruscio}.} \bibinfo{year}{2021}\natexlab{}.
\newblock \showarticletitle{Adversarial Attacks to API Recommender Systems: Time to Wake Up and Smell the Coffee?}. In \bibinfo{booktitle}{\emph{ASE}}. \bibinfo{pages}{253--265}.
\newblock


\bibitem[Nguyen et~al\mbox{.}(2020)]%
        {nguyen2020suggesting}
\bibfield{author}{\bibinfo{person}{Son Nguyen}, \bibinfo{person}{Hung Phan}, \bibinfo{person}{Trinh Le}, {and} \bibinfo{person}{Tien~N Nguyen}.} \bibinfo{year}{2020}\natexlab{}.
\newblock \showarticletitle{Suggesting natural method names to check name consistencies}. In \bibinfo{booktitle}{\emph{ICSE}}. \bibinfo{pages}{1372--1384}.
\newblock


\bibitem[Nguyen et~al\mbox{.}(2017)]%
        {nguyen2017exploring}
\bibfield{author}{\bibinfo{person}{Trong~Duc Nguyen}, \bibinfo{person}{Anh~Tuan Nguyen}, \bibinfo{person}{Hung~Dang Phan}, {and} \bibinfo{person}{Tien~N Nguyen}.} \bibinfo{year}{2017}\natexlab{}.
\newblock \showarticletitle{Exploring API embedding for API usages and applications}. In \bibinfo{booktitle}{\emph{ICSE}}. \bibinfo{pages}{438--449}.
\newblock


\bibitem[Nijkamp et~al\mbox{.}(2022)]%
        {nijkamp2022codegen}
\bibfield{author}{\bibinfo{person}{Erik Nijkamp}, \bibinfo{person}{Bo Pang}, \bibinfo{person}{Hiroaki Hayashi}, \bibinfo{person}{Lifu Tu}, \bibinfo{person}{Huan Wang}, \bibinfo{person}{Yingbo Zhou}, \bibinfo{person}{Silvio Savarese}, {and} \bibinfo{person}{Caiming Xiong}.} \bibinfo{year}{2022}\natexlab{}.
\newblock \showarticletitle{Codegen: An open large language model for code with multi-turn program synthesis}.
\newblock \bibinfo{journal}{\emph{arXiv preprint arXiv:2203.13474}} (\bibinfo{year}{2022}).
\newblock


\bibitem[Niu et~al\mbox{.}(2022)]%
        {niu2022spt}
\bibfield{author}{\bibinfo{person}{Changan Niu}, \bibinfo{person}{Chuanyi Li}, \bibinfo{person}{Vincent Ng}, \bibinfo{person}{Jidong Ge}, \bibinfo{person}{Liguo Huang}, {and} \bibinfo{person}{Bin Luo}.} \bibinfo{year}{2022}\natexlab{}.
\newblock \showarticletitle{SPT-Code: Sequence-to-Sequence Pre-Training for Learning Source Code Representations}. In \bibinfo{booktitle}{\emph{ICSE}}. \bibinfo{pages}{1--13}.
\newblock


\bibitem[Nye et~al\mbox{.}(2019)]%
        {nye2019learning}
\bibfield{author}{\bibinfo{person}{Maxwell Nye}, \bibinfo{person}{Luke Hewitt}, \bibinfo{person}{Joshua Tenenbaum}, {and} \bibinfo{person}{Armando Solar-Lezama}.} \bibinfo{year}{2019}\natexlab{}.
\newblock \showarticletitle{Learning to infer program sketches}. In \bibinfo{booktitle}{\emph{ICML}}. \bibinfo{pages}{4861--4870}.
\newblock


\bibitem[OpenAI(2022)]%
        {ChatGPT}
\bibfield{author}{\bibinfo{person}{OpenAI}.} \bibinfo{year}{2022}\natexlab{}.
\newblock \bibinfo{title}{{ChatGPT}}.
\newblock \bibinfo{howpublished}{\url{https://openai.com/blog/chatgpt/}}.
\newblock


\bibitem[Ott et~al\mbox{.}(2019)]%
        {ott2019fairseq}
\bibfield{author}{\bibinfo{person}{Myle Ott}, \bibinfo{person}{Sergey Edunov}, \bibinfo{person}{Alexei Baevski}, \bibinfo{person}{Angela Fan}, \bibinfo{person}{Sam Gross}, \bibinfo{person}{Nathan Ng}, \bibinfo{person}{David Grangier}, {and} \bibinfo{person}{Michael Auli}.} \bibinfo{year}{2019}\natexlab{}.
\newblock \showarticletitle{fairseq: A Fast, Extensible Toolkit for Sequence Modeling}. In \bibinfo{booktitle}{\emph{NAACL-HLT: Demonstrations}}.
\newblock


\bibitem[Pandi et~al\mbox{.}(2020)]%
        {pandi2020opttyper}
\bibfield{author}{\bibinfo{person}{Irene~Vlassi Pandi}, \bibinfo{person}{Earl~T Barr}, \bibinfo{person}{Andrew~D Gordon}, {and} \bibinfo{person}{Charles Sutton}.} \bibinfo{year}{2020}\natexlab{}.
\newblock \showarticletitle{OptTyper: Probabilistic Type Inference by Optimising Logical and Natural Constraints}.
\newblock \bibinfo{journal}{\emph{arXiv:2004.00348}} (\bibinfo{year}{2020}).
\newblock


\bibitem[Panthaplackel et~al\mbox{.}(2021)]%
        {panthaplackel2021deep}
\bibfield{author}{\bibinfo{person}{Sheena Panthaplackel}, \bibinfo{person}{Junyi~Jessy Li}, \bibinfo{person}{Milos Gligoric}, {and} \bibinfo{person}{Raymond~J Mooney}.} \bibinfo{year}{2021}\natexlab{}.
\newblock \showarticletitle{Deep Just-In-Time Inconsistency Detection Between Comments and Source Code}. In \bibinfo{booktitle}{\emph{AAAI}}, Vol.~\bibinfo{volume}{35}. \bibinfo{pages}{427--435}.
\newblock


\bibitem[Panthaplackel et~al\mbox{.}(2020)]%
        {panthaplackel2020learning}
\bibfield{author}{\bibinfo{person}{Sheena Panthaplackel}, \bibinfo{person}{Pengyu Nie}, \bibinfo{person}{Milos Gligoric}, \bibinfo{person}{Junyi~Jessy Li}, {and} \bibinfo{person}{Raymond Mooney}.} \bibinfo{year}{2020}\natexlab{}.
\newblock \showarticletitle{Learning to Update Natural Language Comments Based on Code Changes}. In \bibinfo{booktitle}{\emph{ACL}}. \bibinfo{pages}{1853--1868}.
\newblock


\bibitem[Peng et~al\mbox{.}(2021)]%
        {peng2021how}
\bibfield{author}{\bibinfo{person}{Dinglan Peng}, \bibinfo{person}{Shuxin Zheng}, \bibinfo{person}{Yatao Li}, \bibinfo{person}{Guolin Ke}, \bibinfo{person}{Di He}, {and} \bibinfo{person}{Tie{-}Yan Liu}.} \bibinfo{year}{2021}\natexlab{}.
\newblock \showarticletitle{How could Neural Networks understand Programs?}. In \bibinfo{booktitle}{\emph{ICML}}, Vol.~\bibinfo{volume}{139}. \bibinfo{pages}{8476--8486}.
\newblock


\bibitem[Poesia et~al\mbox{.}(2022)]%
        {poesia2022synchromesh}
\bibfield{author}{\bibinfo{person}{Gabriel Poesia}, \bibinfo{person}{Alex Polozov}, \bibinfo{person}{Vu Le}, \bibinfo{person}{Ashish Tiwari}, \bibinfo{person}{Gustavo Soares}, \bibinfo{person}{Christopher Meek}, {and} \bibinfo{person}{Sumit Gulwani}.} \bibinfo{year}{2022}\natexlab{}.
\newblock \showarticletitle{Synchromesh: Reliable Code Generation from Pre-trained Language Models}. In \bibinfo{booktitle}{\emph{ICLR}}.
\newblock


\bibitem[Pornprasit et~al\mbox{.}(2021)]%
        {pornprasit2021pyexplainer}
\bibfield{author}{\bibinfo{person}{Chanathip Pornprasit}, \bibinfo{person}{Chakkrit Tantithamthavorn}, \bibinfo{person}{Jirayus Jiarpakdee}, \bibinfo{person}{Michael Fu}, {and} \bibinfo{person}{Patanamon Thongtanunam}.} \bibinfo{year}{2021}\natexlab{}.
\newblock \showarticletitle{PyExplainer: Explaining the Predictions of Just-In-Time Defect Models}. In \bibinfo{booktitle}{\emph{ASE}}. \bibinfo{pages}{407--418}.
\newblock


\bibitem[Pradel et~al\mbox{.}(2020)]%
        {pradel2020typewriter}
\bibfield{author}{\bibinfo{person}{Michael Pradel}, \bibinfo{person}{Georgios Gousios}, \bibinfo{person}{Jason Liu}, {and} \bibinfo{person}{Satish Chandra}.} \bibinfo{year}{2020}\natexlab{}.
\newblock \showarticletitle{Typewriter: Neural type prediction with search-based validation}. In \bibinfo{booktitle}{\emph{ESEC/FSE}}. \bibinfo{pages}{209--220}.
\newblock


\bibitem[Pradel and Sen(2018)]%
        {pradel2018deepbugs}
\bibfield{author}{\bibinfo{person}{Michael Pradel} {and} \bibinfo{person}{Koushik Sen}.} \bibinfo{year}{2018}\natexlab{}.
\newblock \showarticletitle{Deepbugs: A learning approach to name-based bug detection}.
\newblock \bibinfo{journal}{\emph{OOPSLA}}  \bibinfo{volume}{2} (\bibinfo{year}{2018}), \bibinfo{pages}{1--25}.
\newblock


\bibitem[Quiring et~al\mbox{.}(2019)]%
        {quiring2019misleading}
\bibfield{author}{\bibinfo{person}{Erwin Quiring}, \bibinfo{person}{Alwin Maier}, {and} \bibinfo{person}{Konrad Rieck}.} \bibinfo{year}{2019}\natexlab{}.
\newblock \showarticletitle{Misleading authorship attribution of source code using adversarial learning}. In \bibinfo{booktitle}{\emph{USENIX Security 19}}. \bibinfo{pages}{479--496}.
\newblock


\bibitem[Rabin et~al\mbox{.}(2021)]%
        {rabin2021understanding}
\bibfield{author}{\bibinfo{person}{Md. Rafiqul~Islam Rabin}, \bibinfo{person}{Vincent~J. Hellendoorn}, {and} \bibinfo{person}{Mohammad~Amin Alipour}.} \bibinfo{year}{2021}\natexlab{}.
\newblock \showarticletitle{Understanding neural code intelligence through program simplification}. In \bibinfo{booktitle}{\emph{ESEC/FSE}}. \bibinfo{publisher}{{ACM}}, \bibinfo{pages}{441--452}.
\newblock


\bibitem[Rabinovich et~al\mbox{.}(2017)]%
        {rabinovich2017abstract}
\bibfield{author}{\bibinfo{person}{Maxim Rabinovich}, \bibinfo{person}{Mitchell Stern}, {and} \bibinfo{person}{Dan Klein}.} \bibinfo{year}{2017}\natexlab{}.
\newblock \showarticletitle{Abstract Syntax Networks for Code Generation and Semantic Parsing}. In \bibinfo{booktitle}{\emph{ACL}}. \bibinfo{pages}{1139--1149}.
\newblock


\bibitem[Raffel et~al\mbox{.}(2020)]%
        {raffel2020exploring}
\bibfield{author}{\bibinfo{person}{Colin Raffel}, \bibinfo{person}{Noam Shazeer}, \bibinfo{person}{Adam Roberts}, \bibinfo{person}{Katherine Lee}, \bibinfo{person}{Sharan Narang}, \bibinfo{person}{Michael Matena}, {et~al\mbox{.}}} \bibinfo{year}{2020}\natexlab{}.
\newblock \showarticletitle{Exploring the Limits of Transfer Learning with a Unified Text-to-Text Transformer}.
\newblock \bibinfo{journal}{\emph{JMLR}}  \bibinfo{volume}{21} (\bibinfo{year}{2020}), \bibinfo{pages}{1--67}.
\newblock


\bibitem[Ramakrishnan and Albarghouthi(2022)]%
        {ramakrishnan2022backdoors}
\bibfield{author}{\bibinfo{person}{Goutham Ramakrishnan} {and} \bibinfo{person}{Aws Albarghouthi}.} \bibinfo{year}{2022}\natexlab{}.
\newblock \showarticletitle{Backdoors in Neural Models of Source Code}. In \bibinfo{booktitle}{\emph{ICPR}}. \bibinfo{publisher}{{IEEE}}, \bibinfo{pages}{2892--2899}.
\newblock


\bibitem[Ramakrishnan et~al\mbox{.}(2020)]%
        {ramakrishnan2020semantic}
\bibfield{author}{\bibinfo{person}{Goutham Ramakrishnan}, \bibinfo{person}{Jordan Henkel}, \bibinfo{person}{Zi Wang}, \bibinfo{person}{Aws Albarghouthi}, \bibinfo{person}{Somesh Jha}, {and} \bibinfo{person}{Thomas Reps}.} \bibinfo{year}{2020}\natexlab{}.
\newblock \showarticletitle{Semantic robustness of models of source code}.
\newblock \bibinfo{journal}{\emph{arXiv:2002.03043}} (\bibinfo{year}{2020}).
\newblock


\bibitem[Raychev et~al\mbox{.}(2016)]%
        {raychev2016probabilistic}
\bibfield{author}{\bibinfo{person}{Veselin Raychev}, \bibinfo{person}{Pavol Bielik}, {and} \bibinfo{person}{Martin Vechev}.} \bibinfo{year}{2016}\natexlab{}.
\newblock \showarticletitle{Probabilistic model for code with decision trees}.
\newblock \bibinfo{journal}{\emph{ACM SIGPLAN Notices}} \bibinfo{volume}{51}, \bibinfo{number}{10} (\bibinfo{year}{2016}), \bibinfo{pages}{731--747}.
\newblock


\bibitem[Raychev et~al\mbox{.}(2014)]%
        {raychev2014code}
\bibfield{author}{\bibinfo{person}{Veselin Raychev}, \bibinfo{person}{Martin Vechev}, {and} \bibinfo{person}{Eran Yahav}.} \bibinfo{year}{2014}\natexlab{}.
\newblock \showarticletitle{Code completion with statistical language models}. In \bibinfo{booktitle}{\emph{ICPC}}. \bibinfo{pages}{419--428}.
\newblock


\bibitem[Roziere et~al\mbox{.}(2023)]%
        {roziere2023code}
\bibfield{author}{\bibinfo{person}{Baptiste Roziere}, \bibinfo{person}{Jonas Gehring}, \bibinfo{person}{Fabian Gloeckle}, \bibinfo{person}{Sten Sootla}, \bibinfo{person}{Itai Gat}, \bibinfo{person}{Xiaoqing~Ellen Tan}, \bibinfo{person}{Yossi Adi}, \bibinfo{person}{Jingyu Liu}, \bibinfo{person}{Tal Remez}, \bibinfo{person}{J{\'e}r{\'e}my Rapin}, {et~al\mbox{.}}} \bibinfo{year}{2023}\natexlab{}.
\newblock \showarticletitle{Code llama: Open foundation models for code}.
\newblock \bibinfo{journal}{\emph{arXiv preprint arXiv:2308.12950}} (\bibinfo{year}{2023}).
\newblock


\bibitem[Rozi{\`{e}}re et~al\mbox{.}(2020)]%
        {lachaux2020unsupervised}
\bibfield{author}{\bibinfo{person}{Baptiste Rozi{\`{e}}re}, \bibinfo{person}{Marie{-}Anne Lachaux}, \bibinfo{person}{Lowik Chanussot}, {and} \bibinfo{person}{Guillaume Lample}.} \bibinfo{year}{2020}\natexlab{}.
\newblock \showarticletitle{Unsupervised Translation of Programming Languages}. In \bibinfo{booktitle}{\emph{NeurIPS}}.
\newblock


\bibitem[Rozi{\`{e}}re et~al\mbox{.}(2022)]%
        {roziere2022leveraging}
\bibfield{author}{\bibinfo{person}{Baptiste Rozi{\`{e}}re}, \bibinfo{person}{Jie Zhang}, \bibinfo{person}{Fran{\c{c}}ois Charton}, \bibinfo{person}{Mark Harman}, \bibinfo{person}{Gabriel Synnaeve}, {and} \bibinfo{person}{Guillaume Lample}.} \bibinfo{year}{2022}\natexlab{}.
\newblock \showarticletitle{Leveraging Automated Unit Tests for Unsupervised Code Translation}. In \bibinfo{booktitle}{\emph{ICLR}}.
\newblock


\bibitem[Salib(2004)]%
        {salib2004faster}
\bibfield{author}{\bibinfo{person}{Michael Salib}.} \bibinfo{year}{2004}\natexlab{}.
\newblock \showarticletitle{Faster than C: Static type inference with Starkiller}.
\newblock \bibinfo{journal}{\emph{PyCon Proceedings, Washington DC}}  \bibinfo{volume}{3} (\bibinfo{year}{2004}).
\newblock


\bibitem[Santos et~al\mbox{.}(2018)]%
        {santos2018syntax}
\bibfield{author}{\bibinfo{person}{Eddie~Antonio Santos}, \bibinfo{person}{Joshua~Charles Campbell}, \bibinfo{person}{Dhvani Patel}, \bibinfo{person}{Abram Hindle}, {and} \bibinfo{person}{Jos{\'e}~Nelson Amaral}.} \bibinfo{year}{2018}\natexlab{}.
\newblock \showarticletitle{Syntax and sensibility: Using language models to detect and correct syntax errors}. In \bibinfo{booktitle}{\emph{SANER}}. \bibinfo{pages}{311--322}.
\newblock


\bibitem[Schuster et~al\mbox{.}(2021)]%
        {schuster2021you}
\bibfield{author}{\bibinfo{person}{Roei Schuster}, \bibinfo{person}{Congzheng Song}, \bibinfo{person}{Eran Tromer}, {and} \bibinfo{person}{Vitaly Shmatikov}.} \bibinfo{year}{2021}\natexlab{}.
\newblock \showarticletitle{You autocomplete me: Poisoning vulnerabilities in neural code completion}. In \bibinfo{booktitle}{\emph{USENIX Security}}.
\newblock


\bibitem[Severi et~al\mbox{.}(2021)]%
        {severi2021explanation}
\bibfield{author}{\bibinfo{person}{Giorgio Severi}, \bibinfo{person}{Jim Meyer}, \bibinfo{person}{Scott Coull}, {and} \bibinfo{person}{Alina Oprea}.} \bibinfo{year}{2021}\natexlab{}.
\newblock \showarticletitle{Explanation-Guided Backdoor Poisoning Attacks Against Malware Classifiers}. In \bibinfo{booktitle}{\emph{USENIX Security}}.
\newblock


\bibitem[Shahbazi et~al\mbox{.}(2021)]%
        {shahbazi2021API2Com}
\bibfield{author}{\bibinfo{person}{Ramin Shahbazi}, \bibinfo{person}{Rishab Sharma}, {and} \bibinfo{person}{Fatemeh~H. Fard}.} \bibinfo{year}{2021}\natexlab{}.
\newblock \showarticletitle{API2Com: On the Improvement of Automatically Generated Code Comments Using {API} Documentations}. In \bibinfo{booktitle}{\emph{ICPC}}. \bibinfo{publisher}{{IEEE}}, \bibinfo{pages}{411--421}.
\newblock


\bibitem[Sharma et~al\mbox{.}(2022)]%
        {sharma2022exploratory}
\bibfield{author}{\bibinfo{person}{Rishab Sharma}, \bibinfo{person}{Fuxiang Chen}, \bibinfo{person}{Fatemeh~H. Fard}, {and} \bibinfo{person}{David Lo}.} \bibinfo{year}{2022}\natexlab{}.
\newblock \showarticletitle{An exploratory study on code attention in {BERT}}. In \bibinfo{booktitle}{\emph{ICPC}}. \bibinfo{publisher}{{ACM}}, \bibinfo{pages}{437--448}.
\newblock


\bibitem[Shi et~al\mbox{.}(2021)]%
        {shi2021cast}
\bibfield{author}{\bibinfo{person}{Ensheng Shi}, \bibinfo{person}{Yanlin Wang}, \bibinfo{person}{Lun Du}, \bibinfo{person}{Hongyu Zhang}, \bibinfo{person}{Shi Han}, {et~al\mbox{.}}} \bibinfo{year}{2021}\natexlab{}.
\newblock \showarticletitle{CAST: Enhancing Code Summarization with Hierarchical Splitting and Reconstruction of Abstract Syntax Trees}. In \bibinfo{booktitle}{\emph{EMNLP}}. \bibinfo{pages}{4053--4062}.
\newblock


\bibitem[Shi et~al\mbox{.}(2022b)]%
        {shi2022compressing}
\bibfield{author}{\bibinfo{person}{Jieke Shi}, \bibinfo{person}{Zhou Yang}, \bibinfo{person}{Bowen Xu}, \bibinfo{person}{Hong~Jin Kang}, {and} \bibinfo{person}{David Lo}.} \bibinfo{year}{2022}\natexlab{b}.
\newblock \showarticletitle{Compressing Pre-trained Models of Code into 3 {MB}}. In \bibinfo{booktitle}{\emph{ASE}}. \bibinfo{publisher}{{ACM}}, \bibinfo{pages}{24:1--24:12}.
\newblock


\bibitem[Shi et~al\mbox{.}(2022a)]%
        {shi2022are}
\bibfield{author}{\bibinfo{person}{Lin Shi}, \bibinfo{person}{Fangwen Mu}, \bibinfo{person}{Xiao Chen}, \bibinfo{person}{Song Wang}, \bibinfo{person}{Junjie Wang}, \bibinfo{person}{Ye Yang}, \bibinfo{person}{Ge Li}, \bibinfo{person}{Xin Xia}, {and} \bibinfo{person}{Qing Wang}.} \bibinfo{year}{2022}\natexlab{a}.
\newblock \showarticletitle{Are we building on the rock? on the importance of data preprocessing for code summarization}. In \bibinfo{booktitle}{\emph{ESEC/FSE}}. \bibinfo{publisher}{{ACM}}, \bibinfo{pages}{107--119}.
\newblock


\bibitem[Shi et~al\mbox{.}(2022c)]%
        {shi2022how}
\bibfield{author}{\bibinfo{person}{Yucen Shi}, \bibinfo{person}{Ying Yin}, \bibinfo{person}{Zhengkui Wang}, \bibinfo{person}{David Lo}, \bibinfo{person}{Tao Zhang}, \bibinfo{person}{Xin Xia}, \bibinfo{person}{Yuhai Zhao}, {and} \bibinfo{person}{Bowen Xu}.} \bibinfo{year}{2022}\natexlab{c}.
\newblock \showarticletitle{How to better utilize code graphs in semantic code search?}. In \bibinfo{booktitle}{\emph{ESEC/FSE}}. \bibinfo{pages}{722--733}.
\newblock


\bibitem[Shrivastava et~al\mbox{.}(2020)]%
        {shrivastava2020fly}
\bibfield{author}{\bibinfo{person}{Disha Shrivastava}, \bibinfo{person}{Hugo Larochelle}, {and} \bibinfo{person}{Daniel Tarlow}.} \bibinfo{year}{2020}\natexlab{}.
\newblock \showarticletitle{On-the-Fly Adaptation of Source Code Models using Meta-Learning}.
\newblock \bibinfo{journal}{\emph{arXiv:2003.11768}} (\bibinfo{year}{2020}).
\newblock


\bibitem[Shu and Zhang(2017)]%
        {shu2017neural}
\bibfield{author}{\bibinfo{person}{Chengxun Shu} {and} \bibinfo{person}{Hongyu Zhang}.} \bibinfo{year}{2017}\natexlab{}.
\newblock \showarticletitle{Neural Programming by Example}. In \bibinfo{booktitle}{\emph{AAAI}}. \bibinfo{pages}{1539--1545}.
\newblock


\bibitem[Sui et~al\mbox{.}(2020)]%
        {sui2020flow2vec}
\bibfield{author}{\bibinfo{person}{Yulei Sui}, \bibinfo{person}{Xiao Cheng}, \bibinfo{person}{Guanqin Zhang}, {and} \bibinfo{person}{Haoyu Wang}.} \bibinfo{year}{2020}\natexlab{}.
\newblock \showarticletitle{Flow2Vec: value-flow-based precise code embedding}.
\newblock \bibinfo{journal}{\emph{OOPSLA}}  \bibinfo{volume}{4} (\bibinfo{year}{2020}), \bibinfo{pages}{1--27}.
\newblock


\bibitem[Sui and Xue(2016)]%
        {sui2016svf}
\bibfield{author}{\bibinfo{person}{Yulei Sui} {and} \bibinfo{person}{Jingling Xue}.} \bibinfo{year}{2016}\natexlab{}.
\newblock \showarticletitle{SVF: interprocedural static value-flow analysis in LLVM}. In \bibinfo{booktitle}{\emph{Proceedings of the 25th international conference on compiler construction}}. \bibinfo{pages}{265--266}.
\newblock


\bibitem[Sun et~al\mbox{.}(2022a)]%
        {sun2022code}
\bibfield{author}{\bibinfo{person}{Weisong Sun}, \bibinfo{person}{Chunrong Fang}, \bibinfo{person}{Yuchen Chen}, \bibinfo{person}{Guanhong Tao}, \bibinfo{person}{Tingxu Han}, {and} \bibinfo{person}{Quanjun Zhang}.} \bibinfo{year}{2022}\natexlab{a}.
\newblock \showarticletitle{Code Search based on Context-aware Code Translation}. In \bibinfo{booktitle}{\emph{ICSE}}. \bibinfo{pages}{388--400}.
\newblock


\bibitem[Sun et~al\mbox{.}(2022b)]%
        {sun2022heterogeneous}
\bibfield{author}{\bibinfo{person}{Yizhou Sun}, \bibinfo{person}{Jiawei Han}, \bibinfo{person}{Xifeng Yan}, \bibinfo{person}{Philip~S. Yu}, {and} \bibinfo{person}{Tianyi Wu}.} \bibinfo{year}{2022}\natexlab{b}.
\newblock \showarticletitle{Heterogeneous Information Networks: the Past, the Present, and the Future}.
\newblock \bibinfo{journal}{\emph{Proc. {VLDB} Endow.}} \bibinfo{volume}{15}, \bibinfo{number}{12} (\bibinfo{year}{2022}), \bibinfo{pages}{3807--3811}.
\newblock


\bibitem[Sun et~al\mbox{.}(2022c)]%
        {sun2022on}
\bibfield{author}{\bibinfo{person}{Zhensu Sun}, \bibinfo{person}{Li Li}, \bibinfo{person}{Yan Liu}, \bibinfo{person}{Xiaoning Du}, {and} \bibinfo{person}{Li Li}.} \bibinfo{year}{2022}\natexlab{c}.
\newblock \showarticletitle{On the Importance of Building High-quality Training Datasets for Neural Code Search}. In \bibinfo{booktitle}{\emph{ICSE}}. \bibinfo{publisher}{{ACM}}, \bibinfo{pages}{1609--1620}.
\newblock


\bibitem[Sun et~al\mbox{.}(2019)]%
        {sun2019grammar}
\bibfield{author}{\bibinfo{person}{Zeyu Sun}, \bibinfo{person}{Qihao Zhu}, \bibinfo{person}{Lili Mou}, \bibinfo{person}{Yingfei Xiong}, \bibinfo{person}{Ge Li}, {and} \bibinfo{person}{Lu Zhang}.} \bibinfo{year}{2019}\natexlab{}.
\newblock \showarticletitle{A grammar-based structural cnn decoder for code generation}. In \bibinfo{booktitle}{\emph{AAAI}}, Vol.~\bibinfo{volume}{33}. \bibinfo{pages}{7055--7062}.
\newblock


\bibitem[Sun et~al\mbox{.}(2020)]%
        {sun2020treegen}
\bibfield{author}{\bibinfo{person}{Zeyu Sun}, \bibinfo{person}{Qihao Zhu}, \bibinfo{person}{Yingfei Xiong}, \bibinfo{person}{Yican Sun}, \bibinfo{person}{Lili Mou}, {and} \bibinfo{person}{Lu Zhang}.} \bibinfo{year}{2020}\natexlab{}.
\newblock \showarticletitle{TreeGen: {A} Tree-Based Transformer Architecture for Code Generation}. In \bibinfo{booktitle}{\emph{AAAI}}. \bibinfo{pages}{8984--8991}.
\newblock


\bibitem[Svyatkovskiy et~al\mbox{.}(2020)]%
        {svyatkovskiy2020intellicode}
\bibfield{author}{\bibinfo{person}{Alexey Svyatkovskiy}, \bibinfo{person}{Shao~Kun Deng}, \bibinfo{person}{Shengyu Fu}, {and} \bibinfo{person}{Neel Sundaresan}.} \bibinfo{year}{2020}\natexlab{}.
\newblock \showarticletitle{Intellicode compose: Code generation using transformer}. In \bibinfo{booktitle}{\emph{ESEC/FSE}}. \bibinfo{pages}{1433--1443}.
\newblock


\bibitem[Svyatkovskiy et~al\mbox{.}(2021)]%
        {svyatkovskiy2021fast}
\bibfield{author}{\bibinfo{person}{Alexey Svyatkovskiy}, \bibinfo{person}{Sebastian Lee}, \bibinfo{person}{Anna Hadjitofi}, \bibinfo{person}{Maik Riechert}, \bibinfo{person}{Juliana~Vicente Franco}, {and} \bibinfo{person}{Miltiadis Allamanis}.} \bibinfo{year}{2021}\natexlab{}.
\newblock \showarticletitle{Fast and memory-efficient neural code completion}. In \bibinfo{booktitle}{\emph{MSR}}. \bibinfo{pages}{329--340}.
\newblock


\bibitem[Svyatkovskiy et~al\mbox{.}(2019)]%
        {svyatkovskiy2019pythia}
\bibfield{author}{\bibinfo{person}{Alexey Svyatkovskiy}, \bibinfo{person}{Ying Zhao}, \bibinfo{person}{Shengyu Fu}, {and} \bibinfo{person}{Neel Sundaresan}.} \bibinfo{year}{2019}\natexlab{}.
\newblock \showarticletitle{Pythia: Ai-assisted code completion system}. In \bibinfo{booktitle}{\emph{SIGKDD}}. \bibinfo{pages}{2727--2735}.
\newblock


\bibitem[Tang et~al\mbox{.}(2022)]%
        {shen2022ast}
\bibfield{author}{\bibinfo{person}{Ze Tang}, \bibinfo{person}{Xiaoyu Shen}, \bibinfo{person}{Chuanyi Li}, \bibinfo{person}{Jidong Ge}, \bibinfo{person}{Liguo Huang}, \bibinfo{person}{Zheling Zhu}, {and} \bibinfo{person}{Bin Luo}.} \bibinfo{year}{2022}\natexlab{}.
\newblock \showarticletitle{AST-Trans: Code Summarization with Efficient Tree-Structured Attention}. In \bibinfo{booktitle}{\emph{ICSE}}.
\newblock


\bibitem[Tao et~al\mbox{.}(2022)]%
        {tao2022contrastive}
\bibfield{author}{\bibinfo{person}{Chenning Tao}, \bibinfo{person}{Qi Zhan}, \bibinfo{person}{Xing Hu}, {and} \bibinfo{person}{Xin Xia}.} \bibinfo{year}{2022}\natexlab{}.
\newblock \showarticletitle{{C4:} contrastive cross-language code clone detection}. In \bibinfo{booktitle}{\emph{ICPC}}. \bibinfo{publisher}{{ACM}}, \bibinfo{pages}{413--424}.
\newblock


\bibitem[Tarlow et~al\mbox{.}(2020)]%
        {tarlow2020learning}
\bibfield{author}{\bibinfo{person}{Daniel Tarlow}, \bibinfo{person}{Subhodeep Moitra}, \bibinfo{person}{Andrew Rice}, \bibinfo{person}{Zimin Chen}, \bibinfo{person}{Pierre-Antoine Manzagol}, \bibinfo{person}{Charles Sutton}, {and} \bibinfo{person}{Edward Aftandilian}.} \bibinfo{year}{2020}\natexlab{}.
\newblock \showarticletitle{Learning to fix build errors with graph2diff neural networks}. In \bibinfo{booktitle}{\emph{ICSE Workshops}}. \bibinfo{pages}{19--20}.
\newblock


\bibitem[Tian et~al\mbox{.}(2020)]%
        {tian2020evaluating}
\bibfield{author}{\bibinfo{person}{Haoye Tian}, \bibinfo{person}{Kui Liu}, \bibinfo{person}{Abdoul~Kader Kabor{\'e}}, \bibinfo{person}{Anil Koyuncu}, \bibinfo{person}{Li Li}, {et~al\mbox{.}}} \bibinfo{year}{2020}\natexlab{}.
\newblock \showarticletitle{Evaluating representation learning of code changes for predicting patch correctness in program repair}. In \bibinfo{booktitle}{\emph{ASE}}. \bibinfo{pages}{981--992}.
\newblock


\bibitem[Tufano et~al\mbox{.}(2019)]%
        {tufano2019learning}
\bibfield{author}{\bibinfo{person}{Michele Tufano}, \bibinfo{person}{Jevgenija Pantiuchina}, \bibinfo{person}{Cody Watson}, \bibinfo{person}{Gabriele Bavota}, {and} \bibinfo{person}{Denys Poshyvanyk}.} \bibinfo{year}{2019}\natexlab{}.
\newblock \showarticletitle{On learning meaningful code changes via neural machine translation}. In \bibinfo{booktitle}{\emph{ICSE}}. \bibinfo{pages}{25--36}.
\newblock


\bibitem[Tufano et~al\mbox{.}(2018a)]%
        {tufano2018empirical}
\bibfield{author}{\bibinfo{person}{Michele Tufano}, \bibinfo{person}{Cody Watson}, \bibinfo{person}{Gabriele Bavota}, \bibinfo{person}{Massimiliano Di~Penta}, {et~al\mbox{.}}} \bibinfo{year}{2018}\natexlab{a}.
\newblock \showarticletitle{An empirical investigation into learning bug-fixing patches in the wild via neural machine translation}. In \bibinfo{booktitle}{\emph{ASE}}. \bibinfo{pages}{832--837}.
\newblock


\bibitem[Tufano et~al\mbox{.}(2018b)]%
        {tufano2018deep}
\bibfield{author}{\bibinfo{person}{Michele Tufano}, \bibinfo{person}{Cody Watson}, \bibinfo{person}{Gabriele Bavota}, \bibinfo{person}{Massimiliano Di~Penta}, \bibinfo{person}{Martin White}, {and} \bibinfo{person}{Denys Poshyvanyk}.} \bibinfo{year}{2018}\natexlab{b}.
\newblock \showarticletitle{Deep learning similarities from different representations of source code}. In \bibinfo{booktitle}{\emph{MSR}}. \bibinfo{pages}{542--553}.
\newblock


\bibitem[Vasic et~al\mbox{.}(2018)]%
        {vasic2018neural}
\bibfield{author}{\bibinfo{person}{Marko Vasic}, \bibinfo{person}{Aditya Kanade}, \bibinfo{person}{Petros Maniatis}, \bibinfo{person}{David Bieber}, {and} \bibinfo{person}{Rishabh Singh}.} \bibinfo{year}{2018}\natexlab{}.
\newblock \showarticletitle{Neural Program Repair by Jointly Learning to Localize and Repair}. In \bibinfo{booktitle}{\emph{ICLR}}.
\newblock


\bibitem[Vaswani et~al\mbox{.}(2017)]%
        {vaswani2017attention}
\bibfield{author}{\bibinfo{person}{Ashish Vaswani}, \bibinfo{person}{Noam Shazeer}, \bibinfo{person}{Niki Parmar}, \bibinfo{person}{Jakob Uszkoreit}, \bibinfo{person}{Llion Jones}, \bibinfo{person}{Aidan~N Gomez}, \bibinfo{person}{{\L}ukasz Kaiser}, {and} \bibinfo{person}{Illia Polosukhin}.} \bibinfo{year}{2017}\natexlab{}.
\newblock \showarticletitle{Attention is all you need}. In \bibinfo{booktitle}{\emph{NeurIPS}}. \bibinfo{pages}{5998--6008}.
\newblock


\bibitem[VenkataKeerthy et~al\mbox{.}(2020)]%
        {venkatakeerthy2019ir2vec}
\bibfield{author}{\bibinfo{person}{S VenkataKeerthy}, \bibinfo{person}{Rohit Aggarwal}, \bibinfo{person}{Shalini Jain}, \bibinfo{person}{Maunendra~Sankar Desarkar}, \bibinfo{person}{Ramakrishna Upadrasta}, {and} \bibinfo{person}{YN Srikant}.} \bibinfo{year}{2020}\natexlab{}.
\newblock \showarticletitle{Ir2vec: Llvm ir based scalable program embeddings}.
\newblock \bibinfo{journal}{\emph{TACO}} \bibinfo{volume}{17}, \bibinfo{number}{4} (\bibinfo{year}{2020}), \bibinfo{pages}{1--27}.
\newblock


\bibitem[Wan et~al\mbox{.}(2022a)]%
        {wan2022naturalcc}
\bibfield{author}{\bibinfo{person}{Yao Wan}, \bibinfo{person}{Yang He}, \bibinfo{person}{Zhangqian Bi}, \bibinfo{person}{Jianguo Zhang}, \bibinfo{person}{Yulei Sui}, \bibinfo{person}{Hongyu Zhang}, {et~al\mbox{.}}} \bibinfo{year}{2022}\natexlab{a}.
\newblock \showarticletitle{NaturalCC: An Open-Source Toolkit for Code Intelligence}. In \bibinfo{booktitle}{\emph{ICSE, Companion Volume}}.
\newblock


\bibitem[Wan et~al\mbox{.}(2019)]%
        {wan2019multi}
\bibfield{author}{\bibinfo{person}{Yao Wan}, \bibinfo{person}{Jingdong Shu}, \bibinfo{person}{Yulei Sui}, \bibinfo{person}{Guandong Xu}, \bibinfo{person}{Zhou Zhao}, \bibinfo{person}{Jian Wu}, {and} \bibinfo{person}{Philip~S. Yu}.} \bibinfo{year}{2019}\natexlab{}.
\newblock \showarticletitle{Multi-modal Attention Network Learning for Semantic Source Code Retrieval}. In \bibinfo{booktitle}{\emph{ASE}}. \bibinfo{pages}{13--25}.
\newblock


\bibitem[Wan et~al\mbox{.}(2022b)]%
        {wan2022you}
\bibfield{author}{\bibinfo{person}{Yao Wan}, \bibinfo{person}{Shijie Zhang}, \bibinfo{person}{Hongyu Zhang}, \bibinfo{person}{Yulei Sui}, \bibinfo{person}{Guandong Xu}, \bibinfo{person}{Dezhong Yao}, \bibinfo{person}{Hai Jin}, {and} \bibinfo{person}{Lichao Sun}.} \bibinfo{year}{2022}\natexlab{b}.
\newblock \showarticletitle{You see what {I} want you to see: poisoning vulnerabilities in neural code search}. In \bibinfo{booktitle}{\emph{ESEC/FSE}}. \bibinfo{pages}{1233--1245}.
\newblock


\bibitem[Wan et~al\mbox{.}(2022c)]%
        {wan2022what}
\bibfield{author}{\bibinfo{person}{Yao Wan}, \bibinfo{person}{Wei Zhao}, \bibinfo{person}{Hongyu Zhang}, \bibinfo{person}{Yulei Sui}, \bibinfo{person}{Guandong Xu}, {and} \bibinfo{person}{Hai Jin}.} \bibinfo{year}{2022}\natexlab{c}.
\newblock \showarticletitle{What Do They Capture? - {A} Structural Analysis of Pre-Trained Language Models for Source Code}. In \bibinfo{booktitle}{\emph{ICSE}}. \bibinfo{pages}{2377--2388}.
\newblock


\bibitem[Wan et~al\mbox{.}(2018)]%
        {wan2018improving}
\bibfield{author}{\bibinfo{person}{Yao Wan}, \bibinfo{person}{Zhou Zhao}, \bibinfo{person}{Min Yang}, \bibinfo{person}{Guandong Xu}, \bibinfo{person}{Haochao Ying}, \bibinfo{person}{Jian Wu}, {and} \bibinfo{person}{Philip~S Yu}.} \bibinfo{year}{2018}\natexlab{}.
\newblock \showarticletitle{Improving automatic source code summarization via deep reinforcement learning}. In \bibinfo{booktitle}{\emph{ASE}}. \bibinfo{pages}{397--407}.
\newblock


\bibitem[Wang et~al\mbox{.}(2022c)]%
        {wang2022no}
\bibfield{author}{\bibinfo{person}{Chaozheng Wang}, \bibinfo{person}{Yuanhang Yang}, \bibinfo{person}{Cuiyun Gao}, \bibinfo{person}{Yun Peng}, \bibinfo{person}{Hongyu Zhang}, {and} \bibinfo{person}{Michael~R. Lyu}.} \bibinfo{year}{2022}\natexlab{c}.
\newblock \showarticletitle{No more fine-tuning? an experimental evaluation of prompt tuning in code intelligence}. In \bibinfo{booktitle}{\emph{ESEC/FSE}}. \bibinfo{pages}{382--394}.
\newblock


\bibitem[Wang et~al\mbox{.}(2022b)]%
        {wang2022bridging}
\bibfield{author}{\bibinfo{person}{Deze Wang}, \bibinfo{person}{Zhouyang Jia}, \bibinfo{person}{Shanshan Li}, \bibinfo{person}{Yue Yu}, \bibinfo{person}{Yun Xiong}, \bibinfo{person}{Wei Dong}, {and} \bibinfo{person}{Xiangke Liao}.} \bibinfo{year}{2022}\natexlab{b}.
\newblock \showarticletitle{Bridging Pre-trained Models and Downstream Tasks for Source Code Understanding}. In \bibinfo{booktitle}{\emph{ICSE}}. \bibinfo{pages}{287--298}.
\newblock


\bibitem[Wang et~al\mbox{.}(2020c)]%
        {wang2020combining}
\bibfield{author}{\bibinfo{person}{Huanting Wang}, \bibinfo{person}{Guixin Ye}, \bibinfo{person}{Zhanyong Tang}, \bibinfo{person}{Shin~Hwei Tan}, {et~al\mbox{.}}} \bibinfo{year}{2020}\natexlab{c}.
\newblock \showarticletitle{Combining graph-based learning with automated data collection for code vulnerability detection}.
\newblock \bibinfo{journal}{\emph{TIFS}}  \bibinfo{volume}{16} (\bibinfo{year}{2020}), \bibinfo{pages}{1943--1958}.
\newblock


\bibitem[Wang et~al\mbox{.}(2020a)]%
        {wang2020synergy}
\bibfield{author}{\bibinfo{person}{Simin Wang}, \bibinfo{person}{Liguo Huang}, \bibinfo{person}{Jidong Ge}, \bibinfo{person}{Tengfei Zhang}, \bibinfo{person}{Haitao Feng}, \bibinfo{person}{Ming Li}, \bibinfo{person}{He Zhang}, {and} \bibinfo{person}{Vincent Ng}.} \bibinfo{year}{2020}\natexlab{a}.
\newblock \showarticletitle{Synergy between Machine/Deep Learning and Software Engineering: How Far Are We?}
\newblock \bibinfo{journal}{\emph{arXiv:2008.05515}} (\bibinfo{year}{2020}).
\newblock


\bibitem[Wang et~al\mbox{.}(2016)]%
        {wang2016automatically}
\bibfield{author}{\bibinfo{person}{Song Wang}, \bibinfo{person}{Taiyue Liu}, {and} \bibinfo{person}{Lin Tan}.} \bibinfo{year}{2016}\natexlab{}.
\newblock \showarticletitle{Automatically learning semantic features for defect prediction}. In \bibinfo{booktitle}{\emph{ICSE}}. \bibinfo{pages}{297--308}.
\newblock


\bibitem[Wang et~al\mbox{.}(2020b)]%
        {wang2020detecting}
\bibfield{author}{\bibinfo{person}{Wenhan Wang}, \bibinfo{person}{Ge Li}, \bibinfo{person}{Bo Ma}, \bibinfo{person}{Xin Xia}, {and} \bibinfo{person}{Zhi Jin}.} \bibinfo{year}{2020}\natexlab{b}.
\newblock \showarticletitle{Detecting code clones with graph neural network and flow-augmented abstract syntax tree}. In \bibinfo{booktitle}{\emph{SANER}}. \bibinfo{pages}{261--271}.
\newblock


\bibitem[Wang et~al\mbox{.}(2021b)]%
        {wang2021syncobert}
\bibfield{author}{\bibinfo{person}{Xin Wang}, \bibinfo{person}{Yasheng Wang}, \bibinfo{person}{Fei Mi}, \bibinfo{person}{Pingyi Zhou}, \bibinfo{person}{Yao Wan}, \bibinfo{person}{Xiao Liu}, \bibinfo{person}{Li Li}, \bibinfo{person}{Hao Wu}, \bibinfo{person}{Jin Liu}, {and} \bibinfo{person}{Xin Jiang}.} \bibinfo{year}{2021}\natexlab{b}.
\newblock \showarticletitle{SynCoBERT: Syntax-Guided Multi-Modal Contrastive Pre-Training for Code Representation}.
\newblock \bibinfo{journal}{\emph{arXiv:2108.04556}} (\bibinfo{year}{2021}).
\newblock


\bibitem[Wang et~al\mbox{.}(2022a)]%
        {wang2022gypsum}
\bibfield{author}{\bibinfo{person}{Yu Wang}, \bibinfo{person}{Yu Dong}, \bibinfo{person}{Xuesong Lu}, {and} \bibinfo{person}{Aoying Zhou}.} \bibinfo{year}{2022}\natexlab{a}.
\newblock \showarticletitle{GypSum: learning hybrid representations for code summarization}. In \bibinfo{booktitle}{\emph{ICPC}}. \bibinfo{publisher}{{ACM}}, \bibinfo{pages}{12--23}.
\newblock


\bibitem[Wang et~al\mbox{.}(2023)]%
        {wang2023codet5+}
\bibfield{author}{\bibinfo{person}{Yue Wang}, \bibinfo{person}{Hung Le}, \bibinfo{person}{Akhilesh~Deepak Gotmare}, \bibinfo{person}{Nghi~DQ Bui}, \bibinfo{person}{Junnan Li}, {and} \bibinfo{person}{Steven~CH Hoi}.} \bibinfo{year}{2023}\natexlab{}.
\newblock \showarticletitle{Codet5+: Open code large language models for code understanding and generation}.
\newblock \bibinfo{journal}{\emph{arXiv preprint arXiv:2305.07922}} (\bibinfo{year}{2023}).
\newblock


\bibitem[Wang and Li(2021)]%
        {wang2021code}
\bibfield{author}{\bibinfo{person}{Yanlin Wang} {and} \bibinfo{person}{Hui Li}.} \bibinfo{year}{2021}\natexlab{}.
\newblock \showarticletitle{Code completion by modeling flattened abstract syntax trees as graphs}. In \bibinfo{booktitle}{\emph{AAAI}}, Vol.~\bibinfo{volume}{35}. \bibinfo{pages}{14015--14023}.
\newblock


\bibitem[Wang et~al\mbox{.}(2021a)]%
        {wang2021codet5}
\bibfield{author}{\bibinfo{person}{Yue Wang}, \bibinfo{person}{Weishi Wang}, \bibinfo{person}{Shafiq~R. Joty}, {and} \bibinfo{person}{Steven C.~H. Hoi}.} \bibinfo{year}{2021}\natexlab{a}.
\newblock \showarticletitle{CodeT5: Identifier-aware Unified Pre-trained Encoder-Decoder Models for Code Understanding and Generation}. In \bibinfo{booktitle}{\emph{EMNLP}}. \bibinfo{pages}{8696--8708}.
\newblock


\bibitem[Watson et~al\mbox{.}(2020)]%
        {watson2020systematic}
\bibfield{author}{\bibinfo{person}{Cody Watson}, \bibinfo{person}{Ncthan Cooper}, \bibinfo{person}{David~Nader Palacio}, \bibinfo{person}{Kevin Moran}, {and} \bibinfo{person}{Denys Poshyvanyk}.} \bibinfo{year}{2020}\natexlab{}.
\newblock \showarticletitle{A Systematic Literature Review on the Use of Deep Learning in Software Engineering Research}.
\newblock \bibinfo{journal}{\emph{arXiv:2009.06520}} (\bibinfo{year}{2020}).
\newblock


\bibitem[Wei et~al\mbox{.}(2019)]%
        {wei2019code}
\bibfield{author}{\bibinfo{person}{Bolin Wei}, \bibinfo{person}{Ge Li}, \bibinfo{person}{Xin Xia}, \bibinfo{person}{Zhiyi Fu}, {and} \bibinfo{person}{Zhi Jin}.} \bibinfo{year}{2019}\natexlab{}.
\newblock \showarticletitle{Code Generation as a Dual Task of Code Summarization}. In \bibinfo{booktitle}{\emph{NeurIPS}}. \bibinfo{pages}{6559--6569}.
\newblock


\bibitem[Wei et~al\mbox{.}(2020b)]%
        {wei2020retrieve}
\bibfield{author}{\bibinfo{person}{Bolin Wei}, \bibinfo{person}{Yongmin Li}, \bibinfo{person}{Ge Li}, \bibinfo{person}{Xin Xia}, {and} \bibinfo{person}{Zhi Jin}.} \bibinfo{year}{2020}\natexlab{b}.
\newblock \showarticletitle{Retrieve and refine: exemplar-based neural comment generation}. In \bibinfo{booktitle}{\emph{ASE}}. \bibinfo{pages}{349--360}.
\newblock


\bibitem[Wei and Li(2017)]%
        {wei2017supervised}
\bibfield{author}{\bibinfo{person}{Huihui Wei} {and} \bibinfo{person}{Ming Li}.} \bibinfo{year}{2017}\natexlab{}.
\newblock \showarticletitle{Supervised Deep Features for Software Functional Clone Detection by Exploiting Lexical and Syntactical Information in Source Code.}. In \bibinfo{booktitle}{\emph{IJCAI}}. \bibinfo{pages}{3034--3040}.
\newblock


\bibitem[Wei et~al\mbox{.}(2020a)]%
        {wei2020lambdanet}
\bibfield{author}{\bibinfo{person}{Jiayi Wei}, \bibinfo{person}{Maruth Goyal}, \bibinfo{person}{Greg Durrett}, {and} \bibinfo{person}{Isil Dillig}.} \bibinfo{year}{2020}\natexlab{a}.
\newblock \showarticletitle{LambdaNet: Probabilistic Type Inference using Graph Neural Networks}. In \bibinfo{booktitle}{\emph{ICLR}}.
\newblock


\bibitem[Wei et~al\mbox{.}(2022)]%
        {wei2022clear}
\bibfield{author}{\bibinfo{person}{Moshi Wei}, \bibinfo{person}{Nima~Shiri Harzevili}, \bibinfo{person}{Yuchao Huang}, \bibinfo{person}{Junjie Wang}, {and} \bibinfo{person}{Song Wang}.} \bibinfo{year}{2022}\natexlab{}.
\newblock \showarticletitle{CLEAR: contrastive learning for API recommendation}. In \bibinfo{booktitle}{\emph{ICSE}}. \bibinfo{pages}{376--387}.
\newblock


\bibitem[White et~al\mbox{.}(2019)]%
        {white2019sorting}
\bibfield{author}{\bibinfo{person}{Martin White}, \bibinfo{person}{Michele Tufano}, \bibinfo{person}{Matias Martinez}, \bibinfo{person}{Martin Monperrus}, {and} \bibinfo{person}{Denys Poshyvanyk}.} \bibinfo{year}{2019}\natexlab{}.
\newblock \showarticletitle{Sorting and transforming program repair ingredients via deep learning code similarities}. In \bibinfo{booktitle}{\emph{SANER}}. \bibinfo{pages}{479--490}.
\newblock


\bibitem[White et~al\mbox{.}(2016)]%
        {white2016deep}
\bibfield{author}{\bibinfo{person}{Martin White}, \bibinfo{person}{Michele Tufano}, \bibinfo{person}{Christopher Vendome}, {and} \bibinfo{person}{Denys Poshyvanyk}.} \bibinfo{year}{2016}\natexlab{}.
\newblock \showarticletitle{Deep learning code fragments for code clone detection}. In \bibinfo{booktitle}{\emph{ASE}}. \bibinfo{pages}{87--98}.
\newblock


\bibitem[White et~al\mbox{.}(2015)]%
        {white2015toward}
\bibfield{author}{\bibinfo{person}{Martin White}, \bibinfo{person}{Christopher Vendome}, \bibinfo{person}{Mario Linares-V{\'a}squez}, {and} \bibinfo{person}{Denys Poshyvanyk}.} \bibinfo{year}{2015}\natexlab{}.
\newblock \showarticletitle{Toward deep learning software repositories}. In \bibinfo{booktitle}{\emph{MSR}}. \bibinfo{pages}{334--345}.
\newblock


\bibitem[Wu et~al\mbox{.}(2021)]%
        {wu2020sit3}
\bibfield{author}{\bibinfo{person}{Hongqiu Wu}, \bibinfo{person}{Hai Zhao}, {and} \bibinfo{person}{Min Zhang}.} \bibinfo{year}{2021}\natexlab{}.
\newblock \showarticletitle{Code Summarization with Structure-induced Transformer}. In \bibinfo{booktitle}{\emph{Findings of ACL}}. \bibinfo{pages}{1078--1090}.
\newblock


\bibitem[Wu et~al\mbox{.}(2022a)]%
        {wu2022detecting}
\bibfield{author}{\bibinfo{person}{Yueming Wu}, \bibinfo{person}{Siyue Feng}, \bibinfo{person}{Deqing Zou}, {and} \bibinfo{person}{Hai Jin}.} \bibinfo{year}{2022}\natexlab{a}.
\newblock \showarticletitle{Detecting Semantic Code Clones by Building AST-based Markov Chains Model}. In \bibinfo{booktitle}{\emph{ASE}}. \bibinfo{publisher}{{ACM}}, \bibinfo{pages}{34:1--34:13}.
\newblock


\bibitem[Wu et~al\mbox{.}(2019)]%
        {wu2019detectron2}
\bibfield{author}{\bibinfo{person}{Yuxin Wu}, \bibinfo{person}{Alexander Kirillov}, \bibinfo{person}{Francisco Massa}, \bibinfo{person}{Wan-Yen Lo}, {and} \bibinfo{person}{Ross Girshick}.} \bibinfo{year}{2019}\natexlab{}.
\newblock \bibinfo{title}{Detectron2}.
\newblock \bibinfo{howpublished}{\url{https://github.com/facebookresearch/detectron2}}.
\newblock


\bibitem[Wu et~al\mbox{.}(2020)]%
        {wu2020scdetector}
\bibfield{author}{\bibinfo{person}{Yueming Wu}, \bibinfo{person}{Deqing Zou}, \bibinfo{person}{Shihan Dou}, \bibinfo{person}{Siru Yang}, \bibinfo{person}{Wei Yang}, \bibinfo{person}{Feng Cheng}, \bibinfo{person}{Hong Liang}, {and} \bibinfo{person}{Hai Jin}.} \bibinfo{year}{2020}\natexlab{}.
\newblock \showarticletitle{SCDetector: Software Functional Clone Detection Based on Semantic Tokens Analysis}. In \bibinfo{booktitle}{\emph{ASE}}. \bibinfo{pages}{821--833}.
\newblock


\bibitem[Wu et~al\mbox{.}(2022b)]%
        {wu2022vulcnn}
\bibfield{author}{\bibinfo{person}{Yueming Wu}, \bibinfo{person}{Deqing Zou}, \bibinfo{person}{Shihan Dou}, \bibinfo{person}{Wei Yang}, \bibinfo{person}{Duo Xu}, {and} \bibinfo{person}{Hai Jin}.} \bibinfo{year}{2022}\natexlab{b}.
\newblock \showarticletitle{VulCNN: An Image-inspired Scalable Vulnerability Detection System}. In \bibinfo{booktitle}{\emph{ICSE}}. \bibinfo{pages}{2365--2376}.
\newblock


\bibitem[Xia et~al\mbox{.}(2023)]%
        {xia2023automated}
\bibfield{author}{\bibinfo{person}{Chunqiu~Steven Xia}, \bibinfo{person}{Yuxiang Wei}, {and} \bibinfo{person}{Lingming Zhang}.} \bibinfo{year}{2023}\natexlab{}.
\newblock \showarticletitle{Automated program repair in the era of large pre-trained language models}. In \bibinfo{booktitle}{\emph{Proceedings of the 45th International Conference on Software Engineering (ICSE 2023). Association for Computing Machinery}}.
\newblock


\bibitem[Xie et~al\mbox{.}(2022)]%
        {xie2022lowresource}
\bibfield{author}{\bibinfo{person}{Rui Xie}, \bibinfo{person}{Tianxiang Hu}, \bibinfo{person}{Wei Ye}, {and} \bibinfo{person}{Shikun Zhang}.} \bibinfo{year}{2022}\natexlab{}.
\newblock \showarticletitle{Low-Resources Project-Specific Code Summarization}. In \bibinfo{booktitle}{\emph{ASE}}. \bibinfo{publisher}{{ACM}}, \bibinfo{pages}{68:1--68:12}.
\newblock


\bibitem[Xie et~al\mbox{.}(2021)]%
        {xie2021exploiting}
\bibfield{author}{\bibinfo{person}{Rui Xie}, \bibinfo{person}{Wei Ye}, \bibinfo{person}{Jinan Sun}, {and} \bibinfo{person}{Shikun Zhang}.} \bibinfo{year}{2021}\natexlab{}.
\newblock \showarticletitle{Exploiting Method Names to Improve Code Summarization: {A} Deliberation Multi-Task Learning Approach}. In \bibinfo{booktitle}{\emph{ICPC}}. \bibinfo{publisher}{{IEEE}}, \bibinfo{pages}{138--148}.
\newblock


\bibitem[Xu et~al\mbox{.}(2020)]%
        {xu2020incorporating}
\bibfield{author}{\bibinfo{person}{Frank~F. Xu}, \bibinfo{person}{Zhengbao Jiang}, \bibinfo{person}{Pengcheng Yin}, \bibinfo{person}{Bogdan Vasilescu}, {and} \bibinfo{person}{Graham Neubig}.} \bibinfo{year}{2020}\natexlab{}.
\newblock \showarticletitle{Incorporating External Knowledge through Pre-training for Natural Language to Code Generation}. In \bibinfo{booktitle}{\emph{ACL}}. \bibinfo{pages}{6045--6052}.
\newblock


\bibitem[Yamaguchi et~al\mbox{.}(2014)]%
        {yamaguchi2014modeling}
\bibfield{author}{\bibinfo{person}{Fabian Yamaguchi}, \bibinfo{person}{Nico Golde}, \bibinfo{person}{Daniel Arp}, {and} \bibinfo{person}{Konrad Rieck}.} \bibinfo{year}{2014}\natexlab{}.
\newblock \showarticletitle{Modeling and discovering vulnerabilities with code property graphs}. In \bibinfo{booktitle}{\emph{S\&P}}. \bibinfo{pages}{590--604}.
\newblock


\bibitem[Yang et~al\mbox{.}(2022a)]%
        {yang2022dualsc}
\bibfield{author}{\bibinfo{person}{Guang Yang}, \bibinfo{person}{Xiang Chen}, \bibinfo{person}{Yanlin Zhou}, {and} \bibinfo{person}{Chi Yu}.} \bibinfo{year}{2022}\natexlab{a}.
\newblock \showarticletitle{DualSC: Automatic Generation and Summarization of Shellcode via Transformer and Dual Learning}. In \bibinfo{booktitle}{\emph{SANER}}. \bibinfo{pages}{361--372}.
\newblock


\bibitem[Yang et~al\mbox{.}(2022c)]%
        {yang2021survey}
\bibfield{author}{\bibinfo{person}{Yanming Yang}, \bibinfo{person}{Xin Xia}, \bibinfo{person}{David Lo}, {and} \bibinfo{person}{John Grundy}.} \bibinfo{year}{2022}\natexlab{c}.
\newblock \showarticletitle{A Survey on Deep Learning for Software Engineering}.
\newblock \bibinfo{journal}{\emph{ACM Comput. Surv.}} \bibinfo{volume}{54}, \bibinfo{number}{10s}, Article \bibinfo{articleno}{206} (\bibinfo{date}{sep} \bibinfo{year}{2022}), \bibinfo{numpages}{73}~pages.
\newblock


\bibitem[Yang et~al\mbox{.}(2021)]%
        {yang2021multimodal}
\bibfield{author}{\bibinfo{person}{Zhen Yang}, \bibinfo{person}{Jacky Keung}, \bibinfo{person}{Xiao Yu}, \bibinfo{person}{Xiaodong Gu}, \bibinfo{person}{Zhengyuan Wei}, \bibinfo{person}{Xiaoxue Ma}, {and} \bibinfo{person}{Miao Zhang}.} \bibinfo{year}{2021}\natexlab{}.
\newblock \showarticletitle{A Multi-Modal Transformer-based Code Summarization Approach for Smart Contracts}. In \bibinfo{booktitle}{\emph{ICPC}}. \bibinfo{publisher}{{IEEE}}, \bibinfo{pages}{1--12}.
\newblock


\bibitem[Yang et~al\mbox{.}(2022b)]%
        {yang2022natural}
\bibfield{author}{\bibinfo{person}{Zhou Yang}, \bibinfo{person}{Jieke Shi}, \bibinfo{person}{Junda He}, {and} \bibinfo{person}{David Lo}.} \bibinfo{year}{2022}\natexlab{b}.
\newblock \showarticletitle{Natural Attack for Pre-trained Models of Code}. In \bibinfo{booktitle}{\emph{ICSE}}. \bibinfo{publisher}{{ACM}}, \bibinfo{pages}{1482--1493}.
\newblock


\bibitem[Yao et~al\mbox{.}(2019)]%
        {yao2019coacor}
\bibfield{author}{\bibinfo{person}{Ziyu Yao}, \bibinfo{person}{Jayavardhan~Reddy Peddamail}, {and} \bibinfo{person}{Huan Sun}.} \bibinfo{year}{2019}\natexlab{}.
\newblock \showarticletitle{Coacor: Code annotation for code retrieval with reinforcement learning}. In \bibinfo{booktitle}{\emph{The World Wide Web Conference}}. \bibinfo{pages}{2203--2214}.
\newblock


\bibitem[Yasunaga and Liang(2020)]%
        {yasunaga2020graph}
\bibfield{author}{\bibinfo{person}{Michihiro Yasunaga} {and} \bibinfo{person}{Percy Liang}.} \bibinfo{year}{2020}\natexlab{}.
\newblock \showarticletitle{Graph-based, self-supervised program repair from diagnostic feedback}. In \bibinfo{booktitle}{\emph{ICML}}. \bibinfo{pages}{10799--10808}.
\newblock


\bibitem[Ye et~al\mbox{.}(2020)]%
        {ye2020leveraging}
\bibfield{author}{\bibinfo{person}{Wei Ye}, \bibinfo{person}{Rui Xie}, \bibinfo{person}{Jinglei Zhang}, \bibinfo{person}{Tianxiang Hu}, \bibinfo{person}{Xiaoyin Wang}, {and} \bibinfo{person}{Shikun Zhang}.} \bibinfo{year}{2020}\natexlab{}.
\newblock \showarticletitle{Leveraging code generation to improve code retrieval and summarization via dual learning}. In \bibinfo{booktitle}{\emph{Proceedings of The Web Conference 2020}}. \bibinfo{pages}{2309--2319}.
\newblock


\bibitem[Yefet et~al\mbox{.}(2020)]%
        {yefet2020adversarial}
\bibfield{author}{\bibinfo{person}{Noam Yefet}, \bibinfo{person}{Uri Alon}, {and} \bibinfo{person}{Eran Yahav}.} \bibinfo{year}{2020}\natexlab{}.
\newblock \showarticletitle{Adversarial examples for models of code}.
\newblock \bibinfo{journal}{\emph{OOPSLA}}  \bibinfo{volume}{4} (\bibinfo{year}{2020}), \bibinfo{pages}{1--30}.
\newblock


\bibitem[Yin and Neubig(2017)]%
        {yin2017syntactic}
\bibfield{author}{\bibinfo{person}{Pengcheng Yin} {and} \bibinfo{person}{Graham Neubig}.} \bibinfo{year}{2017}\natexlab{}.
\newblock \showarticletitle{A Syntactic Neural Model for General-Purpose Code Generation}. In \bibinfo{booktitle}{\emph{ACL}}. \bibinfo{pages}{440--450}.
\newblock


\bibitem[Yu et~al\mbox{.}(2018a)]%
        {yu2018syntaxsqlnet}
\bibfield{author}{\bibinfo{person}{Tao Yu}, \bibinfo{person}{Michihiro Yasunaga}, \bibinfo{person}{Kai Yang}, \bibinfo{person}{Rui Zhang}, \bibinfo{person}{Dongxu Wang}, \bibinfo{person}{Zifan Li}, {and} \bibinfo{person}{Dragomir~R. Radev}.} \bibinfo{year}{2018}\natexlab{a}.
\newblock \showarticletitle{SyntaxSQLNet: Syntax Tree Networks for Complex and Cross-Domain Text-to-SQL Task}. In \bibinfo{booktitle}{\emph{EMNLP}}. \bibinfo{pages}{1653--1663}.
\newblock


\bibitem[Yu et~al\mbox{.}(2019a)]%
        {yu2019cosql}
\bibfield{author}{\bibinfo{person}{Tao Yu}, \bibinfo{person}{Rui Zhang}, \bibinfo{person}{Heyang Er}, \bibinfo{person}{Suyi Li}, \bibinfo{person}{Eric Xue}, \bibinfo{person}{Bo Pang}, \bibinfo{person}{Xi~Victoria Lin}, {et~al\mbox{.}}} \bibinfo{year}{2019}\natexlab{a}.
\newblock \showarticletitle{CoSQL: {A} Conversational Text-to-SQL Challenge Towards Cross-Domain Natural Language Interfaces to Databases}. In \bibinfo{booktitle}{\emph{EMNLP}}. \bibinfo{pages}{1962--1979}.
\newblock


\bibitem[Yu et~al\mbox{.}(2018b)]%
        {yu2018spider}
\bibfield{author}{\bibinfo{person}{Tao Yu}, \bibinfo{person}{Rui Zhang}, \bibinfo{person}{Kai Yang}, \bibinfo{person}{Michihiro Yasunaga}, \bibinfo{person}{Dongxu Wang}, \bibinfo{person}{Zifan Li}, {et~al\mbox{.}}} \bibinfo{year}{2018}\natexlab{b}.
\newblock \showarticletitle{Spider: {A} Large-Scale Human-Labeled Dataset for Complex and Cross-Domain Semantic Parsing and Text-to-SQL Task}. In \bibinfo{booktitle}{\emph{EMNLP}}. \bibinfo{pages}{3911--3921}.
\newblock


\bibitem[Yu et~al\mbox{.}(2019b)]%
        {yu2019sparc}
\bibfield{author}{\bibinfo{person}{Tao Yu}, \bibinfo{person}{Rui Zhang}, \bibinfo{person}{Michihiro Yasunaga}, \bibinfo{person}{Yi~Chern Tan}, \bibinfo{person}{Xi~Victoria Lin}, \bibinfo{person}{Suyi Li}, \bibinfo{person}{Heyang Er}, \bibinfo{person}{Irene Li}, \bibinfo{person}{Bo Pang}, \bibinfo{person}{Tao Chen}, {et~al\mbox{.}}} \bibinfo{year}{2019}\natexlab{b}.
\newblock \showarticletitle{SParC: Cross-Domain Semantic Parsing in Context}. In \bibinfo{booktitle}{\emph{ACL}}. \bibinfo{pages}{4511--4523}.
\newblock


\bibitem[Zhang et~al\mbox{.}(2020a)]%
        {zhang2020generating}
\bibfield{author}{\bibinfo{person}{Huangzhao Zhang}, \bibinfo{person}{Zhuo Li}, \bibinfo{person}{Ge Li}, \bibinfo{person}{Lei Ma}, \bibinfo{person}{Yang Liu}, {and} \bibinfo{person}{Zhi Jin}.} \bibinfo{year}{2020}\natexlab{a}.
\newblock \showarticletitle{Generating adversarial examples for holding robustness of source code processing models}. In \bibinfo{booktitle}{\emph{AAAI}}, Vol.~\bibinfo{volume}{34}. \bibinfo{pages}{1169--1176}.
\newblock


\bibitem[Zhang et~al\mbox{.}(2021b)]%
        {zhang2021disentangled}
\bibfield{author}{\bibinfo{person}{Jingfeng Zhang}, \bibinfo{person}{Haiwen Hong}, \bibinfo{person}{Yin Zhang}, \bibinfo{person}{Yao Wan}, \bibinfo{person}{Ye Liu}, {and} \bibinfo{person}{Yulei Sui}.} \bibinfo{year}{2021}\natexlab{b}.
\newblock \showarticletitle{Disentangled Code Representation Learning for Multiple Programming Languages}. In \bibinfo{booktitle}{\emph{Findings of ACL}}. \bibinfo{pages}{4454--4466}.
\newblock


\bibitem[Zhang et~al\mbox{.}(2022a)]%
        {zhang2022coditt5}
\bibfield{author}{\bibinfo{person}{Jiyang Zhang}, \bibinfo{person}{Sheena Panthaplackel}, \bibinfo{person}{Pengyu Nie}, \bibinfo{person}{Junyi~Jessy Li}, {and} \bibinfo{person}{Milos Gligoric}.} \bibinfo{year}{2022}\natexlab{a}.
\newblock \showarticletitle{CoditT5: Pretraining for Source Code and Natural Language Editing}. In \bibinfo{booktitle}{\emph{37th {IEEE/ACM} International Conference on Automated Software Engineering, {ASE} 2022, Rochester, MI, USA, October 10-14, 2022}}. \bibinfo{publisher}{{ACM}}, \bibinfo{pages}{22:1--22:12}.
\newblock


\bibitem[Zhang et~al\mbox{.}(2020c)]%
        {zhang2020retrieval}
\bibfield{author}{\bibinfo{person}{Jian Zhang}, \bibinfo{person}{Xu Wang}, \bibinfo{person}{Hongyu Zhang}, \bibinfo{person}{Hailong Sun}, {and} \bibinfo{person}{Xudong Liu}.} \bibinfo{year}{2020}\natexlab{c}.
\newblock \showarticletitle{Retrieval-based neural source code summarization}. In \bibinfo{booktitle}{\emph{ICSE}}. \bibinfo{pages}{1385--1397}.
\newblock


\bibitem[Zhang et~al\mbox{.}(2019)]%
        {zhang2019novel}
\bibfield{author}{\bibinfo{person}{Jian Zhang}, \bibinfo{person}{Xu Wang}, \bibinfo{person}{Hongyu Zhang}, \bibinfo{person}{Hailong Sun}, \bibinfo{person}{Kaixuan Wang}, {and} \bibinfo{person}{Xudong Liu}.} \bibinfo{year}{2019}\natexlab{}.
\newblock \showarticletitle{A novel neural source code representation based on abstract syntax tree}. In \bibinfo{booktitle}{\emph{ICSE}}. \bibinfo{pages}{783--794}.
\newblock


\bibitem[Zhang et~al\mbox{.}(2021a)]%
        {zhang2021interpretable}
\bibfield{author}{\bibinfo{person}{Tianyi Zhang}, \bibinfo{person}{Zhiyang Chen}, \bibinfo{person}{Yuanli Zhu}, \bibinfo{person}{Priyan Vaithilingam}, \bibinfo{person}{Xinyu Wang}, {and} \bibinfo{person}{Elena~L Glassman}.} \bibinfo{year}{2021}\natexlab{a}.
\newblock \showarticletitle{Interpretable Program Synthesis}. In \bibinfo{booktitle}{\emph{Proceedings of the 2021 CHI Conference on Human Factors in Computing Systems}}. \bibinfo{pages}{1--16}.
\newblock


\bibitem[Zhang et~al\mbox{.}(2020b)]%
        {zhang2020adversarial}
\bibfield{author}{\bibinfo{person}{Wei~Emma Zhang}, \bibinfo{person}{Quan~Z Sheng}, \bibinfo{person}{Ahoud Alhazmi}, {and} \bibinfo{person}{Chenliang Li}.} \bibinfo{year}{2020}\natexlab{b}.
\newblock \showarticletitle{Adversarial attacks on deep-learning models in natural language processing: A survey}.
\newblock \bibinfo{journal}{\emph{TIST}} \bibinfo{volume}{11}, \bibinfo{number}{3} (\bibinfo{year}{2020}), \bibinfo{pages}{1--41}.
\newblock


\bibitem[Zhang et~al\mbox{.}(2022b)]%
        {zhang2022diet}
\bibfield{author}{\bibinfo{person}{Zhaowei Zhang}, \bibinfo{person}{Hongyu Zhang}, \bibinfo{person}{Beijun Shen}, {and} \bibinfo{person}{Xiaodong Gu}.} \bibinfo{year}{2022}\natexlab{b}.
\newblock \showarticletitle{Diet code is healthy: simplifying programs for pre-trained models of code}. In \bibinfo{booktitle}{\emph{ESEC/FSE}}. \bibinfo{pages}{1073--1084}.
\newblock


\bibitem[Zhao and Huang(2018)]%
        {zhao2018deepsim}
\bibfield{author}{\bibinfo{person}{Gang Zhao} {and} \bibinfo{person}{Jeff Huang}.} \bibinfo{year}{2018}\natexlab{}.
\newblock \showarticletitle{Deepsim: deep learning code functional similarity}. In \bibinfo{booktitle}{\emph{ESEC/FSE}}. \bibinfo{pages}{141--151}.
\newblock


\bibitem[Zheng et~al\mbox{.}(2023)]%
        {zheng2023codegeex}
\bibfield{author}{\bibinfo{person}{Qinkai Zheng}, \bibinfo{person}{Xiao Xia}, \bibinfo{person}{Xu Zou}, \bibinfo{person}{Yuxiao Dong}, \bibinfo{person}{Shan Wang}, \bibinfo{person}{Yufei Xue}, \bibinfo{person}{Zihan Wang}, \bibinfo{person}{Lei Shen}, \bibinfo{person}{Andi Wang}, \bibinfo{person}{Yang Li}, {et~al\mbox{.}}} \bibinfo{year}{2023}\natexlab{}.
\newblock \showarticletitle{Codegeex: A pre-trained model for code generation with multilingual evaluations on humaneval-x}.
\newblock \bibinfo{journal}{\emph{arXiv preprint arXiv:2303.17568}} (\bibinfo{year}{2023}).
\newblock


\bibitem[Zhong et~al\mbox{.}(2017)]%
        {zhong2017seq2sql}
\bibfield{author}{\bibinfo{person}{Victor Zhong}, \bibinfo{person}{Caiming Xiong}, {and} \bibinfo{person}{Richard Socher}.} \bibinfo{year}{2017}\natexlab{}.
\newblock \showarticletitle{Seq2sql: Generating structured queries from natural language using reinforcement learning}.
\newblock \bibinfo{journal}{\emph{arXiv:1709.00103}} (\bibinfo{year}{2017}).
\newblock


\bibitem[Zhou et~al\mbox{.}(2021)]%
        {zhou2021assessing}
\bibfield{author}{\bibinfo{person}{Xin Zhou}, \bibinfo{person}{DongGyun Han}, {and} \bibinfo{person}{David Lo}.} \bibinfo{year}{2021}\natexlab{}.
\newblock \showarticletitle{Assessing Generalizability of CodeBERT}. In \bibinfo{booktitle}{\emph{ICSME}}. \bibinfo{publisher}{{IEEE}}, \bibinfo{pages}{425--436}.
\newblock


\bibitem[Zhou et~al\mbox{.}(2019)]%
        {zhou2019devign}
\bibfield{author}{\bibinfo{person}{Yaqin Zhou}, \bibinfo{person}{Shangqing Liu}, \bibinfo{person}{Jing~Kai Siow}, \bibinfo{person}{Xiaoning Du}, {and} \bibinfo{person}{Yang Liu}.} \bibinfo{year}{2019}\natexlab{}.
\newblock \showarticletitle{Devign: Effective Vulnerability Identification by Learning Comprehensive Program Semantics via Graph Neural Networks}. In \bibinfo{booktitle}{\emph{NeurIPS}}. \bibinfo{pages}{10197--10207}.
\newblock


\bibitem[Zhou et~al\mbox{.}(2022)]%
        {zhou2022adversarial}
\bibfield{author}{\bibinfo{person}{Yu Zhou}, \bibinfo{person}{Xiaoqing Zhang}, \bibinfo{person}{Juanjuan Shen}, \bibinfo{person}{Tingting Han}, \bibinfo{person}{Taolue Chen}, {and} \bibinfo{person}{Harald~C. Gall}.} \bibinfo{year}{2022}\natexlab{}.
\newblock \showarticletitle{Adversarial Robustness of Deep Code Comment Generation}.
\newblock \bibinfo{journal}{\emph{TOSEM}} \bibinfo{volume}{31}, \bibinfo{number}{4} (\bibinfo{year}{2022}), \bibinfo{pages}{60:1--60:30}.
\newblock


\bibitem[Zhu et~al\mbox{.}(2020)]%
        {zhu2020ocor}
\bibfield{author}{\bibinfo{person}{Qihao Zhu}, \bibinfo{person}{Zeyu Sun}, \bibinfo{person}{Xiran Liang}, \bibinfo{person}{Yingfei Xiong}, {and} \bibinfo{person}{Lu Zhang}.} \bibinfo{year}{2020}\natexlab{}.
\newblock \showarticletitle{OCoR: an overlapping-aware code retriever}. In \bibinfo{booktitle}{\emph{ASE}}. \bibinfo{pages}{883--894}.
\newblock


\bibitem[Zhu et~al\mbox{.}(2021)]%
        {zhu2021syntax}
\bibfield{author}{\bibinfo{person}{Qihao Zhu}, \bibinfo{person}{Zeyu Sun}, \bibinfo{person}{Yuan{-}an Xiao}, \bibinfo{person}{Wenjie Zhang}, \bibinfo{person}{Kang Yuan}, \bibinfo{person}{Yingfei Xiong}, {and} \bibinfo{person}{Lu Zhang}.} \bibinfo{year}{2021}\natexlab{}.
\newblock \showarticletitle{A syntax-guided edit decoder for neural program repair}. In \bibinfo{booktitle}{\emph{ESEC/FSE}}. \bibinfo{pages}{341--353}.
\newblock


\bibitem[Zhu et~al\mbox{.}(2022)]%
        {zhu2022simple}
\bibfield{author}{\bibinfo{person}{Xiaoning Zhu}, \bibinfo{person}{Chaofeng Sha}, {and} \bibinfo{person}{Junyu Niu}.} \bibinfo{year}{2022}\natexlab{}.
\newblock \showarticletitle{A Simple Retrieval-based Method for Code Comment Generation}. In \bibinfo{booktitle}{\emph{SANER}}. \bibinfo{pages}{1089--1100}.
\newblock


\bibitem[Zou et~al\mbox{.}(2019)]%
        {zou2019muvuldeepecker}
\bibfield{author}{\bibinfo{person}{Deqing Zou}, \bibinfo{person}{Sujuan Wang}, \bibinfo{person}{Shouhuai Xu}, \bibinfo{person}{Zhen Li}, {and} \bibinfo{person}{Hai Jin}.} \bibinfo{year}{2019}\natexlab{}.
\newblock \showarticletitle{$\mu$VulDeePecker: A deep learning-based system for multiclass vulnerability detection}.
\newblock \bibinfo{journal}{\emph{TDSC}} (\bibinfo{year}{2019}).
\newblock


\bibitem[Zou et~al\mbox{.}(2021)]%
        {zou2021interpreting}
\bibfield{author}{\bibinfo{person}{Deqing Zou}, \bibinfo{person}{Yawei Zhu}, \bibinfo{person}{Shouhuai Xu}, \bibinfo{person}{Zhen Li}, \bibinfo{person}{Hai Jin}, {and} \bibinfo{person}{Hengkai Ye}.} \bibinfo{year}{2021}\natexlab{}.
\newblock \showarticletitle{Interpreting deep learning-based vulnerability detector predictions based on heuristic searching}.
\newblock \bibinfo{journal}{\emph{TOSEM}} \bibinfo{volume}{30}, \bibinfo{number}{2} (\bibinfo{year}{2021}), \bibinfo{pages}{1--31}.
\newblock


\bibitem[Z{\"{u}}gner et~al\mbox{.}(2021)]%
        {zugner2021language}
\bibfield{author}{\bibinfo{person}{Daniel Z{\"{u}}gner}, \bibinfo{person}{Tobias Kirschstein}, \bibinfo{person}{Michele Catasta}, \bibinfo{person}{Jure Leskovec}, {and} \bibinfo{person}{Stephan G{\"{u}}nnemann}.} \bibinfo{year}{2021}\natexlab{}.
\newblock \showarticletitle{Language-Agnostic Representation Learning of Source Code from Structure and Context}. In \bibinfo{booktitle}{\emph{ICLR}}.
\newblock


\end{thebibliography}


\begin{thebibliography}{10}

\bibitem{hochreiter1997long2}
Sepp Hochreiter and J{\"u}rgen Schmidhuber.
\newblock Long short-term memory.
\newblock {\em Neural computation}, 9(8):1735--1780, 1997.

\bibitem{chung2015gated2}
Junyoung Chung, Caglar Gulcehre, Kyunghyun Cho, and Yoshua Bengio.
\newblock Gated feedback recurrent neural networks.
\newblock In {\em ICML}, pages 2067--2075, 2015.

\bibitem{bahdanau2014neural2}
Dzmitry Bahdanau, Kyunghyun Cho, and Yoshua Bengio.
\newblock Neural machine translation by jointly learning to align and translate.
\newblock In {\em ICLR}, 2015.

\bibitem{husain2019codesearchnet2}
Hamel Husain, Ho-Hsiang Wu, Tiferet Gazit, Miltiadis Allamanis, and Marc Brockschmidt.
\newblock Codesearchnet challenge: Evaluating the state of semantic code search.
\newblock {\em arXiv:1909.09436}, 2019.

\bibitem{huang2021cosqa2}
Junjie Huang, Duyu Tang, Linjun Shou, Ming Gong, Ke~Xu, Daxin Jiang, Ming Zhou, and Nan Duan.
\newblock Cosqa: 20, 000+ web queries for code search and question answering.
\newblock In {\em ACL}, pages 5690--5700, 2021.

\bibitem{raychev2016probabilistic2}
Veselin Raychev, Pavol Bielik, and Martin Vechev.
\newblock Probabilistic model for code with decision trees.
\newblock {\em ACM SIGPLAN Notices}, 51(10):731--747, 2016.

\bibitem{barone2017parallel2}
Antonio Valerio~Miceli Barone and Rico Sennrich.
\newblock A parallel corpus of python functions and documentation strings for automated code documentation and code generation.
\newblock In {\em IJCNLP}, pages 314--319, 2017.

\bibitem{wan2018improving2}
Yao Wan, Zhou Zhao, Min Yang, Guandong Xu, Haochao Ying, Jian Wu, and Philip~S Yu.
\newblock Improving automatic source code summarization via deep reinforcement learning.
\newblock In {\em ASE}, pages 397--407, 2018.

\bibitem{hu2018deep2}
Xing Hu, Ge~Li, Xin Xia, David Lo, and Zhi Jin.
\newblock Deep code comment generation.
\newblock In {\em ICPC}, pages 200--20010, 2018.

\bibitem{liu2020retrieval2}
Shangqing Liu, Yu~Chen, Xiaofei Xie, Jing~Kai Siow, and Yang Liu.
\newblock Retrieval-augmented generation for code summarization via hybrid gnn.
\newblock In {\em ICLR}, 2020.

\bibitem{hellendoorn2018deep2}
Vincent~J Hellendoorn, Christian Bird, Earl~T Barr, and Miltiadis Allamanis.
\newblock Deep learning type inference.
\newblock In {\em ESEC/FSE}, pages 152--162, 2018.

\bibitem{puri2021project2}
Ruchir Puri, David~S. Kung, Geert Janssen, Wei Zhang, Giacomo Domeniconi, et~al.
\newblock Project codenet: {A} large-scale {AI} for code dataset for learning a diversity of coding tasks.
\newblock {\em CoRR}, abs/2105.12655, 2021.

\bibitem{agashe2019juice2}
Rajas Agashe, Srinivasan Iyer, and Luke Zettlemoyer.
\newblock Juice: {A} large scale distantly supervised dataset for open domain context-based code generation.
\newblock In {\em EMNLP}, pages 5435--5445, 2019.

\bibitem{cummins2020programl2}
Chris Cummins, Zacharias Fisches, Tal Ben-Nun, Torsten Hoefler, Michael O'Boyle, and Hugh Leather.
\newblock {ProGraML: A Graph-based Program Representation for Data Flow Analysis and Compiler Optimizations}.
\newblock In {\em ICML}, 2021.

\bibitem{leclair2019recommendations2}
Alexander LeClair and Collin McMillan.
\newblock Recommendations for datasets for source code summarization.
\newblock In {\em NAACL}, pages 3931--3937, 2019.

\bibitem{hasan2021codesc2}
Masum Hasan, Tanveer Muttaqueen, Abdullah~Al Ishtiaq, Kazi~Sajeed Mehrab, Md. Mahim~Anjum Haque, Tahmid Hasan, Wasi~Uddin Ahmad, Anindya Iqbal, and Rifat Shahriyar.
\newblock Codesc: {A} large code-description parallel dataset.
\newblock In {\em Findings of ACL}, pages 210--218, 2021.

\bibitem{ahmad2021avatar2}
Wasi~Uddin Ahmad, Md~Golam~Rahman Tushar, Saikat Chakraborty, and Kai-Wei Chang.
\newblock Avatar: A parallel corpus for java-python program translation.
\newblock {\em arXiv:2108.11590}, 2021.

\bibitem{bahrami2021pytorrent2}
Mehdi Bahrami, NC~Shrikanth, Shade Ruangwan, Lei Liu, Yuji Mizobuchi, Masahiro Fukuyori, Wei-Peng Chen, Kazuki Munakata, and Tim Menzies.
\newblock Pytorrent: A python library corpus for large-scale language models.
\newblock {\em arXiv:2110.01710}, 2021.

\bibitem{yao2018staqc2}
Ziyu Yao, Daniel~S Weld, Wei-Peng Chen, and Huan Sun.
\newblock Staqc: A systematically mined question-code dataset from stack overflow.
\newblock In {\em Proceedings of the 2018 World Wide Web Conference}, pages 1693--1703, 2018.

\bibitem{liu2021codeqa2}
Chenxiao Liu and Xiaojun Wan.
\newblock Codeqa: {A} question answering dataset for source code comprehension.
\newblock In {\em Findings of EMNLP}, pages 2618--2632, 2021.

\bibitem{lu2021codexglue2}
Shuai Lu, Daya Guo, Shuo Ren, Junjie Huang, Alexey Svyatkovskiy, et~al.
\newblock Codexglue: {A} machine learning benchmark dataset for code understanding and generation.
\newblock In {\em NeurIPS Datasets and Benchmarks}, 2021.

\bibitem{papineni2002bleu}
K.~Papineni, S.~Roukos, T.~Ward, and W.~J. Zhu.
\newblock Bleu: a method for automatic evaluation of machine translation.
\newblock In {\em ACL}, pages 311--318, 2002.

\bibitem{banerjee2005meteor}
S.~Banerjee and A.~Lavie.
\newblock Meteor: An automatic metric for mt evaluation with improved correlation with human judgments.
\newblock In {\em Proceedings of the acl workshop on intrinsic and extrinsic evaluation measures for machine translation and/or summarization}, volume~29, pages 65--72, 2005.

\bibitem{lin2004rouge}
C.~Y. Lin.
\newblock Rouge: A package for automatic evaluation of summaries.
\newblock {\em Text Summarization Branches Out}, 2004.

\end{thebibliography}

\clearpage
    


\section*{\Large \bfseries Supplementary Materials}
\vspace{0.6em}

\appendix

\setcounter{section}{0}

\section{Surveyed Venues}
Table~\ref{table_venue} summarizes a list of top-tier conferences and journals surveyed in this paper.

\section{Neural Network Modules and Techniques}
Benefiting from the powerful representation ability of deep learning, recently deep neural networks have been widely used to represent source code as distributed vector representations.
Here we quickly review several major neural network modules and techniques, e.g., RNN, CNN, attention mechanism, Transformer, graph neural network, and model pre-training.
Let $\mathcal{C}=\{c_1, c_2,\ldots, c_n\}$ denote a code corpus, which is composed of a collection of code snippets.
For each code snippet $c_i$, it is composed of a sequence of code tokens, i.e., $c_i=\{x_1, x_2,\ldots,x_{|c_i|}\}$, where $|\cdot|$ denotes the length of the sequence.
Let $\mathbf{x}_i=w(x_i)$ denote the word embedding corresponding to the $i$-th token in the code snippet.

\noindent{\textbf{RNN.}}
Recurrent Neural Networks (RNNs) are neural networks designed to handle the sequential inputs with variable lengths.
At time step $t$, the hidden state $\mathbf{h}_t$ of RNNs is updated as follows:
\begin{equation}
	\mathbf{h}_t = f(\mathbf{x}_t, \mathbf{h}_{t-1})\,,
\end{equation}
where 
$f$ is the non-linear mapping function, which is usually a hyperbolic tangent, i.e., $f(\mathbf{x}_t, \mathbf{h}_t)=\operatorname{tanh}(\mathbf{W}_x\mathbf{x}_t+\mathbf{W}_h\mathbf{h}_{t-1}+\mathbf{b})$.

There have been many variants of RNNs. To alleviate the gradient vanishing issue of RNNs, the Long Short-Term Memory (LSTM)~\citesecondary{hochreiter1997long2} technology with a gate mechanism is proposed to determine the information accumulation, where the input gate, forget gate and output gate control the input, output and forget part of the entire network through weights and activation function.
GRU (Gate Recurrent Unit)~\citesecondary{chung2015gated2} is a simplified version of LSTM with fewer parameters by combining the forget and input gates into a single update gate by merging the cell state and hidden state.
Furthermore, bi-directional RNNs are used to represent the sequential contents in both forward and backward directions, with two parallel RNNs and combined outputs.

Although RNNs are effective in representing sequential texts, it is difficult to be parallelized since sequential computation inhibits parallelization. That means, the computation of the current time step is dependent on the output of the previous time step.
Furthermore, the RNNs are also difficult to handle long-range dependency when processing long sequences.

\noindent{\textbf{CNN.}}
Convolutional Neural Networks (CNNs) which are originally designed to process the pixels of images, have also been introduced to model the sequential texts.
A CNN is composed of convolutional layers and pooling layers. 
The convolutional layer uses convolution operation to extract meaningful local patterns of inputs, and the pooling layer reduces the parameters and computation to make the networks deeper.
A CNN first concatenates the word embedding of each code token as
$	\mathbf{x}_{1:n}=\mathbf{x}_1\oplus \mathbf{x}_2\oplus  \cdots \oplus \mathbf{x}_n
$, where $\oplus $ is the concatenation operator. For a window of words $\mathbf{x}_{i:i+h-1}$, a feature $\mathbf{h}_i$ is calculated by the filter $f$ as $\mathbf{h}_i = f(\mathbf{W}\mathbf{x}_{i:i+h-1} + \mathbf{b})$, 
where $\mathbf{b}$ is a bias term and $f$ is a non-linear function like hyperbolic tangent.
After applying the filter to all possible windows of words, a feature map will be obtained, i.e.,
$\mathbf{h} = [\mathbf{h}_1,\mathbf{h}_2,\ldots,\mathbf{h}_{n-h+1}]$.
Then a pooling layer can be applied to obtain the final representation of code, i.e., $\hat{\mathbf{h}} = \operatorname{Pooling}(\mathbf{h})$.

The CNNs that exploit local dependencies using convolution operations are easy to parallelize.
However, they require many layers to model the long-distance dependencies when handling long sequences.

\noindent{\textbf{Attention Mechanism.}}
The attention mechanism is a simple yet effective module in deep neural networks. 
It was first introduced by~\citesecondary{bahdanau2014neural2} in Neural Machine Translation (NMT) to mitigate the issue 
of information loss in compressing long sentences into a fixed-length vector, as well as the alignment between input and output sequences.
Since then, the attention mechanism has become a widely used component to improve the performance of various models in NLP and computer vision.
In NMT, the attention mechanism is designed to include an additional context vector 
that allows the decoder to access the entire encoded input sequence $\{\mathbf{h}_1, \mathbf{h}_2,\ldots, \mathbf{h}_m \}$. Especially, it aims to learn the attention weights $\alpha_{ij}$, which captures the relevance between the encoder hidden state $\mathbf{h}_i$ and decoder hidden state $\mathbf{s}_j$. Consequently, the context vector can be formulated as $c_j=\sum_{i=1}^{T}\alpha_{ij}\mathbf{h}_i$.
In general, the attention can also be seen as a mapping of keys $\mathbf{K}$ to an attention distribution $\alpha$ according to query $\mathbf{q}$, i.e., $\alpha(\mathbf{q}, \mathbf{K}) = \operatorname{softmax}(g(\mathbf{q}, \mathbf{K}))$,
where 
$g$ is the attention score function which measures the similarity between the query and key. 
In some cases, there is also an additional input of values $V$ on which the attention distribution is applied. Hence a generalized attention with a set of key-value pairs $(\mathbf{K}, \mathbf{V})$ and query $\mathbf{Q}$ is formulated as $\alpha(\mathbf{Q}, \mathbf{K}, \mathbf{V}) = \operatorname{softmax}(g(\mathbf{Q}, \mathbf{K}))\cdot \mathbf{V}$.

\begin{table*}[t!]
	\centering
	\caption{A list of top-tier conferences and journals surveyed in this paper.}
	\label{table_venue}
	\setlength{\tabcolsep}{6pt} 
	\resizebox{1.\linewidth}{!}{
		\begin{tabular}{l|l|l|l}
			\toprule
			&No.&\textbf{Venue}	&	\textbf{Venue (Full name)}	\\
			\midrule
			\multirow{9}{*}{\rotatebox[origin=c]{90}{SE}}&1	&ICSE	&ACM/IEEE International Conference on Software Engineering	\\
			&2	&ASE	&IEEE/ACM International Conference Automated Software Engineering	\\
			&3	&FSE/ESEC	&ACM SIGSOFT Symposium on the Foundation of Software Engineering/European Software Engineering Conference	\\
			&4	&SANER	&IEEE International Conference on Software Analysis, Evolution and Reengineering	\\
			&5	&ICSME	&IEEE International Conference on Software Maintenance and Evolution	\\
			&6	&ICPC	&IEEE International Conference on Program Comprehension	\\
			&7	&ESEM	&International Symposium on Empirical Software Engineering and Measurement	\\
   			&8	&MSR	&International Conference on Mining Software Repositories \\
			&9	&TSE	&IEEE Transactions on Software Engineering	\\
			&10	&TOSEM	&ACM Transactions on Software Engineering and Methodology	\\
			\hline
			\multirow{7}{*}{\rotatebox[origin=c]{90}{PL/Compiler}}&1	&POPL	&ACM SIGPLAN-SIGACT Symposium on Principles of Programming Languages	\\
			&2	&PLDI	&ACM SIGPLAN Conference on Programming Language Design \& Implementation	\\
			&3	&OOPSLA	&Conference on Object-Oriented Programming Systems, Languages, and Applications	\\
			&4	&CGO	&Code Generation and Optimization	\\
			&5	&PACT	&International Conference on Parallel Architectures and Compilation Techniques	\\
			&6	&TOPLAS	&ACM Transactions on Programming Languages \& Systems	\\
			&7	&TACO	&ACM Transactions on Architecture and Code Optimization	\\
			\hline
			\multirow{7}{*}{\rotatebox[origin=c]{90}{AI}}&1	&NeurIPS	&Annual Conference on Neural Information Processing Systems	\\
			&2	&ICML	&International Conference on Machine Learning	\\
			&3	&ICLR	& International Conference on Learning Representations	\\
			&4	&AAAI	&AAAI Conference on Artificial Intelligence	\\
			&5	&IJCAI	&International Joint Conference on Artificial Intelligence	\\
			&6	&SIGKDD	&ACM Knowledge Discovery and Data Mining	\\
			&7	&WWW	&The Web Conference	\\
			\hline
			\multirow{3}{*}{\rotatebox[origin=c]{90}{NLP}}&1	&ACL	&Annual Meeting of the Association for Computational Linguistics	\\
			&2	&EMNLP	&Conference on Empirical Methods in Natural Language Processing	\\
			&3	&NAACL	&The Annual Conference of the North American Chapter of the Association for Computational Linguistics	\\
			\hline
			\multirow{5}{*}{\rotatebox[origin=c]{90}{Security}}&1	&CCS	&ACM Conference on Computer and Communications Security	\\
			&2	&S\&P	&IEEE Symposium on Security and Privacy	\\
			&3	&USENIX Security	&Usenix Security Symposium	\\
			&4	&NDSS	&ISOC Network and Distributed System Security Symposium	\\
			&5	&TIFS	&IEEE Transactions on Information Forensics and Security	\\
			\bottomrule
		\end{tabular}
	}
\end{table*}

\noindent{\textbf{Transformer.}}
Since sequential computation in RNNs inhibits parallelization, recently, a new state-of-the-art network Transformer has been designed for parallel processing of sequences~\cite{vaswani2017attention}. 
The Transformer is mainly composed of multiple self-attention blocks.
When feeding a sequence of code tokens $c_i=\{x_1, x_2,\ldots,x_{|c_i|}\}$ into Transformer, the Transformer block at layer $l$ will produce a sequence of hidden states for each code token, i.e., $\mathbf{H}^{l}=[ \mathbf{h}_1^l, \mathbf{h}_2^l,\ldots,\mathbf{h}_{|c_i|}^l ]$.
For each layer, the layer representation $\mathbf{H}^{l}$ is computed by the $l$-th layer Transformer block $\mathbf{H}^{l} = \mathrm{ Transformer}_{l}(\mathbf{H}^{l-1})$, where $l \in \{1,2,\ldots,L\}$.  
In each Transformer block, multiple self-attention heads are used to aggregate the output vectors of the previous layer.
A general attention mechanism can be formulated as the weighted sum of the value vector $\mathbf{V}$, using the query vector $\mathbf{Q}$ and the key vector $\mathbf{K}$:
\begin{equation}
	\alpha(\mathbf{Q}, \mathbf{K}, \mathbf{V})=\operatorname{softmax}\left(\frac{\mathbf{Q} \mathbf{K}^{T}}{\sqrt{d_{\text {model}}}}\right) \cdot \mathbf{V}\,,
\end{equation}
where $d_{\rm model}$ represents the dimension of hidden representations. For self-attention, $\mathbf{Q}$, $\mathbf{K}$, and $\mathbf{V}$ are mappings of previous hidden representations by different linear functions, {\em i.e.,}
$\mathbf{Q} =\mathbf{H}^{l-1} \mathbf{W}_{Q}^{l}$, $\mathbf{K} =\mathbf{H}^{l-1} \mathbf{W}_{K}^{l}$, and  $\mathbf{V} =\mathbf{H}^{l-1} \mathbf{W}_{V}^{l}$, respectively. 
At last, the encoder produces a final contextual representation $\mathbf{H}^{L} = [\mathbf{h}^{L}_{1}, \ldots, \mathbf{h}^{L}_{n}]$, which is obtained from the last Transformer block.

\noindent{\textbf{Graph Neural Network.}}
The Graph Neural Networks (GNNs) take a graph as input and the goal is to learn the representations of each node by recursively updating until convergence, which aggregates the information of each neighborhood node. 
Let $\mathcal{G} = (\mathcal{V}, \mathcal{E}, \mathbf{A})$ denote a graph, where $\mathcal{V}$ is a set of nodes, $\mathcal{E}$ is a set of edges, and $\mathbf{A}$ is a adjacency matrix. In a graph, let $v_i \in \mathcal{V}$ denote a node and $e_{ij} = (v_i, v_j) \in \mathcal{E}$ denote an edge. 
For graph $\mathcal{G}$, the hidden representation $\mathbf{h}_v$ for each node $v$ can be updated as follows:
\begin{equation}
	\mathbf{h}_{v,t+1} = f(\ell_v, \ell_e, \mathbf{h}_{v',t}, \ell_{v'})\,, 
\end{equation}
where $f$ is a local transition function (shared among all nodes), which updates the node state according to the input neighborhood.
$\ell_v$ denotes the label attributes of node $v$, $\ell_e$ denotes the label attributes of the corresponding edges of node $v$, $\mathbf{h}_{v',t}$ denotes the hidden representations of node $v$'s neighbors at time step $t$, and $\ell_{v'}$ denotes the label attributes of node $v$'s neighbors. 
For the options of $f$, there are several approaches attempting to use the gate mechanism like GRU~\citesecondary{chung2015gated} or LSTM~\citesecondary{hochreiter1997long} in the propagation step to diminish the restrictions in the former GNN models, thereby improving the long-term propagation of information across the graph structure. If the local transition function $f$ is GRU, the model is called Gated Graph Neural Network (GGNN)~\citesecondary{li2015gated}.

\section{Public Datasets}
\begin{table*}[t!]
	\centering
	\caption{A summary of datasets that have been used for code intelligence.}
	\label{table_dataset}
	\setlength{\tabcolsep}{3pt} 
	\resizebox{\linewidth}{!}{
		\begin{tabular}{l|c|c|c|c|c|c|c}
			\toprule
			\textbf{Dataset}  &\textbf{Language}& \textbf{Train}	& \textbf{Valid} & \textbf{Test} & \textbf{Doc.} & \textbf{Granularity} & \textbf{AST} \\
			\midrule
			CSN-Java	&Java &454,451 &15,328 &26,909 & Y & Function& Y  \\
			CSN-JavaScript	&JavaScript &123,889 &8,253 &6,483 & Y & Function& Y \\
			CSN-Python	&Python &412,178 &23,107 &22,176 & Y & Function& Y \\
			CSN-PHP	&PHP &523,712 &26,015 &28,391 & Y & Function& Y\\
			CSN-Go	&Go &317,832 &14,242 &14,291 & Y & Function& Y \\
			CSN-Ruby	&Ruby &48,791 &2,209 &2279 & Y & Function& Y \\
			\hline
			CoSQA-QA	&Java, Python &20,000 & 604 &- & Y & Function & N \\
			CoSQA-Search	&Java, Python &19,604 & 500 &500 & Y & Function & N \\
			\hline
			Py150	&Python &100,000 &- &50,000 & N & File& Y \\
			python-code-docstring	&Python	&55,538 &18,505 &18,502 & Y & Function& Y \\
			DeepCom-Java	&Java &69,708 &8,714 &8,714 & Y & Function& Y \\
			CCSD	&C &1,275,937 &425,312 &425,312 & Y & Function & N \\
			Typescript	&Typescript & 49,850 & 7,854 & 4,650 & N & File & N \\
			CodeNet	&mainly C++ & 1,235,000 &-  &- & N & File & N \\
			JuICe	&Python &1,518,049 &1,744 &1,981 & Y & File & N \\
			ProGraML	&C, C++, Fortran, Haskell, Swift &250,428 &- &- & N & File & N \\
			FunCom	&Java &1,937,136 &106,153 &105,832 & Y & Function & N \\
			CoDesc	&Java &3,369,218  &421,149 &421,149 & Y & Function & N \\
			APPS	&Python &117,232 &- &115,212 & N & File & N \\
			AVATAR	&Python, Java &40,423 &848 &1,699 & N &File &N \\
			PyTorrent   &Python &2,273,157 &284,145 &284,144 & Y &Function &Y \\
		    StaQC	&Python, SQL	&267,065 &- &- &N & File &N \\
			\hline
			CodeQA-Python	&Python &56,084 &6,999 &6,999 & N & Function & N \\
			CodeQA-Java	&Java &95,777 &11,999 &11,999 & N & Function & N \\
			\hline
			CodeXGLUE-BCB	&Java &901,028 &415,416 &415,416 & N & Function & N \\
			CodeXGLUE-POJ104	&C/C++ &32,000 &8,000 &12,000 & N & Function & N \\
			CodeXGLUE-Devign	&C &21,854 &2,732 &2,732 & N & Function & N \\
			CodeXGLUE-Bugs2Fix	&Java &52,364 &6,545 &6,545 & N & Function & N \\
			CodeXGLUE-CodeTrans	&Java/C\# &10,295 &499 &1,000 & N & Function & N \\
			CodeXGLUE-WebQuery	&Python &251,820 &9,604 &- & Y & Function & N \\
			CodeXGLUE-Concode	&Java &100,000 &2,000 &2,000 & Y & Function & N \\
			CodeXGLUE-MicoDocs	&mainly English &155,926 &4,000 &4,00 & Y & Docs & N \\
			\bottomrule
		\end{tabular}
	}
\end{table*}

Many datasets have been collected for different tasks of code intelligence.
In the literature, many pre-processing steps have been conducted before using the dataset for evaluation.
\begin{itemize}
	\item \textbf{CodeSearchNet (CSN)}\footnote{https://github.com/github/CodeSearchNet}~\citesecondary{husain2019codesearchnet2} is a dataset collected from GitHub, originally for code search. It contains about 6 million functions, where 2 millions of them are 
	annotated
	with natural language descriptions, covering six programming languages (i.e., Go, Java, JavaScript, PHP, Python, and Ruby). Since the code snippets in this dataset are paired with natural language descriptions and have the characteristic of multi-linguality, this dataset can also be used for code summarization and cross-language-related tasks.
	In addition, this dataset has been used for large-scale pre-training, attributed to its large scale.
	\item \textbf{CoSQA}\footnote{https://github.com/Jun-jie-Huang/CoCLR}~\citesecondary{huang2021cosqa2} is a dataset of paired natural-language queries and code snippets designed for code search and question answering, where the queries are from the real search logs of Microsoft Bing and the code snippets are from GitHub. It consists of 20,604 pairs of data samples and each pair is annotated by at least 3 crowd-sourced workers.
	\item \textbf{Py150}\footnote{https://eth-sri.github.io/py150}~\citesecondary{raychev2016probabilistic2} is a collection of 150$k$ Python source code files from GitHub, which has been widely used for evaluating code completion.
 	\item \textbf{python-code-docstring}\footnote{https://github.com/wanyao1992/code\_summarization\_public}~\citesecondary{barone2017parallel2,wan2018improving2} is a dataset of parallel Python code snippets with corresponding descriptions, which has been widely adopted for code summarization. 
	\item \textbf{DeepCom-Java}\footnote{https://github.com/xing-hu/DeepCom}~\citesecondary{hu2018deep2} is a dataset collected from the open source repositories in GitHub. In this dataset, only repositories are implemented in Java, and those with at least 10 stars are considered. Finally, 588,108 pairs of Java methods and their corresponding Javadoc are obtained.
	\item \textbf{CCSD-benchmark-for-code-summarization}\footnote{https://github.com/shangqing-liu/CCSD-benchmark-for-code-summarization}~\citesecondary{liu2020retrieval2} is a C benchmark dataset for code summarization, which is crawled from GitHub, and contains 95$k$+ functions as well as natural-language descriptions.
	\item \textbf{Typescript}\footnote{https://github.com/DeepTyper/DeepTyper}~\citesecondary{hellendoorn2018deep2} is a dataset of Typescript code snippets, which contain the type information of variables and can be used for type inference evaluation.
	\item \textbf{CodeNet}\footnote{https://github.com/IBM/Project\_CodeNet}~\citesecondary{puri2021project2} is a large-scale dataset released by IBM for code intelligence, consisting of 14 million code samples and about 500 million lines of code in 55 different programming languages.
	\item \textbf{JuICe}\footnote{https://github.com/rajasagashe/juice}~\citesecondary{agashe2019juice2} is large-scale dataset for open-domain, context-based code generation, which consists of 1.5 million examples with a curated test set of 3.7$k$ instances from online programming assignments.
	\item \textbf{ProGraML}\footnote{https://github.com/ChrisCummins/ProGraML}~\citesecondary{cummins2020programl2} is a benchmark dataset that converts programs into graphs based on LLVM-IRs, consisting of 250$k$ LLVM-IR files covering six programming languages. It can be used for evaluating GNNs for program optimization and analysis.
	\item \textbf{FunCom}\footnote{http://leclair.tech/data/funcom/}~\citesecondary{leclair2019recommendations2} provides a curated dataset for code summarization, which excludes auto-generated source code files, as well all functions that are lack of associated comments.
	\item \textbf{CoDesc}\footnote{https://github.com/code-desc/CoDesc}~\citesecondary{hasan2021codesc2} is a large-scale parallel dataset composed of 4.2 million Java methods and natural language descriptions, based on CodeSearchNet, DeepCom, CONCODE and FunCom, with comprehensive data cleaning processes and manually check.
	\item \textbf{APPS}\footnote{https://github.com/hendrycks/apps}~\citesecondary{hendrycks2021measuring} is a benchmark dataset for code generation in Python based on natural language specification, consisting of 10,000 problems at various levels of difficulty, covering simple introductory problems, interview-level problems, and coding competition challenges.
	\item \textbf{AVATAR}\footnote{https://github.com/wasiahmad/AVATAR}~\citesecondary{ahmad2021avatar2} is a parallel corpus for Java-Python program translation. It is composed of 8,475 programming problems and their solutions implemented in Java and Python, collected from open source programming contest sites (e.g., AtCoder, Google Code Jam, and Codeforces), and online platforms (e.g., GeeksforGeeks, LeetCode, and Project Euler).
    \item \textbf{PyTorrent}~\citesecondary{bahrami2021pytorrent2} is a Python dataset of paired source code and natural-language comment, which is similar to the CodeSearchNet dataset. The difference is that it is collected from Python libraries such as PyPI and Anaconda packages, rather than GitHub projects.
	\item \textbf{StaQC}~\citesecondary{yao2018staqc2} is a collection of about 148$k$ Python and 120$k$ SQL question-code pairs, mined from Stack Overflow.
	\item \textbf{CodeQA}\footnote{https://github.com/jadecxliu/CodeQA}~\citesecondary{liu2021codeqa2} is a free-form question-answering dataset for code comprehension, in which a code snippet and a question are given, and a textual answer is required to be generated. It is composed of a Java dataset with 119,778 question-answer pairs and a Python dataset with 70,085 question-answer pairs. 
	\item \textbf{CodeXGLUE}\footnote{https://github.com/microsoft/CodeXGLUE}~\citesecondary{lu2021codexglue2} is a benchmark dataset for code understanding and generation, which can support multiple code intelligence tasks, such as code summarization, code completion, code search, and program translation. 
\end{itemize}

\section{Evaluation Metrics}
We have evaluated the performance of our proposed model based on four widely-used evaluation criteria in the area of neural machine translation, i.e., BLEU~\citesecondary{papineni2002bleu}, METEOR~\citesecondary{banerjee2005meteor}, and ROUGE~\citesecondary{lin2004rouge}. 
BLEU measures the average n-gram precision on a set of reference sentences, with a penalty for short sentences. METEOR is recall-oriented and measures how well our model captures the contents from the references in our output. ROUGE considers sentence-level structure similarity and identifies the longest co-occurring in sequence n-grams automatically. 
	
\noindent\textbf{BLEU (Bilingual Evaluation Understudy) }
is a classical evaluation metric in neural machine translation, which measures the average n-gram precision on a set of reference sentences, with a penalty for short sentences. 
\begin{equation}
    p_n = \frac{\sum_{w_n\in a} \min \left ( c_a(w_n), \underset{j=1,\cdots,|n|}{\max}c_{b_j}(w_n) \right )}{\sum_{w_n\in a} c_a(w_n)},
\end{equation}
where $a$ and $b$ represent a candidate sentence and its corresponding reference, respectively. $w_n$ represents the n-gram word, $c_a(w_n)$ represents the count of n-gram $w_n$ in sentence $a$. Since $p_n$ encourages to prediction of short sentences, thus, we introduce a Brevity Penalty (BP) defined as follows:
\begin{equation}
    \operatorname{BP}=\left\{\begin{matrix}
	1 & if\ c>r\\ 
	e^{(1-\frac{r}{c})}& if \ c \leq r,		
    \end{matrix}\right.
\end{equation}
where $r$ is the reference sentence length, $c$ the length of the candidate sentence. Finally, the BLEU is calculated as follows:
\begin{equation}
    \operatorname{BLEU} = \operatorname{BP}* \exp \left(\sum_{n=1}^{N}\alpha_n\log p_n\right),
\end{equation}
where $N=1,2,3,4$, $p_n$ is the n-gram precision of n-grams up to $N$, $\alpha_n$ is positive weight for each gram.
	
\noindent\textbf{METEOR} is recall-oriented and measures how well our model captures content from the references in our output. METEOR matches candidate sentences and reference sentence words one by one, and then calculates the harmonic average of their accuracy and recall rates between the corresponding candidate sentence and the reference sentence. It is calculated as follows:
\begin{equation}
    \operatorname{METEOR} = \underset{j=1,\cdots,|b|}{\max}\left ( \frac{10PR}{R+9P} \right )\left ( 1-\frac{1}{2} \left ( \frac{\#\text{chunks}}{\#\text{matched\ uni-gram}} \right )\right ),
\end{equation}
where $P$ = uni-gram precision, $R$ =uni-gram recall, $\#chunks$ is the number of matched chunks between two sentences.  
$\#\operatorname{matched\ unigram}$ is the number of matched uni-grams between the candidate sentence and the reference sentence.

\noindent\textbf{ROUGE} is one of the general metrics in the automatic text summarization, it is a measurement based on the longest common sub-sequence (Longest Common Sub-sequence, LCS).
ROUGE takes into account sentence-level structure similarity naturally and identifies the longest co-occurring in sequence n-grams automatically. It is calculated as follows:
\begin{equation}
    \operatorname{ROUGE} = \frac{\sum_{j=1}^{|b|} \sum_{w_n\in b_j} \min \left ( c_a(w_n), c_{b_j}(w_n) \right )}{\sum_{j=1}^{|b|} \sum_{w_n\in b_j} c_{b_j}(w_n)},
\end{equation}

\noindent\textbf{MRR (Mean Reciprocal Rank)} serves as a metric for assessing processes that generate lists of potential responses to a set of queries, arranged based on their likelihood of correctness. This metric computes the average reciprocal ranks for the outcomes associated with a given set of queries. The reciprocal rank of a response to a query corresponds to the inverse of the rank assigned to the first correct answer within the list.
The MRR is formulated as follows:
\begin{equation}
\operatorname{MRR} = \frac{1}{Q} \sum_{i=1}^{Q} \frac{1}{rank_i}\,,
\end{equation}
where $Q$ is the number of queries, and 
$rank_i$ is the rank position of the first correct answer for the $i$-th query.

\noindent\textbf{Accuracy} stands as a pivotal metric for evaluating the efficacy of a classification model. This metric is delineated by the ratio of correct predictions to the total number of predictions generated by the model, as follows:
\begin{equation}
\operatorname{Accuracy} = \frac{\text{Number of Correct Predictions}}{\text{Total Number of Predictions Made}}\,.
\end{equation}

\bibliographystylesecondary{unsrt} 
\bibliographysecondary{refappendix} 


\end{document}